\documentclass[a4]{elsart}

\usepackage{epsfig}
\usepackage{subfigure}
\usepackage{wrapfig}

\usepackage[breaklinks]{hyperref}

\usepackage{amsmath}
\usepackage{amssymb}
\usepackage{cite}
\usepackage{array}

\usepackage{textcomp}

\usepackage{setup} 

\usepackage{fancyhdr}
\renewcommand{\headsep}{1cm}
\newcommand{\ul}{\underline}

\numberwithin{equation}{section}
\numberwithin{figure}{section}

\newcommand{\beqs}{\begin{equation*}}
\newcommand{\beq}{\begin{equation}}

\newcommand{\eeqs}{\end{equation*}}
\newcommand{\eeq}{\end{equation}}

\newcommand{\beqas}{\begin{eqnarray*}}
\newcommand{\beqa}{\begin{eqnarray}}

\newcommand{\eeqas}{\end{eqnarray*}}
\newcommand{\eeqa}{\end{eqnarray}}

\newcommand{\seq}[5]{\newline 
\parbox{#1}{\begin{eqnarray*} #2 \end{eqnarray*}} \hfill
\parbox{#3}{\begin{eqnarray*} #4 \end{eqnarray*}} \hfill
\parbox{1cm}{\begin{equation} \label{#5} \end{equation}} 
\newline}

\newcommand{\eq}[2]{\begin{equation} #1 \label{#2} \end{equation}}

\newcommand{\eqa}[2]{\begin{eqnarray} #1 \label{#2} \end{eqnarray}}

\newcommand{\meq}[2]{\begin{multline} #1 \label{#2} \end{multline}}

\newcommand{\eps}{\varepsilon}
\newcommand{\al}{\alpha}
\newcommand{\be}{\beta}
\newcommand{\ga}{\gamma}
\newcommand{\de}{\delta}
\newcommand{\om}{\omega}
\newcommand{\ka}{\kappa}
\newcommand{\la}{\lambda}
\newcommand{\si}{\sigma}

\newcommand{\Om}{\Omega}

\newcommand{\La}{\Lambda}

\newcommand{\blist}{\begin{itemize}}

\newcommand{\elist}{\end{itemize}}

\providecommand{\href}[2]{#2}

\DeclareFontFamily{OT1}{rsfs}{}
\DeclareFontShape{OT1}{rsfs}{m}{n}{ <-7> rsfs5 <7-10> rsfs7 <10->rsfs10}{} 
\DeclareMathAlphabet{\mycal}{OT1}{rsfs}{m}{n}
\newcommand{\scri}{{\mycal I}}

\DeclareMathAlphabet{\matheurm}{U}{eur}{m}{n}  
\DeclareMathAlphabet{\matheubf}{U}{eur}{b}{n}  
\usepackage[mathscr]{eucal}                    

\usepackage{xspace} 

\newcommand{\gcycl}{\operatorname{gcycl}}

\newcommand{\LC}{\Omega} 

\newcommand{\BMf}{\mathcal{M}}

\newcommand{\Poisson}{\mathcal{P}}

\newcommand{\casimir}{c}

\newcommand{\Action}{L}

\DeclareMathAlphabet{\matheurm}{U}{eur}{m}{n}  

\renewcommand{\^}{{}^}
\renewcommand{\_}{\!{}_}

\newcommand{\rvec}[1]{\overset{\,
    \raisebox{-0.5ex}{$\scriptscriptstyle \rightarrow$}}{#1}{}}

\newcommand{\rpartial}{\rvec{\partial}}

\newcommand{\mtrx}[4]{\left(
    \begin{array}{cc}
      #1 & #2 \\
      #3 & #4
    \end{array}\right)
  }

\begin{document}


\begin{frontmatter}
 
\title{Dilaton Gravity in Two Dimensions}

\author[Wien]{D. Grumiller},\ead{grumil@hep.itp.tuwien.ac.at}
\author[Wien]{W. Kummer},\ead{wkummer@tph.tuwien.ac.at}
\author[Leipzig,SPb]{D.V.\ Vassilevich}\ead{vassil@itp.uni-leipzig.de}
\address[Wien]{Institut f\"ur Theoretische Physik, TU Wien, 
                 Wiedner Hauptstr.\  8--10, A-1040 Wien, Austria}
\address[Leipzig]{Institut f\"ur Theoretische Physik, Universit\"at Leipzig,
                 Augustusplatz 10, D-04109 Leipzig, Germany}
\address[SPb]{V.A.\ Fock Insitute of Physics, St. Petersburg University, 198904
St.\ Petersburg, Russia}

\begin{abstract}
The study of general two dimensional models of gravity allows to tackle 
basic questions of quantum gravity, bypassing important technical 
complications which make the treatment in higher dimensions difficult. 
As the physically important examples of spherically symmetric Black Holes, 
together with string inspired models, belong to this class, 
valuable knowledge can also be gained for these
systems in the quantum case. In the last decade new insights regarding 
the exact quantization of the geometric part of such theories have been 
obtained. They allow a systematic quantum field theoretical treatment, 
also in interactions with matter, without explicit introduction of a 
specific classical background geometry. The present review tries to 
assemble these results in a coherent manner, putting them at 
the same time into the perspective of the quite large literature on 
this subject.
\end{abstract}

\begin{keyword}
dilaton gravity \sep quantum gravity \sep black holes \sep two dimensional models

\PACS 04.60.-w \sep 04.60.Ds \sep 04.60.Gw \sep 04.60.Kz \sep 04.70.-s \sep 04.70.Bw \sep 04.70.Dy \sep  11.10.Lm \sep 97.60.Lf
\end{keyword}
\end{frontmatter}

\newpage
\thispagestyle{plain}
\tableofcontents
\thispagestyle{plain}
\listoffigures
\thispagestyle{plain}
\newpage


\thispagestyle{fancy}
\newpage

\section{Introduction}\label{se:1}

%
%


The fundamental difficulties encountered in the numerous attempts to merge quantum
theory with General Relativity by now are well-known even far outside the narrow
circle of specialists in these fields. Despite many valiant efforts and new
approaches like loop quantum gravity \cite{Rovelli:1998yv} or string theory\footnote{%
The recent book \cite{Polchinski:1998rq} can be recommended.
} a final solution is not in sight. However, even many special questions search
an answer\footnote{%
A brief history of quantum gravity can be found in ref. \cite{Rovelli:1998yv}.
}. 

Of course, at energies which will be accessible experimentally in the foreseeable
future, due to the smallness of Newton's constant, respectively the large value
of the Planck mass, an effective quantum theory of gravity can be constructed
\cite{donoghue:1994dn} in a standard way which in its infrared asymptotical
regime as an \textit{effective} quantum theory may well describe our low energy
world. Its extremely small corrections to classical General Relativity (GR)
are in full agreement with experimental limits \cite{Will:2001mx}. However,
the fact that Newton's constant carries a dimension, inevitably makes perturbative
quantum gravity inconsistent at energies of the order of the Planck mass.

In a more technical language, starting from a fixed classical background, already
a long time ago perturbation theory has shown that although pure gravity is
one-loop renormalizable \cite{'tHooft:1974bx} this renormalizability breaks
down at two loops \cite{Goroff:1986th}, but already at one-loop when matter
interactions are taken into account. Supergravity was only able to push the
onset of non-renormalizability to higher loop order (cf. e.g. 
\cite{Howe:1984sr,Bern:1998ug,Deser:1998jz}).
It is often argued that a full treatment of the metric, including non-perturbative
effects from the backreaction of matter, may solve the problem but to this day
this remains a conjecture\footnote{%
For a recent argument in favor of this conjecture using Weinberg's argument
of {}``asymptotic safety{}'' cf.\ e.g.\  \cite{Lauscher:2001rz}.
}. A basic conceptual problem of a theory like gravity is the double role of
geometric variables which are not only fields but also determine the (dynamical)
background upon which the physical variables live. This is e.g.\ of special
importance for the uncertainty relation at energies above the Planck scale leading
to Wheeler's notion of {}``space-time-foam{}'' \cite{Wheeler:1964}.

Another question which has baffled theorists is the problem of time. In ordinary
quantum mechanics the time variable is set apart from the {}``observables{}'',
whereas in the straightforward quantum formulation of gravity (the so-called
Wheeler-deWitt equation \cite{Wheeler:1968,DeWitt:1967yk}) a variable like
time must be introduced more or less by hand through {}``time-slicing{}'',
a multi-fingered time etc.\ \cite{Isham:1991mm}. Already at the classical level
of GR {}``time{}'' and {}``space{}'' change their roles when passing through a
horizon which leads again to considerable complications in a Hamiltonian 
approach \cite{Ashtekar:1974, Kuchar:1980ht}.

Measuring the ``observables'' of usual quantum mechanics one realizes that the 
genuine measurement process is related always to a determination of the matrix 
element of some scattering operator with asymptotically defined ingoing and 
outgoing states. For a gauge theory like gravity, existing proofs of 
gauge-independence for the $S$-matrix \cite{Kummer:2001ip} may be applicable for 
asymptotically flat quantum gravity systems. But the problem of other 
experimentally accessible (gauge independent!) genuine observables is 
open, when the dynamics of the geometry comes into play in a nontrivial 
manner, affecting e.g. the notion what is meant by asymptotics.

The quantum properties of black holes (BH) still pose many questions.
Because of the emission of Hawking radiation 
\cite{Hawking:1975sw,Unruh:1976db},
a semi-classical effect, a BH should successively lose energy. If there is
no remnant of its previous existence at the end of its lifetime, the 
information of pure states swallowed
by it will have only turned into the mixed state of Hawking radiation, 
violating basic notions of quantum mechanics.
Thus, of special interest (and outside the range of methods based upon the 
fixed background of a large BH) are the last stages of BH evaporation. 

Other open problems -- related to BH physics and more generally to quantum 
gravity -- have been the virtual BH appearing as an intermediate stage in 
scattering processes, the (non-)existence of a well-defined $S$-matrix and 
$CPT$ (non-)invariance. When the metric of the BH is quantized its 
fluctuations may include {}``negative{}'' volumes. Should those fluctuations 
be allowed or excluded? The intuitive notion of ``space-time foam'' seems to 
suggest quantum gravity induced topology fluctuations. Is it possible to 
extract such processes from a model without {\em ad hoc} assumptions? 
From experience of quantum field theory in Minkowski space one may hope that a 
classical singularity like the one in the Schwarzschild BH may be eliminated 
by quantum effects -- possibly at the price of a necessary renormalization 
procedure. Of course, the latter may just reflect the fact that interactions 
with further fields (e.g. other modes in string theory) are not taken into 
account properly. Can this hope be fulfilled?

In attempts to find answers to these questions it seems very reasonable to 
always try to proceed as far as possible with the known laws of quantum 
mechanics applied to GR. This is extremely difficult\footnote{%
A recent survey of the present situation is the one of Carlip 
\cite{Carlip:2001wq}. 
} in \( D=4 \). Therefore, for many years a rich literature developed on lower
dimensional models of gravity. The \( 2D \) Einstein-Hilbert action is just
the Gauss-Bonnet term. Therefore, intrinsically \( 2D \) models are locally
trivial and a further structure is introduced. This is provided by the dilaton
field which naturally arises in all sorts of compactifications from higher 
dimensions.
Such models, the most prominent being the one of Jackiw and Teitelboim (JT), 
were thoroughly investigated during the 1980-s 
\cite{Barbashov:1979bm,D'Hoker:1982er,Teitelboim:1983ux,D'Hoker:1983is, 
D'Hoker:1983ef,Jackiw:1985je,Katanaev:1986wk,Katanaev:1990qm,Mann:1990gh,
Sikkema:1991ib}. An excellent summary (containing also a more comprehensive
list of references on literature before 1988) is contained in the textbook of
Brown \cite{Brown:1988}. Among those models spherically reduced gravity (SRG),
the truncation of \( D=4 \) gravity to its s-wave part, possesses perhaps the
most direct physical motivation. One can either treat this system directly in
\( D=4 \) and impose spherical symmetry in the equations of motion (e.o.m.-s)
\cite{Kuchar:1994zk} or impose spherical symmetry already in the action
\cite{Berger:1972pg,Unruh:1976db,Benguria:1977in,Thomi:1984na,Hajicek:1984mz,
Mignemi:1989qc,Thiemann:1993jj,Kastrup:1994br,Kuchar:1994zk,Lau:1996fr,
Grumiller:1999rz}, thus obtaining a dilaton theory\footnote{%
The dilaton appears due to the {}``warped product{}'' structure of the metric.
For details of the spherical reduction procedure we refer to appendix 
\ref{app:A}.}. Classically, both approaches are equivalent.

The rekindled interest in generalized dilaton theories in \( D=2 \) (henceforth
GDTs) started in the early 1990-s, triggered by the string inspired \cite{Mandal:1991tz,Elitzur:1991cb,Witten:1991yr,Dijkgraaf:1992ba, McGuigan:1992qp,Ishibashi:1991wh,DeAlwis:1991vz,Khastgir:1991ip}
dilaton black hole model\footnote{%
A textbook-like discussion of this model can be found in refs. 
\cite{Giddings:1994pj,Strominger:1994tn}.
}, studied in the influential paper of Callan, Giddings, Harvey and Strominger
(CGHS) \cite{Callan:1992rs}. At approximately the same time it was realized
that \( 2D \) dilaton gravity can be treated as a non-linear gauge-theory \cite{Verlinde:1991rf,Ikeda:1993aj}.

As already suggested by earlier work, all GDTs considered so far could be extracted
from the dilaton action \cite{Russo:1992yg,Odintsov:1991qu}\begin{equation}
\label{eq:GDT}
L^{(\textrm{dil})}=\int d^{2}x\, \sqrt{-g}\; \left[ X \frac{R}{2}-\frac{U(X)}{2}\; (\nabla X)^{2}+V(X)\; \right] +L^{(m)}\, ,
\end{equation}
 where \( R \) is the Ricci-scalar, \( X \) the dilaton, \( U(X) \) and \( V(X) \)
arbitrary functions thereof, \( g \) is the determinant of the metric \( g_{\mu \nu } \),
and \( L^{(m)} \) contains eventual matter fields.

When $U(X)=0$ the e.o.m. for the dilaton from (\ref{eq:GDT}) is algebraic. For
invertible $V'(X)$  
the dilaton field can be eliminated altogether, and the Lagrangian
density is given by an arbitrary function of the Ricci-scalar. A recent review
on the classical solution of such models is ref. \cite{Schmidt:1999wb}. In
comparison with that, the literature on such models generalized to depend also\footnote{
For the definition of the Lorentz scalar formed by torsion and of the curvature
scalar, both expressed in terms of Cartan variables zweibeine \( e^{a}_{\mu } \)
and spin connection \( \omega ^{ab}_{\mu } \) we refer to sect.
\ref{se:2.1} below.
} on torsion $T^a$ is relatively scarce. It mainly consists of elaborations based upon a theory
proposed by Katanaev and Volovich (KV) which is quadratic in curvature and torsion
\cite{Katanaev:1986wk,Katanaev:1990qm}, also known as {}``Poincar\'{e} gauge
gravity{}'' \cite{Mielke:1993nc}.

A common feature of these classical treatments of models with and without 
torsion is the almost exclusive use\footnote{%
A notable exception is Polyakov \cite{Polyakov:1987zb}.} of the gauge-fixing 
for the \( D=2 \) metric familiar from string theory, namely the conformal 
gauge. Then the e.o.m.-s become complicated partial differential
equations. The determination of the solutions, which turns out to be always
possible in the matterless case (\( L^{(m)}=0 \) in (\ref{eq:GDT})), for 
nontrivial dilaton field dependence usually requires considerable mathematical 
effort. The same had been true for the first papers on theories with torsion 
\cite{Katanaev:1986wk,Katanaev:1990qm}.
However, in that context it was realized soon that gauge-fixing is not necessary,
because the invariant quantities \( R \) and \( T^{a}T_{a} \) themselves may
be taken as variables in the KV-model \cite{Solodukhin:1993bn,Solodukhin:1993bs, Solodukhin:1993xs,Solodukhin:1994sv}.
This approach has been extended to general theories with torsion\footnote{%
A recent review of this approach is provided by Obukhov and Hehl \cite{Obukhov:1997uc}.
}. 

As a matter of fact, in GR many other gauge-fixings for the metric have been
well-known for a long time: the Eddington-Finkelstein (EF) gauge, the 
Painlev\'{e}-Gullstrand gauge, the Lema{\capitalcircumflex{\i}}tre gauge 
etc.\,. As compared to the {}``diagonal{}'' gauges like
the conformal and the Schwarzschild type gauge, they possess the advantage
that coordinate singularities can be avoided, i.e.\ the singularities in those
metrics are essentially related to the {}``physical{}'' ones in the curvature.
It was shown for the first time in \cite{Kummer:1992bg} that the use of a 
temporal gauge for the Cartan variables (cf.\ eq. (\ref{eq:a2}) below) in the 
(matterless)
KV-model made the solution extremely simple. This gauge corresponds to the EF
gauge for the metric. Soon afterwards it was realized that the solution could
be obtained even without previous gauge-fixing, either by guessing the Darboux
coordinates \cite{Schaller:1994np} or by direct solution of the e.o.m.-s
\cite{Kummer:1995qv} (cf. sect. \ref{se:2.3.1}). Then the 
temporal gauge of \cite{Kummer:1992bg} merely
represents the most natural gauge fixing within this gauge-independent setting.
The basis of these results had been a first order formulation of \( D=2 \)
covariant theories by means of a covariant Hamiltonian action in terms of the 
Cartan variables and further auxiliary fields \( X^{a} \) which (beside the 
dilaton field \( X \)) take the role of canonical momenta (cf.\ eq. 
(\ref{2.62}) below). They cover a very general class of theories comprising 
not only the KV-model, but also more general theories with torsion\footnote{%
In that case there is the restriction that it must be possible to eliminate
\textit{all} auxiliary fields \( X^{a} \) and \( X \) (see sect. 
\ref{se:2.1.3}).}. The most attractive feature of theories of type (\ref{2.62})
is that an important subclass of them is in a one-to-one correspondence with 
the GDT-s (\ref{eq:GDT}). This dynamical equivalence, including the essential 
feature that also the global properties are exactly identical, seems to have 
been noticed first in \cite{Katanaev:1996bh} and used extensively in studies 
of the corresponding quantum theory 
\cite{Kummer:1997hy,Kummer:1998zs,Kummer:1998jj}.

Generalizing the formulation (\ref{2.62}) to the much more comprehensive class
of {}``Poisson-Sigma models{}'' \cite{Schaller:1994es,Strobl:1994yk} on the
one hand helped to explain the deeper reasons of the advantages from the use
of the first oder version, on the other hand led to very interesting 
applications in other fields \cite{Alekseev:1995py}, including especially also 
string theory \cite{Schomerus:1999ug,Seiberg:1999vs}. Recently this approach 
was shown to represent a very direct route to $2D$ dilaton supergravity 
\cite{Ertl:2000si} without auxiliary fields.

Apart from the dilaton BH \cite{Callan:1992rs} where an exact (classical) solution
is possible also when matter is included, general solutions for generic \( D=2 \)
gravity theories with matter cannot be obtained. This has been possible only
in restricted cases, namely when fermionic matter is chiral\footnote{%
This solution was rediscovered in ref. \cite{Solodukhin:1995ux}.
} \cite{Kummer:1992ef} or when the interaction with (anti)selfdual scalar matter
is considered \cite{Pelzer:1998ea}.

Semi-classical treatments of GDT-s take the one loop correction from matter
into account when the classical e.o.m.-s are solved. They have been used mainly
in the CGHS-model and its generalizations \cite{Bilal:1993kv,DeAlwis:1991vz,
Russo:1992ht,Bose:1995pz,Cruz:1996zt,Fabbri:1998hs,Kim:1999wa,
Zaslavsky:1999zh,Zaslavsky:1998ca,Hamada:1993hh,Hamada:1993zw,Vaz:1997kh}. 
In our present report we concentrate only upon Hawking radiation as a 
quantum effect of matter on a \textit{fixed}
(classical) geometrical background, because just during the last years
interesting insight has been obtained there, although by no means all 
problems have been settled.

Finally we turn to the full quantization of GDTs. It was believed by several
authors (cf. e.g. \cite{Russo:1992yg,Odintsov:1991qu,Kantowski:1992dv,
Elizalde:1994qq,Elizalde:1995zj}) that even in the absence of interactions 
with matter nontrivial quantum corrections exist and can be computed by a 
perturbative path integral on some fixed background. Again the evaluation in 
the temporal gauge \cite{Kummer:1992bg}, at first for the KV-model showed that 
the use of other gauges just obscures a very simple mechanism. Actually all 
divergent counter-terms can be absorbed into one compact expression.
After subtracting that in the absence of matter the solution of the classical
theory represents an exact {}``quantum{}'' result. Later this perturbative
argument has been reformulated as an exact path integral, first again for the
KV-model \cite{Haider:1994cw} and then for general theories of gravity in 
\( D=2 \) \cite{Kummer:1997hy,Kummer:1998zs,Kummer:1998jj,Grumiller:2000ah,
Fischer:2001vz,Grumiller:2001rg}.

In our present review we concentrate on the path integral approach, with Dirac
quantization only referred to for sake of comparison. In any case, the common
starting point is the Hamiltonian analysis which in a theory formulated
in terms of Cartan variables in $D=2$ possesses substantial technical advantages.
The constraints, even in the presence of matter interactions, form an algebra
with momentum-dependent structure constants. Despite that nonlinearity the
simplest version of the Batalin-Vilkovisky procedure \cite{Batalin:1977pb} suffices, namely
the one also applicable to ordinary nonabelian gauge theories in Minkowski space.
With a temporal gauge fixing for the Cartan variables also used in the quantized
theory, the geometric part of the action yields the exact path integral. Possible
background geometries appear naturally as homogeneous solutions of differential
equations which coincide with the classical ones, reflecting {}``local quantum
triviality{}'' of $2D$ gravity theories in the absence of matter, a property 
which had been observed as well before
in the Dirac quantization of the KV-model \cite{Schaller:1994np}.

These features are very difficult to locate in the GDT-formulation 
(\ref{eq:GDT}), but become evident in the equivalent first order version with 
a {}``Hamiltonian{}'' action.  

Of course, non-renormalizability persists in the perturbation expansion when
the matter fields are integrated out. But as an effective theory in cases like
spherically reduced gravity, specific processes can be calculated, relying on
the (gauge-independent) concept of $S$-matrix elements. With this method, 
scattering of s-waves in spherically 
reduced gravity has provided a very direct way to create a {}``virtual{}'' 
BH as an intermediate state without further assumptions \cite{Fischer:2001vz}.

The structure of our present report is determined essentially by the approach
described in the last paragraphs. One reason is the fact that a very 
comprehensive overview of very general classical and quantum theories in 
\( D=2 \) is made possible in this manner. Also a presentation seems to be 
overdue in which results, scattered now
among many different original papers can be integrated into a coherent picture.
Parallel developments and differences to other approaches will be
included in the appropriate places.

\subsection{Structure of this review}

This review is organized as follows: 
\blist
\item Section 1 in its remaining part contains a short primer on differential 
geometry (with special emphasis on \( D=2 \)). \emph{En passant} most of our 
notations are fixed in that subsection. 
\item Section 2 motivates the study of GDTs and introduces
its action in the three most frequently used forms (dilaton action, first order
action, and Poisson-Sigma action) and describes the relations between them. 
\item Section 3 gives all classical solutions of GDTs in the
absence of matter. The global structure of such theories is discussed using
Schwarzschild space-time as a simple example. As a further illustration we consider
a family of dilaton models describing a single black hole in Minkowski, Rindler
or de Sitter space-time. 
\item Section 4 extends the discussion to additional gauge-fields, supergravity and
(bosonic or fermionic) matter fields. 
\item Section 5 considers the role of energy in GDTs. In particular,
the ADM mass, quasilocal energy, an absolute conservation law and its corresponding
N\"{o}ther symmetry are discussed. 
\item Section 6 leaves the classical realm providing a concise treatment of (semi-classical)
Hawking radiation for minimally and non-minimally coupled matter. 
\item Section 7 is devoted to non-perturbative path integral quantization of the geometric
sector of GDTs with (scalar) matter, giving rise to
a non-local and non-polynomial effective action depending solely on the matter
fields and external sources. The matter sector is treated perturbatively. 
\item Section 8 shows some consequences of the previously developed perturbation theory:
the virtual black hole phenomenon, the appearance of non-local vertices, and
\( S \)-matrix elements for \( s \)-wave gravitational scattering. 
\item Section 9 describes the status of Dirac quantization for a typical 
example of that approach.
\item Section 10 concludes with a brief summary and an outlook regarding open questions. 
\item Appendix A recalls the spherical reduction procedure in the Cartan formalism. 
\item Appendix B collects some basic properties of the heat kernel expansion needed
in Section 6. 
\elist
Several topics are closely related to the subject of this
review, but are \textit{not} included: 
\begin{enumerate}
\item Various calculations and explanations of the BH entropy \cite{Frolov:1998vs,Peet:1998es}
became a large and rather independent field of research which shows, however,
overlaps \cite{Frolov:1999my,Frolov:1996hd}
with the general treatment of the dilaton theories presented in
this review. We do not cover approaches which imply further physical 
assumptions which transgress the orthodox application of quantum theory to 
gravity \cite{Berezin:1998fn,Berezin:1998xf,Bojowald:1999ex,Barvinsky:1996hr,
Bekenstein:1974}.
\item The ideas of the holographic principle \cite{'tHooft:1993gx,Susskind:1995vu}
and of the AdS/CFT correspondence \cite{Maldacena:1998re,Gubser:1998bc,Witten:1998qj}
are now being actively applied to BH physics (see, e.g. \cite{Sachs:2001qb}
and references therein). 
\item There exist different approaches to integrability of gravity models in 
two dimensions \cite{Nicolai:1996pd,Korotkin:1998fi,Korotkin:1998ps,
Nicolai:1998gi}. In particular, a rather sophisticated technique has been 
applied to solve the effective $2D$ models emerging after toroidal reduction 
(instead of the spherical reduction considered in this review) of the 
four-dimensional Einstein equations \cite{Bernard:1999yn,Varzugin:2000gh}.
Recently again interesting developments should be noted in Liouville gravity
\cite{Faddeev:2000if,Teschner:2001rv}. Some relations between $2D$ dilaton 
gravity and the theory of solitons were discussed in \cite{Cadoni:1998ej,
Navarro:1998hc}. 
\end{enumerate}
Each of these topics deserves a separate review, and in some cases such reviews
exist. Therefore, we have restricted ourselves in those fields to just a few
(somewhat randomly selected) references which hopefully will permit further
orientation.

\subsection{Differential geometry}

\label{se:2.1}

\subsubsection{Short primer for general dimensions}

\label{se:2.1.1}

In the comprehensive approach advocated for \( D=2 \) gravity the use of Cartan
variables (zweibeine, spin-connection) plays a pivotal role. As an introduction
and in order to fix our notations we shall review briefly this formalism. For
details we refer to the mathematical literature (cf. e.g. \cite{nakaharageometry}).

On a manifold with \( D \) dimensions in each point one introduces vielbeine
\( e^{\mu }_{a}(x) \), where Greek indices refer to the (holonomic) coordinates
\( x^{\mu }=(x^{0},x^{1},\dots ,x^{D-1}) \) and Latin indices denote the ones
related to a (local) Lorentz frame with metric 
$\eta=\mbox{diag}~(1,-1,\dots,-1)$. 
The dual vector space is spanned by the inverse vielbeine\footnote{%
For simplicity we shall use indiscriminately the term {}``vielbein{}''
for the vielbein, the inverse vielbein and the dual basis of 1-forms (the components
of which are given by the inverse vielbein) whenever the meaning is clear either
from the context or from the position of indices.
} \( e^{a}_{\mu }(x) \): \begin{equation}
\label{2.29}
e^{\mu a}e_{\mu }^{b}=\eta ^{ab}
\end{equation}
\( SO(1,D-1) \) matrices \( L^{a}{}_{b}(x) \) of the (local) Lorentz 
transformations obey 
\begin{equation}
\label{2.1}
{L^{a}}_{c}\, {L_{b}}^{c}\; =\; \delta^a_b\; .
\end{equation}
 A Lorentz vector \( V^{a}=e^{a}_{\mu }V^{\mu } \) transforms under local Lorentz
transformations as \begin{equation}
\label{2.2}
{V'}^{a}(x)={L^{a}}_{b}(x)V^{b}(x)
\end{equation}
 This implies a covariant derivative \begin{equation}
\label{2.3}
{(D_{\mu })^{a}}_{b}\; =\; \delta ^{a}_{b}\, \partial _{\mu }+\omega _{\mu }{}^{a}{}_{b}\; ,
\end{equation}
 if the spin-connection \( \omega _{\mu }{}^{a}{}_{b} \) is introduced as the
appropriate gauge field with transformation \begin{equation}
\label{2.4}
\omega'_{\mu}{\, }^{a}{}_{b}\; =-L{}_{b}{}^{d}\; (\partial _{\mu }L^{a}{}_{d})+L^{a}{}_{c}\, \omega _{\mu }{}^{c}{}_{d}\, L_{b}{}^{d}\; .
\end{equation}
 The infinitesimal version of (\ref{2.4}) follows from \( L^{a}{}_{b}=\delta ^{a}{}_{b}+l^{a}{}_{b}+\mathcal{O}(l^{2}) \)
where \( l^{a}{}_{b}=-l_{b}{}^{a} \) .

Formally also diffeomorphisms 
\begin{equation}
\label{2.4a}
{\bar{x}}^{\mu }(x)=x^{\mu }-\xi ^{\mu }(x)+\mathcal{O}(\xi ^{2})
\end{equation}
can be interpreted, at least locally, as gauge transformations, when the
Lie variation is employed which implies a transformation referring to the same
point. In \begin{equation}
\label{2.4b}
\frac{\partial {\bar{x}}^{\mu }}{\partial x^{\nu }}=\delta _{\nu }^{\mu }-\xi ^{\mu }_{,\, \nu }\; ,\hspace {0.5cm}\frac{\partial x^{\nu }}{\partial {\bar{x}}^{\mu }}=\delta ^{\nu }_{\mu }+\xi ^{\nu }_{,\, \mu }
\end{equation}
 partial derivatives with respect to \( x^{\nu } \) have been abbreviated by
the index after a comma.

For instance, for the Lie variation of a tensor of first order \( {\bar{V}}_{\mu }(\bar{x})=\frac{\partial x^{\nu }}{\partial {\bar{x}}^{\mu }}\, V_{\nu }(x) \)
one obtains \begin{equation}
\label{2.5}
\delta _{\xi }V_{\mu }(x)={\bar{V}}_{\mu }(x)-V_{\mu }(x)={\xi ^{\nu }},{}_{\mu }\, V_{\nu }+\xi ^{\nu }V_{\mu },{}_{\nu }\; .
\end{equation}
For the dual to the tangential space, e.g.\ \( V^{\mu }\partial _{\mu }=V^{\mu }(\partial _{\mu }{\bar{x}}^{\nu })\bar{\partial}_{\nu }={\bar{V}}^{\mu }{\bar{\partial }}_{\mu } \)
one derives the analogous transformation \begin{equation}
\label{2.6}
\delta _{\xi }V^{\mu }={\bar{V}}^{\mu }(x)-V^{\mu }(x)=-{\xi ^{\mu }},{}_{\nu }V^{\nu }+\xi ^{\nu }{V^{\mu }},{}_{\nu }\; .
\end{equation}

The metric \( g_{\mu \nu } \) in the line element is a quadratic expression
of the vielbeine \begin{equation}
\label{2.7}
(ds)^{2}=g_{\mu \nu }\, dx^{\mu }dx^{\nu }=e_{\mu }^{a}e_{\nu }^{b}\, \eta _{ab}\, dx^{\mu }dx^{\nu }\, ,
\end{equation}
and, therefore, a less elementary variable. Also the re\-para\-metriza\-tion
invariant volume element \begin{multline}
\sqrt{(-)^{D-1}g}\; d^{D}\, x = \sqrt{(-)^{D-1}\det g_{\mu\nu}} \, d^{D}\, x = \\
= \sqrt{(-)^{D-1}(\det e_{\mu}^{a})^{2}\, \det \eta}\, d^{D}\, x = \vert \det e_{\mu}^{a} 
\vert \, d^{D}{}x  = |e|\, d^{D}{}x
\label{2.10}
\end{multline} is of polynomial form if expressed in vielbein components.

The advantage of the form calculus \cite{nakaharageometry} is that diffeomorphism
invariance is automatically implied, when the Cartan variables are converted
into one forms \begin{equation}
\label{2.11}
e_{\mu }^{a}\; \rightarrow \; e^{a}=e^{a}_{\mu }dx^{\mu },\hspace {0.5cm}\omega _{\mu }{}^{a}{}_{b}\; \rightarrow \; \omega ^{a}{}_{b}=\omega _{\mu }{}^{a}{}_{b}\, dx^{\mu }
\end{equation}
 which are special cases of p-forms \begin{equation}
\label{2.12}
\Omega _{p}=\frac{1}{p!}\, \Omega _{\mu _{1}\, \dots \; \mu _{p}}\, dx^{\mu _{1}}\wedge dx^{\mu _{2}}\wedge \dots \wedge dx^{\mu _{p}}\; .
\end{equation}
 Due to the antisymmetry of the wedge product \( dx^{\mu }\wedge dx^{\nu }=dx^{\mu }\otimes dx^{\nu }-dx^{\nu }\otimes dx^{\mu }=-dx^{\nu }\wedge dx^{\mu } \)
all totally antisymmetric tensors \( \Omega _{\mu _{1}\dots \mu _{p}} \) are
described in this way. Clearly \( \Omega _{p}=0 \) for \( p>D \). The action
of the \( (p+q) \)-form \( \Om _{q}\wedge \Om _{q} \) on \( p+q \) vectors
is defined by \begin{equation}
\label{2.12a}
\Omega _{p}\wedge \Xi _{q}(V_{1},\dots ,V_{p+q})=\frac{1}{p!q!}\sum _{\pi }\de _{\pi }\Omega (V_{\pi (1)},\dots ,V_{\pi (p)})\Xi (V_{\pi (p+1)},\dots ,V_{\pi (p+q)}),
\end{equation}
 where the sum is taken over all permutations \( \pi  \) of \( 1,\dots ,p+q \)
and \( \de _{\pi } \) is \( +1 \) for an even number of transpositions and
\( -1 \) for an odd number of transpositions. It is convenient at this point
to introduce the condensed notation for (anti)symmetrization: \eq{
\alpha_{[\mu_{1} \dots \mu_{p}]}:=\frac{1}{p!}\sum_{\pi }\de_{\pi }\alpha_{a_{\pi(1)}
\dots a_{\pi(p)}} ,\hspace{0.5cm} \sigma_{(\mu_{1} \dots \mu_{p})}:=\frac{1}{p!}
\sum_{\pi }\sigma_{a_{\pi(1)}\dots a_{\pi(p)}},
}{2.asy} where the sum is taken over all permutations \( \pi  \) of \( 1,\dots ,p \)
and \( \de _{\pi } \) is defined as before. In the volume form \begin{equation}
\label{2.15}
\Omega _{p=D}=\frac{1}{D!}a_{[\mu _{1}\dots \mu _{D}]}\; \tilde{\epsilon }^{\mu _{1}\, \dots \; \mu _{D}}\, d^{D}x=\frac{1}{D!}a_{[\mu _{1}\, \dots \; \mu _{D}]}\, |e|\epsilon ^{\mu _{1}\, \dots \; \mu _{D}}d^{D}x
\end{equation}
 the product of differentials must be proportional to the totally antisymmetric
Levi-Civit\'{a} symbol \( \tilde{\epsilon }^{01\dots (D-1)}=-1 \) or, alternatively,
to the tensor \( \epsilon =|e|^{-1}\tilde{\epsilon } \) (cf.\ (\ref{2.10})).
The integral of the volume form \( \int _{\mathcal{M}_{D}}\Omega _{D} \) on
the manifold \( \mathcal{M}_{D} \) contains the scalar \( a=a_{[\mu _{1}\, \dots \; \mu _{D}]}\, \epsilon ^{\mu _{1}\, \dots \; \mu _{D}} \)
which is the starting point to construct diffeomorphism invariant Lagrangians.

 By means of the metric (\ref{2.7}) a mixed \( \epsilon  \)-tensor \begin{equation}
\label{2.16}
{\epsilon _{\mu _{1}\, \dots \; \mu _{p}}}{}^{\mu _{p+1}\, \dots \; \mu _{p+q}}\; =\; g_{\mu _{1}\nu _{1}}\, g_{\mu _{2}\nu _{2}}\, \dots \, g_{\mu _{p}\nu _{p}}\, \epsilon ^{\nu _{1}\, \dots \, \nu _{p}\mu _{p+1}\dots \mu _{p+q}}
\end{equation}
 can be defined which allows the introduction of the Hodge dual of \( \Omega _{p} \)
as a \( D-p \) form \begin{equation}
\label{2.17}
{}^{\ast }\, \Omega _{p}\; =\; {\Omega' }_{D-p}=\frac{1}{p!(D-p)!}\epsilon _{\mu _{1}\, \dots \, \mu _{D-p}}{}^{\nu _{1}\, \dots \, \nu _{p}}\; \Omega _{\nu _{1}\, \dots \, \nu _{p}}\; dx^{\mu _{1}}\, \wedge \, \dots \, \wedge dx^{\mu _{D-p}}\; .
\end{equation}
 In \( D=\rm even\) and for Lorentzian signature we obtain for a p-form \eq{
\ast\ast\Om_{p}=(-1)^{p+1}\Om_{p}.
}{2.starstar} The exterior differential one form \( d=dx^{\mu }\partial _{\mu } \)
with \( d^{2}=0 \) increases the form degree by one: \begin{equation}
\label{2.18}
d\Omega _{p}=\frac{1}{p!}\partial _{\mu }\Omega _{\mu _{1}\dots \mu _{p}}dx^{\mu }\wedge dx^{\mu _{1}}\wedge \dots \wedge dx^{\mu _{p}}
\end{equation}
 Onto a product of forms \( d \) acts as \begin{equation}
\label{2.20}
d\left( \Omega _{p}\wedge \Omega _{q}\right) =d\Omega _{p}\wedge \Omega _{q}+(-1)^{p}\Omega _{p}\wedge d\Omega _{q}\; .
\end{equation}
 We shall need little else from the form calculus \cite{nakaharageometry} except
the Poincar\'{e} Lemma which says that for a closed form, obeying \( d\Omega _{p}=0 \),
in a certain ({}``star-shaped{}'') neighborhood of a point \( x^{\mu } \)
on a manifold \( \mathcal{M} \), \( \Omega _{p} \) is exact, i.e.\ can be
written as \( \Omega _{p}=d{\Omega' }_{p-1} \). 

In order to simplify our notation we shall drop the \( \wedge  \) symbol whenever
the meaning is clear from the context.

The Cartan variables expressed as one forms (\ref{2.11}) in view of their Lorentz-tensor
properties are examples of algebra valued forms. This is also the case for the
covariant derivative (\ref{2.3}), now written as \begin{equation}
\label{2.21}
D^{a}{}_{b}\; =\; \delta ^{a}_{b}d+\omega ^{a}{}_{b}\; ,
\end{equation}
 when it acts on a Lorentz vector.

From (\ref{2.11}) and (\ref{2.21}) the two natural quantities to be defined
on a manifold are the torsion two-form \begin{equation}
\label{2.22}
T^{a}\; =\; D^{a}{}_{b}\, e^{b}
\end{equation}
 ({}``First Cartan's structure equation{}'') and the curvature two-form \begin{equation}
\label{2.23}
R^{a}{}_{b}=D^{a}{}_{c}\, \omega ^{c}{}_{b}\; 
\end{equation}
 ({}``Second Cartan's structure equation{}''). From (\ref{2.21}) immediately
follows \begin{equation}
\label{2.24}
(D^{2})^{a}{}_{b}=D^{a}{}_{c}D^{c}{}_{b}=R^{a}{}_{b}\, ,
\end{equation}
 Bianchi's first identity. Using (\ref{2.24}) \( D^{3} \) can be written in
two equivalent ways, \begin{equation}
\label{2.25}
D^{a}{}_{b}\, R^{b}{}_{c}-R^{a}{}_{b}\, D^{b}{}_{c}=0,
\end{equation}
 corresponding to Bianchi's second identity
\begin{equation}
\label{2.26}
(dR^{ab})+\omega ^{a}{}_{c}R^{cb}+\omega ^{b}{}_{c}R^{ac}\; =:\; (DR)^{ab}=0\; .
\end{equation}
 The l.h.s.\ defines the action of the covariant derivative (\ref{2.21}) on
\( R^{ab} \), a Lorentz tensor with two indices. The brackets indicate that
those derivatives only act upon the quantity \( R \) and not further to the
right. 
The structure
equations together with the Bianchi identities show that the covariant action
for any gravity action in \( D \) dimensions depending on \( e^{a},\om ^{a}{}_{b} \)
can be constructed as a volume form depending solely on \( R^{ab},\, T^{a} \)
and \( e^{a} \). The most prominent example is Einstein gravity in \( D=4 \)
\cite{Einstein:1915by,Einstein:1915ca} which in the Palatini formulation reads
\cite{Palatini:1919}\begin{equation}
\label{2.27}
L_{HEP}\; \propto \; \int\limits _{\mathcal{M}_{4}}\; R^{ab}e^{c}e^{d}\; \epsilon _{abcd}\; ,
\end{equation}
 having used the definition \( \epsilon _{abcd}=\epsilon_{\mu \nu \si \tau }e_{a}^{\mu }e_{b}^{\nu }e_{c}^{\si }e_{d}^{\tau } \).
The condition of vanishing torsion \( T^{a}=0 \) for this special case already
follows from varying \( \omega^a{}_b  \) independently in (\ref{2.27}).

In the usual textbook formulations of Einstein gravity, in terms of the metric,
the affine connection \( \Gamma _{\mu \nu }{}^{\rho } \) appears as the only
variable in the covariant derivative, e.g.\ for a contravariant vector \( X^{\nu } \)\begin{equation}
\label{2.28}
X^{\nu }_{;\, \mu }:=\nabla _{\mu }\, X^{\nu }=(\nabla _{\mu })^{\nu }{}_{\rho }\, X^{\rho }=(\partial _{\mu }\delta ^{\nu }_{\rho }+\Gamma _{\mu \rho }{}^{\nu })\, X^{\rho }.
\end{equation}
 In the vielbein basis \( e_{\mu }^{a} \) we relate \( X^{b}=e^{b}_{\rho }\, X^{\rho } \)
and let (\ref{2.3}) act onto that \( X^{b} \). Multiplying by the inverse
vielbein (\ref{2.29}) and comparing with (\ref{2.28}) yields \begin{equation}
\label{2.30}
\Gamma _{\mu \nu }{}^{\rho }\; =\; e_{a}{}^{\rho }\, \left[ (D_{\mu })^{a}{}_{b}\, e^{b}_{\nu }\, \right] \; .
\end{equation}
 The same identification follows, of course, from the covariant derivative of
a covariant vector: \begin{equation}
\label{2.31}
X_{\nu ;\, \mu }:=\partial _{\mu }X_{\nu }-\Gamma _{\mu \nu }{}^{\rho }\; X_{\rho }
\end{equation}

Covariant derivatives may be constructed easily also for tensors with mixed
space-time and local Lorentz indices. For instance, that derivative acting upon
the vielbeine \( e^{\rho }_{c} \)\begin{equation}
\label{2.32}
(\mathcal{D}_{\mu }\, e)_{a}^{\nu }\; =\; \left[ (\mathcal{D}_{\mu })^{\nu }{}_{\rho }\, \right] _{a}{}^{c}\; e_{c}^{\rho }\; :=\; (\nabla _{\mu })^{\nu }{}_{\rho }\, e^{\rho }_{a}+(\omega _{\mu })_{a}{}^{c}\, e_{c}^{\nu }=0
\end{equation}
 is seen to vanish. By (\ref{2.29}) this implies the same result for analogously
defined vielbeine \( e^{a}_{\rho } \)\begin{equation}
\label{2.33}
\left( \mathcal{D}_{\mu }e\right) _{\rho }^{a}=0\; .
\end{equation}
 From (\ref{2.33}) and the antisymmetry of \( \omega ^{a}{}_{b}=-\omega _{b}{}^{a} \)
(one version of metricity) corresponding to its property as a Lorentz generator
of \( SO(1,D-1) \) immediately \begin{equation}
\label{2.34}
\nabla _{\mu }g_{\rho \sigma }=0
\end{equation}
 can be derived, the version of the metricity usually employed in torsionless
theories.

Comparing the antisymmetrized part of the affine connection \( \Gamma _{[\mu \nu ]}{}^{\rho }=\frac{1}{2}\left( \Gamma _{\mu \nu }{}^{\rho }-\Gamma _{\nu \mu }{}^{\rho }\right)  \)
of (\ref{2.30}) with the components of the torsion (\ref{2.22}), multiplied
by the inverse vielbein, shows that the expressions are identical: \begin{equation}
\label{2.35}
e_{a}^{\rho }\, T_{\mu \nu }^{a}=\Gamma _{[\mu \nu ]}{}^{\rho }\; .
\end{equation}
 This allows to express the full affine connection \begin{equation}
\label{2.36}
\Gamma _{\mu \nu }{}^{\rho }=\Gamma _{(\mu \nu )}{}^{\rho }+T_{\mu \nu }{}^{\rho }
\end{equation}
 in terms of Christoffel symbols \( \{\mu ,\nu ,\rho \} \) and the contorsion
\( \mathcal{K} \)\begin{equation}
\label{2.37}
\Gamma _{(\mu \nu )\rho }=g_{\rho \sigma }\, \Gamma _{(\mu \nu )}{}^{\sigma }=\{\, \mu ,\nu ,\rho \}+\mathcal{K}_{(\mu \nu )\rho }
\end{equation}
 by the standard trick of considering (\ref{2.34}) in the form \begin{equation}
\label{2.38}
g_{\nu \rho ,\mu }=\Gamma _{\mu \nu }{}^{\lambda }\, g_{\lambda \rho }+\Gamma _{\mu \rho }{}^{\lambda }\, g_{\lambda \nu }
\end{equation}
 with (\ref{2.37}) and by taking the linear combination of the identity (\ref{2.38})
minus the one for \( g_{\mu \nu ,\rho } \) plus the one for \( g_{\rho \mu ,\nu } \).
In this way the Christoffel symbol \begin{equation}
\label{2.39}
\{\mu ,\nu ,\rho \}=\frac{1}{2}\; \left( g_{\nu \rho ,\mu }+g_{\mu \rho ,\nu }-g_{\mu \nu ,\rho }\right) \; ,
\end{equation}
 but also the additional contorsion contribution \( \mathcal{K} \) from the
nonvanishing torsion in (\ref{2.37}) \begin{equation}
\label{2.40}
\mathcal{K}_{(\mu \nu )\rho }=T_{[\rho \mu ]\nu }+T_{[\rho \nu ]\mu }
\end{equation}
 can be found. Nonvanishing torsion and thus also a nonvanishing contorsion 
are important for
the determination of the global properties of a certain solution of a generic
theory of gravity. 

In contrast to ordinary Minkowski space field theories, the variables of 
gravity -- in the most general case the independent Cartan variables
\( e \) and \( \omega  \) -- in the dynamical evolution also determine the
non-Minkowski dynamical background upon which the theory lives. Thus, for the
investigation of that background a device must be found which acts like a test
charge in an electromagnetic field. The simplest possibility in gravity is to
add the Lagrangian of a point particle with path \( x^{\mu }=\bar{x}^{\mu }(\tau ) \)
to the original action (\( \dot{\bar{x}}^{\mu }=d\bar{x}^{\mu }/d\tau  \) with
the affine parameter \( \tau  \)), \begin{equation}
\label{2.41}
L^{(p)}=-m\, \int\limits _{\tau _{1}}^{\tau _{2}}ds=-m\int\limits _{\tau _{1}}^{\tau _{2}}\, \sqrt{g_{\mu \nu }(\bar{x})\, \dot{\bar{x}}^{\mu }\, \dot{\bar{x}}^{\nu }}\; d\tau \, ,
\end{equation}
 with a mass \( m \), small enough to be of negligible gravitational influence.
Variation of \( L^{(p)} \) with respect to \( \bar{x}^{\mu } \) leads to the
usual geodesic equation \begin{equation}
\label{2.42}
\ddot{\bar{x}}^{\mu }+\widetilde{\Gamma }_{(\rho \sigma )}{}^{\mu }\, \dot{\bar{x}}^{\rho }\, \dot{\bar{x}}^{\sigma }=0\, ,
\end{equation}
 where, by construction from (\ref{2.41}), \( \widetilde{\Gamma }_{(\rho \sigma )}{}^{\mu }=g^{\mu \alpha }\, \{\rho ,\sigma ,\alpha \} \)
only {}``feels{}'' the Christoffel part (\ref{2.39}) of the affine connection
and not the contorsion (\ref{2.40}). Alternatively, also the full affine connection
\( \Gamma  \) may be considered in (\ref{2.42}) ({}``autoparallels{}'')
\cite{Hehl:1976kj, Hehl:1995ue}. For that modified geodesic equation for \( \bar{x}^{\alpha }(\tau ) \)
also a (non-local) action replacing (\ref{2.41}) can be found in the literature
\cite{Fiziev:1996te,Kleinert:1996yi}. In order to explore the local and topological
properties of a certain manifold which corresponds to a solution of a generic
gravity theory all points must be connected which can be reached by a device
like the geodesic (\ref{2.42}) by means of a time-like, but also space-like or light-like
path. The classification of possible extensions of a certain patch uses the
notion of {}``geodesic{}'' incompleteness: a geodesic which has only a finite
range of affine parameter, but which is inextendible\footnote{%
This means the corresponding geodesic must have (at least) one endpoint. For
details we refer to \cite{Hawking:1973,waldgeneral}.
} in at least one direction is called incomplete. A spacetime with at least one
incomplete (time/space/light-like) geodesic is called (time/space/light-like)
geodesically incomplete. The notion of incompleteness also yields the most satisfactory
classification of (geometric) singularities. For example, a singularity like
the one in the Schwarzschild metric \cite{Schwarzschild:1916uq} can be reached
by at least one (time- or light-like) geodesic with finite affine parameter
(i.e. with finite proper time for massive test particles).

For \( D=2 \) theories a complete discussion of {}``geodesic topology{}''
for any generic theory can be carried out (cf.\ sect.\ \ref{se:2.3.2}.) \cite{Walker:1970,Klosch:1996qv}.
Here we just want to emphasize the importance of the \textit{type} of device
to be used for the determination of the {}``effective{}'' topology of the
manifold which, in principle, may be different for geodesics, autoparallels, 
spinning particles etc. .

\subsubsection{Two dimensions}

\label{se:2.1.2}

In \( D=2 \) the Lorentz transformations (\ref{2.1}),(\ref{2.2}) simply reduce
to a boost with velocity \( v \)\begin{equation}
\label{2.43}
L^{a}{}_{b}=\left( \begin{array}{cc}
\cosh v & \sinh v\\
\sinh v & \cosh v
\end{array}\right) ^{\! \! a}_{\, \; b}\; =\; \delta ^{a}_{b}+\epsilon ^{a}{}_{b}\, v+\mathcal{O}(v^{2})\, ,
\end{equation}
 where in local Lorentz indices with metric \( \eta _{ab}=\eta ^{ab} \) (\( \eta =\mbox {diag}(+1,-1) \))
the Levi-Civit\'{a} symbol \( \epsilon ^{a}{}_{b}=\eta ^{ac}\, \epsilon _{cb} \)
(\( \epsilon _{01}=-\epsilon^{01}=+1 \)) coincides with the tensor. It is related to the tensor
\( \epsilon _{\mu \nu } \) in holonomic coordinates (cf.\  (\ref{2.15})) by
(explicit values of Lorentz indices in (\ref{2.45}) are underlined) \eq{
\epsilon = -\frac{1}{2}\epsilon_{ab} e^{a} \wedge e^{b}
}{2.44a}\begin{equation}
\label{2.44}
\epsilon _{\mu \nu }=e_{\mu }^{a}e_{\nu }^{b}\epsilon _{ab}=|e|\tilde{\epsilon }_{\mu \nu }=|e|^{-1}g_{\mu \rho }g_{\nu \sigma }\, \tilde{\epsilon }^{\rho \sigma }\, ,
\end{equation}
\begin{equation}
\label{2.45}
|e|=\det \, e_{\mu }^{a}=e_{0}^{\ul {0}}\, e_{1}^{\ul {1}}-e_{0}^{\ul {1}}\, e_{1}^{\ul {0}}\; .
\end{equation}
It should be noted that in (1.45) we choose the sign, which differs from (1.14), in order to be consistent with some original literature.
As there is only one generator \( \varepsilon ^{a}{}_{b} \) in \( SO(1,1) \)
(cf.\ (\ref{2.43})) the spin connection one-form simplifies to a single term
\( \omega ^{a}{}_{b}=\omega \, \epsilon ^{a}{}_{b} \) and hence the one quadratic in
\( \om  \) of \( R_{ab} \) (\ref{2.23}) vanishes: \begin{equation}
\label{2.46}
R^{a}{}_{b}=\epsilon ^{a}{}_{b}\, d\omega \; .
\end{equation}
 From now on for simplicity we shall refer to the 1-form \( \om  \) as the
{}``spin connection{}''.

This shows that the curvature in \( D=2 \) only possesses one independent component
which we take to be the Ricci-scalar\footnote{%
Our convention corresponds to the contraction \( R_{\mu \nu }{}^{\nu \mu }=R_{\mu \nu \, ab}e^{a\nu }e^{b\nu } \)
where \( R_{\mu \nu \, ab} \) are the tensor components of \( R_{ab} \). \( R_{\mu \nu }{}^{\rho \sigma } \)
then coincides with the usual textbook definition \cite{waldgeneral}.
}:
\begin{equation}
\label{2.47}
R=2\, \ast d\omega =2|e|^{-1}\tilde{\epsilon }^{\rho \sigma }\partial _{\rho }\omega _{\sigma }\; .
\end{equation}
 It is clear from this expression that the Hilbert-Einstein action in two dimensions
is a total divergence. In (compact) Euclidean space (\( \sqrt{-g}\rightarrow \sqrt{g} \))
without boundaries it becomes the Euler characteristic of a \( 2D \) Riemannian
space with genus~\( \gamma  \)\begin{equation}
\label{2.61}
\int\limits _{\mathcal{M}_{\gamma }}\, d^{2}x\; \sqrt{g}\, R=8\pi (1-\gamma )\; .
\end{equation}
 Also the torsion simplifies to a volume form
\begin{equation}
\label{2.48}
T^{a}=\frac{1}{2}T_{\mu \nu }{}^{a}dx^{\mu }\wedge dx^{\nu }\; ,\hspace {0.5cm}T_{\mu \nu }{}^{a}=(D_{\mu }e_{\nu })^{a}-(D_{\nu }e_{\mu })^{a}\; ,
\end{equation}
 with \begin{equation}
\label{2.49}
\left( D_{\mu }\right) ^{a}{}_{b}=\partial _{\mu }\delta ^{a}{}_{b}+\omega _{\mu }\epsilon ^{a}{}_{b}\; .
\end{equation}
 The Hodge dual of \( T^{a} \) here is a diffeomorphism scalar: \begin{equation}
\label{2.50}
\tau^a:=\ast \; T^{a}=|e|^{-1}\, \tilde{\epsilon }^{\mu \nu }\; (D_{\mu }e_{\nu}^{a})
\end{equation}
 In \( D=2 \) the inverse of the zweibeine from (\ref{2.29}) obeys the simple
relation \begin{equation}
\label{2.51}
e_{a}^{\mu }=-|e|^{-1}\tilde{\epsilon }^{\mu \nu }\, \epsilon _{ab}\, e^{b}_{\nu }\; .
\end{equation}

The formula for the change of the Ricci scalar under a conformal transformation
of the metric \( \hat{g}_{\mu \nu }=e^{2\rho }\, g_{\mu \nu } \) is most easily
derived from a transformation \( \hat{e}_{\mu }^{a}=e^{\rho }\, e^{a}_{\mu } \)
of the zweibeine for vanishing torsion \( T^{a}=0 \), i.e.\ with \( \omega =\widetilde{\omega }=e_{a}\ast de^{a} \)
in the Ricci scalar (\ref{2.47}) \begin{equation}
\label{2.52}
|e|\, R=2|e|\ast d(e_{a}\ast de^{a})=2\, \tilde{\epsilon }^{\tau \sigma }\partial _{\tau }\left( e_{\sigma }{}^{a}\frac{\tilde{\epsilon }^{\mu \nu }}{|e|}\partial _{\mu }e_{a\nu }\right) \; .
\end{equation}
 Remembering \( \hat{e}=|e|e^{2\rho } \) and using (\ref{2.51}) for 
\( e^{\mu }_{a}e^{\nu }_{b}\eta ^{ab}=g^{\mu \nu } \) 
yields (\( (\hat{e})=\sqrt{-\hat{g}} \)) an important identity:
\begin{equation}
\label{2.53}
\sqrt{-\hat{g}}\, \hat{R}=\sqrt{-g}R-2\partial _{\tau }(\sqrt{-g}g^{\tau \sigma }\, \partial _{\sigma }\rho )\; .
\end{equation}

Light-cone Lorentz vectors are especially useful in \( D=2 \), \begin{equation}
\label{2.54}
X^{\pm }:=\frac{1}{\sqrt{2}}\, (X^{\ul {0}}\pm X^{\ul {1}})\, ,
\end{equation}
 yielding \( X^{2}=X^{a}X_{a}=X^{\bar{a}}X_{\bar{a}}=\eta _{\bar{a}\bar{b}}\, X^{\bar{a}}X^{\bar{b}}=2X^{+}X^{-} \)
with metric 
\begin{equation}
\eta_{\bar{a}\bar{b}} = \left(
\begin{array}{cc}
0 & 1 \\
1 & 0 
\end{array}
\right)
\label{2.55}
\end{equation}
and the corresponding Lorentz $\epsilon$-tensor 
$\epsilon^{\bar{a}}{}_{\bar{b}} = \eta^{\bar{a}\,\bar{c}}\, 
\epsilon_{\bar{c}\,\bar{b}}$ with $\eps^\pm{}_\pm=\pm 1$. 
The light cone components of the torsion (\ref{2.48}) become 
\begin{equation}
T^\pm = (d \pm \omega)\, e^\pm .
\label{2.56}
\end{equation}

Since we are going to discuss fermionic matter (as well as supergravity) we
have to fix our spinor notation. The $\gamma^a$-matrices are defined in a 
local Lorentz frame 
\begin{equation}
\left\lbrace \, \gamma^a, \gamma^b \, \right\rbrace = 
2 \eta^{ab} 
\label{eq:a72}
\end{equation}
\begin{equation}
\begin{split}
\gamma^{\ul{0}} &= \left(
\begin{array}{cc}
0 & 1 \\
1 & 0
\end{array} \right)\, , \qquad
\gamma^{\ul{1}} = 
\left(
\begin{array}{cc}
0 & 1 \\
-1 & 0
\end{array} \right)\\
\gamma_\ast & := - \gamma^{\ul{0}} \gamma^{\ul{1}} = 
\left(
\begin{array}{cc}
1 & 0 \\
0 & -1
\end{array} \right) = -\frac{1}{2}\left[\ga^{\ul{0}},\ga^{\ul{1}}\right].
\end{split}
\label{eq:73}
\end{equation}
In light cone components we obtain
a representation in terms of nilpotent matrices 
\eq{
\gamma^{+} = \sqrt{2} \left(
\begin{array}{cc}
0 & 1 \\
0 & 0
\end{array} \right)\, , \qquad
\gamma^{-} = \sqrt{2} \left(
\begin{array}{cc}
0 & 0 \\
1 & 0
\end{array} \right)
}{2.56a} The covariant derivative acting on two-dimensional Dirac fermions \begin{equation}
\label{eq:74}
D_{\mu }=\partial _{\mu }-\frac{1}{2}\, \gamma_\ast\, \omega _{\mu }
\end{equation}
 is determined by the Lorentz generator for spinors \( [\ga ^{\ul {0}},\ga ^{\ul {1}}]/4=-\gamma_\ast/2 \). 


\clearpage

\section{Models in 1 + 1 Dimensions}

\label{se:2}\label{se:2.2}

There are (at least) four different motivations to study generalized dilaton
theories (GDT) in $D=2$: \blist
\item Starting from Einstein gravity in \( D\geq 4 \) and imposing spherical symmetry
one reproduces a certain GDT 
\item A certain limit of (super-)string theory yields a particular GDT as effective
action 
\item GDTs can be viewed as toy models for quantization of gravity and as a laboratory
for studying BH evaporation 
\item In a first order formulation the underlying Poisson structure reveals relations
to non-commutative geometry and deformation quantization. Again, GDTs are a
convenient laboratory to elucidate these new concepts and techniques. \elist
Moreover, a result obtained along one route is of course also valid for all
other approaches after having translated the jargon from one field to the others.
In this sense, GDTs may even serve as a link between general relativity (GR), string theory, BH physics and non-commutative geometry.

We base our discussion on the first (somewhat more phenomenological) route and
show the links to the other fields in this section.

\subsection{Generalized Dilaton Theories}

\label{se:2.2.1}

\subsubsection{Spherically reduced gravity}\label{se:2.2.1.1} 

The introduction of dilaton fields allows the treatment of the dynamics for
a generic higher dimensional \( (D>2) \) theory of gravity in an effective
theory at lower dimension \( D_{1}<D \), which is still diffeomorphism invariant.
In certain special cases the isometry group of the \( D \)-dimensional metric
is such that it allows for a reduction to \( D_{1}=2 \). Important examples
for \( D=4 \) are toroidal reduction \cite{Kuchar:1971xm, Gowdy:1971jh,Geroch:1972yt,Husain:1989qq,Brodbeck:1999ib}
and spherical reduction \cite{Berger:1972pg,Unruh:1976db,Benguria:1977in, Thomi:1984na,Hajicek:1984mz,Mignemi:1989qc,Thiemann:1993jj,Kastrup:1994br,
Kuchar:1994zk,Lau:1996fr,Grumiller:1999rz}. The latter is of special importance,
because it covers the Schwarzschild BH. Therefore, we concentrate on that 
example.

Splitting locally the \( D \)-dimensional manifold \( \mathcal{M}_{D} \) into
a direct product \( \mathcal{M}_{2}\times {S}^{D-2} \) the line element 
becomes
\begin{equation}
\label{2.57}
(ds)^{2}_{(D)}=g_{\mu \nu }(x)dx^{\mu }dx^{\nu }-\lambda ^{-2}X^{\frac{2}{D-2}}\, (d\Omega )^{2}_{S^{D-2}}
\end{equation}
 where \( (d\Omega )^{2}_{S^{D-2}} \) is the surface element of the \( (D-2) \)-dimensional
sphere, \( x^{\mu }=\{x^{0},x^{1}\} \) are the coordinates in \( \mathcal{M}_{2} \),
and \( \lambda  \) is a parameter of mass dimension one. A straightforward
calculation (cf. e.g. \cite{Grumiller:2001ea}; explicit formulae for the curvature
2-form, the ensuing Ricci-scalar and the Euler- and Pontryagin-class can be 
found in appendix \ref{app:A}) for the \( D \)-dimensional 
Hilbert-Einstein action
\( L_{HE}=\int d^{D}x\, \sqrt{-g_{(D)}}\, R_{(D)} \) yields (\( (\nabla X)^{2}=g^{\mu \nu }\partial _{\mu }X\partial _{\nu }X \))
\begin{multline}
L^{\rm (SRG)} =  \frac{\mathcal{O}_{D-2}}{\lambda^{D-2}16\pi G_{N}} \, 
\int d^{2} x\, \sqrt{-g}\, \\
\left[ XR + \frac{D-3}{D-2} \, \frac{(\nabla X)^{2}}{X} \,
- \lambda^{2} (D-2) (D-3)\, X^{\frac{D-4}{D-2}} \,  \right]\; .
\label{2.58}
\end{multline} In the prefactor, which will be dropped consistently in the following,
\( \mathcal{O}_{D-2} \) denotes the surface of the unit sphere \( S^{D-2} \).
Fixing the \( 2D \) diffeomorphisms (partially) as \( X=(\lambda r)^{D-2} \)
(the radius \( r \) representing one of the coordinates and \( \lambda >0 \))
eq. (\ref{2.57}) yields the usual spherically symmetric line element in which
\( r>0 \) is required.

Another way to obtain a $2D$ theory from a higher dimensional one
is to suppose that the \( D \)-dimensional manifold is a direct product \( \mathcal{M}_{D}=\mathcal{M}_{2}\otimes {T}^{D-2} \),
where \( {T}^{D-2} \) is a torus, and that all fields are independent of the
\( D-2 \) extra coordinates. This procedure is called dimensional reduction.
It also produces a dilaton theory in $2D$ if the higher dimensional
theory already contains the dilaton \cite{Gibbons:1988ps}.

\subsubsection{Dilaton gravity from strings}

\label{se:2.2.1.5}

Developments in string theory contributed much to the increase of interest in
dilaton gravity in the 1990s. The simplest way to obtain it from strings
is to consider the conditions for world-sheet conformal invariance 
\cite{Callan:1985ia}.

The starting point is the non-linear sigma model action for the closed bosonic 
string, 
\begin{equation}
\label{esLs}
L^{(\sigma)}=\frac{1}{4\pi \alpha' }\int d^{2}\xi \sqrt{-h}\left[ g_{\mu \nu }h^{ij}\partial _{i}X^{\mu }\partial _{j}X^{\nu }+\alpha' \Phi {\mathcal{R}}\right] \, ,
\end{equation}
 where \( \xi  \) is a coordinate on the string world-sheet, \( h^{ij} \)
is a metric\footnote{%
This metric should not be confused with $g_{\mu\nu}$ restricted to \( D=2 \) in
(\ref{esLd}).
} there, \( \mathcal{R} \) represents the corresponding scalar curvature. The
other symbols denote: the target space coordinates (\( X^{\mu } \)), the 
target space metric (\( g_{\mu \nu } \)),
and the dilaton field (\( \Phi  \)). As usual, \( \alpha'  \) is the inverse
string tension. The antisymmetric \( B \)-field is set to zero.

It is essential for string consistency that, as a quantum field theory, the
sigma model be locally scale invariant. This is equivalent to the requirement
that the trace of the $2D$ world-sheet energy-momentum tensor vanishes. Its
general structure is
\begin{equation}
\label{esTii}
2\pi T_{i}^{i}=\beta ^{\Phi }{\mathcal{R}}+\beta ^{g}_{\mu \nu }h^{ij}\partial _{i}X^{\mu }\partial _{j}X^{\nu }\, ,
\end{equation}
 where the {}``beta functions{}'' \( \beta ^{\Phi } \) and \( \beta ^{g}_{\mu \nu } \)
are local functionals of the couplings \( g_{\mu \nu } \) and \( \Phi  \),
usually calculated in the form of a power series in \( \alpha'  \).
Note that the first term in \( L^{(\sigma)} \) is conformally invariant and
contributes to the \( \beta  \)-functions at the quantum level only through
the conformal anomaly. It corresponds to \( \mathcal{O}(\alpha' )^{0} \). The second term
in (\ref{esLs}) breaks local scale invariance already at the classical level.
Due to the factor \( \alpha'  \) its contributions to the trace (\ref{esTii})
also start with the zeroth power of \( \alpha'  \). The leading terms in \( \beta ^{\Phi } \)
and \( \beta ^{g}_{\mu \nu } \) were calculated in ref. \cite{Callan:1985ia}.
With our sign conventions they read: 
\begin{eqnarray}
 &  & \frac{\beta ^{\Phi }}{\alpha' }=-\frac{\lambda ^{2}}{4\pi ^{2}}-\frac{1}{16\pi ^{2}}\left( 4(\nabla\Phi )^{2}-4\nabla ^{\mu }\nabla _{\mu }\Phi -R\right) \, ,\label{esbp} \\
 &  & \beta ^{g}_{\mu \nu }=R_{\mu \nu }+2\nabla _{\mu }\nabla _{\nu }\Phi \, ,\label{esbg} 
\end{eqnarray}
 where \( \nabla _{\mu } \) is the covariant derivative in target space, \( R \)
is the scalar curvature of the target space manifold. The constant \( \lambda  \)
depends on the central charge. For the bosonic string it is \begin{equation}
\label{esl2}
\lambda ^{2}=\frac{26-D}{12\alpha' }\, .
\end{equation}
 This constant vanishes for critical strings.

The key observation regarding the beta functions (\ref{esbp}) and (\ref{esbg})
is that the conditions of conformal invariance \( \beta ^{\Phi }=0 \) and \( \beta ^{g}_{\mu \nu }=0 \)
are equivalent to the e.o.m.-s to be derived from the dilaton gravity
action \begin{equation}
\label{esLd}
L^{\rm (dil)}=\int d^{D}X\sqrt{-g}e^{-2\Phi }\left[ R+4(\nabla \Phi )^{2}-4\lambda ^{2}\right] \, .
\end{equation}
 In particular, the dilaton e.o.m. is equivalent to \( \beta ^{\Phi }=0 \).
The Einstein equations are given by a combination of the two beta functions,
\( \beta _{\mu \nu }^{g}-8\pi ^{2}g_{\mu \nu }\beta ^{\Phi }/\alpha' =0 \).

For \( D=2 \) the action (\ref{esLd}) describes the geometric part of the {}``string inspired{}''
dilaton (CGHS) model \cite{Callan:1992rs} which has been studied since the
early 1990-s \cite{Mikovic:1992id,Cangemi:1996yz,Benedict:1996qy, Mikovic:1997de,Kuchar:1997zm,Varadarajan:1998qz}.
It is intimately related to the \( SO(2,1)/U(1) \)-WZW exact conformal
field theory\footnote{%
The non-compact form is \( SO(2,1)/SO(1,1) \). An early review on \( 2D \)
gravity and \( 2D \) string theory from the stringy point of view is ref. 
\cite{Ginsparg:1993is}.
} \cite{Mandal:1991tz,Elitzur:1991cb,Witten:1991yr,Dijkgraaf:1992ba}. 

An amusing feature of  (\ref{esLd}) with \( D=2 \) is that after the identification
\( X=e^{-2\Phi } \) it can be obtained from (\ref{2.58}) by taking there the
limit \( D\rightarrow \infty  \) keeping 
\( \lambda ^{2}(D-2)(D-3)\rightarrow {\rm const.}=4\la^2\)\,. This corresponds
to the classical limit $\al'\to\infty$.

\subsubsection{Generalized dilaton theories -- the action}\label{se:2.1.3}

A result like (\ref{2.58}) or (\ref{esLd}) suggests the consideration of GDTs 
\begin{equation}
\label{2.59}
L^{\rm (dil)}=\int d^{2}x\, \sqrt{-g}\, \left[ \frac{R}{2}\, X-\frac{U(X)}{2}\; (\nabla X)^{2}+V(X)\, \right] \, ,
\end{equation}
 where the overall factor has been chosen for later convenience. Clearly an
even more general action could contain still another arbitrary function \( Z(X) \),
replacing \( X \) in the first term of the square bracket \cite{Banks:1991mk,Odintsov:1991qu}. However,
we assume that \( Z(X) \) is invertible for the range of \( X \) to be considered\footnote{%
To the best of our knowledge there is no literature on nontrivial models where
\( Z(X) \) is not invertible (cf. also \cite{Strobl:1999wv}). By a suitable
redefinition a different simplification with \( U(X)=1 \), \( Z(X)\neq 1 \)
was proposed in ref. \cite{Russo:1992yg}.
}. This allows the inversion \( X=Z^{-1}\, (\widetilde{X}) \) and the reduction
to the form (\ref{2.59}). Indeed the {}``physical{}'' applications seem to be 
always of that type. The BH singularity of SRG reveals itself in the singular 
factor \( U \) of the dynamical term for the dilaton field. This is the first 
hint to the fact that the {}``strength{}'' of that singularity in the solution
of (\ref{2.58}) is not fixed by the action; it will actually turn out to be
a {}``constant of motion{}'' which for the BH coincides with the ADM mass
(cf. sect. \ref{se:5.1}).

An alternative representation is suggested by (\ref{esLd}): \eq{
L^{\rm (dil)}=\frac{1}{2}\int d^{2}x\sqrt{-g}e^{-2\Phi } \left[ R 
- \tilde{U}(\Phi) (\nabla \Phi )^{2} 
+ 2\tilde{V}(\Phi)\right] \, ,
}{2.59.5} 
with $\tilde{U}(\Phi)=4\exp{(-2\Phi)}U(\exp{(-2\Phi)})$ and 
$\tilde{V}(\Phi)=\exp{(2\Phi)}V(\exp{(-2\Phi)})$.
Eqs. (\ref{2.59}) and (\ref{2.59.5}) are related by the redefinition of the
dilaton field
\eq{
X=e^{-2\Phi},
}{e23exp} 
explicitly taking into account positivity of \( X \) which is required in many
models.

Among the GDTs (\ref{2.59}) with \( U(X)=0 \) the
simplest nontrivial choice of refs. \cite{Barbashov:1979bm, D'Hoker:1982er,
Teitelboim:1983ux,D'Hoker:1983is,D'Hoker:1983ef,Jackiw:1995qh} 
\eq{
V_{JT} = \Lambda X,\hspace{0.5cm}U_{JT} = 0\,, 
}{eq:JT} 
the Jackiw-Teitelboim (JT) model,  
has played a decisive role for the understanding of \( 2D \) (lineal)
gravity \cite{Jackiw:1985je}. Depending on the sign of \( \Lambda  \) it describes
a \( 2D \) (anti-) de Sitter manifold with constant positive or negative curvature.
The symmetry properties of the model are related to the Lie algebra \( SO(1,2) \).
It has been explored in detail in the quoted references. Below this algebra
will turn out to represent the special linear case of some, in general, nonlinear
(finite \( W \)-) algebra \cite{deBoer:1996nu} associated with a generic dilaton
theory (\ref{2.59}) (cf.\  Sect.\ \ref{se:2.2.3}).

More complicated models with \( U(X)=0 \), but \( V(X) \) exhibiting a
singularity in \( X \), among others may also involve solutions with space-time
structure of a BH or its generalizations. E.g.\ the choice\footnote{%
Solving the general theory in Sect.\ \ref{se:2.3.1} we shall find that the
potentials \( U \) and \( V \) as in (\ref{2.59}) determining a dilaton action
can even be `designed', starting from a given line-element.
}\begin{equation}
\label{2.60}
V_{RN}=-\frac{2M}{X^{2}}+\frac{Q^{2}}{4X^{3}}
\end{equation}
 produces a line element like the one for the Reissner-Nordstr\"{o}m BH with
charge \( Q \) and mass \( M \) \cite{Reissner:1916,Nordstrom:1916}. Evidently
in this case the singularities are kept fixed by parameters of
the action. They cannot be related to the conservation law referred to already
above for a {}``dynamical{}'' model with singular nonvanishing \( U(X) \)
and regular \( V(X) \). A final remark for the case \( U=0 \) concerns the
possibility to eliminate the dilaton field altogether by means of the \textit{algebraic}
equations of motion produced by varying \( X \) in (\ref{2.59}), \( V'\, (X)=-R/2 \).
If this equation can be inverted, the dilaton Lagrangian for \( U=0 \) turns
into a Lagrangian depending on the function of \( R \) alone \cite{Schmidt:1991ws, Frolov:1992xx,Schmidt:1999wb,Strobl:1999wv}:
\eq{
L = \int d^{2} x \; \sqrt{-g} \, f (R)
}{eq:Rgravity} As compared to such theories (\ref{eq:Rgravity}), the literature
on models generalized so as to depend also on torsion (cf. (\ref{2.50}))
\begin{equation}
\label{eq:RTgravity}
L=\int d^{2}x\; \sqrt{-g}\; h\, (R,\tau^{a}\tau_{a})
\end{equation}
 is relatively scarce. It mainly consists of elaborations based upon the model
of Katanaev and Volovich \cite{Katanaev:1986wk,Katanaev:1990qm} where the function
\( h \) in (\ref{eq:RTgravity}) is quadratic in \( R \) and linear in \( \tau^{a}\tau_{a} \),
also known as {}``Poincar\'{e} gauge gravity{}'' \cite{Solodukhin:1993bn,Solodukhin:1993bs,Solodukhin:1993xs,Solodukhin:1994sv,Obukhov:1997uc,Mielke:1993nc}.

Models with \( U(X)\neq 0 \) and different assumptions for that function and
\( V(X) \) have been studied extensively (cf. e.g. \cite{Mann:1990gh,Banks:1991mk, Odintsov:1991qu,Odintsov:1992fm,Russo:1992ht,Russo:1992yg,Bilal:1993kv,
deAlwis:1993zy,Mann:1993yv,Gegenberg:1993rg,Louis-Martinez:1994cc,Louis-Martinez:1994eh,Lemos:1994py}).
For their solution throughout these works the conformal or the Schwarzschild
gauge have been used, leading to complicated e.o.m.-s, the solution of which
often requires considerable mathematical effort. Because we shall avoid this
complication altogether (sect.\  \ref{se:2.3}) no explicit examples of this
approach will be given here.

\subsubsection{Conformally related theories}\label{se:2.1.4}

Sometimes, it is convenient \cite{Louis-Martinez:1997dx,Navarro:1997gr,
Kiem:1998ay,Cruz:1998jz,Kunstatter:1998my,Cavaglia:1998xj,Cruz:1998mn,
CassemiroF.F.:1998ba,Cavaglia:2000uw} to use a conformal transformation
(\ref{2.53}) with \( \rho (X)=-1/2\, \int ^{X}\, U(y)\, dy \) in (\ref{2.59})
to simplify the dynamics by the transition to a new theory with \( \widetilde{U}=0 \)
and \( \widetilde{V}(X)=V(X)\exp (-2\rho ) \). One has to keep in mind, however,
that the two theories need \textit{not} be equivalent physically. To interpret the
results one must always return to the original theory. This subtlety was sometimes
ignored. One source of this misunderstanding seems to be that in field theory
the transformation of field variables in a fixed flat Minkowski background is
allowed, as long as such a transformation is regular. For a GDT
(\ref{2.59}) with singular \( U \) this has to fail for two reasons.
The first one is that such a conformal transformation must be singular in order
to compensate for a singularity in \( U(X) \). Still one could argue that locally
such a transformation should be permissible. However, and this is the second
crucial reason, in gravity the field theory in its variables at the same time
determines the (dynamical!) manifold upon which it lives. For a singular conformal
transformation the new manifold can possess completely different topological
properties.

An extreme example is the CGHS model (\ref{esLd})\cite{Callan:1992rs} which from
a Schwarz\-schild-like topology may be transformed into flat (Minkowski) space.
The reason can be seen most easily in the transformation behavior of geodesics:
only null geodesics are mapped onto (in general non-affinely parameterized)
null geodesics and their corresponding affine parameters are related by \cite{waldgeneral}:
\eq{
\frac{d\tilde{\tau}}{d\tau} \propto e^{2\rho}
}{2.60a} 
If \( \rho  \) approaches infinity at a certain point, by such a singular 
conformal transformation geometric properties like geodesic (in)completeness 
can be altered\footnote{Since the usual conformal transformation involved in 
this context is proportional to the integral of \( U(X) \) and the latter has 
a singularity in practically all physically interesting models there will be 
at least one such singular point in addition to the (asymptotic) singularity 
at \( X\rightarrow \infty  \) \cite{Grumiller:2000hp}.}.

In fact, this misunderstanding had been clarified already half a century ago
\cite{Fierz:1956} in connection with the Jordan-Brans-Dicke theory in \( D=4 \)
\cite{jordanschwerkraft,Jordan:1959eg, Brans:1961sx}. There already in \( D=4 \)
a {}``Jordan-field{}'' \( X \) in a \( D=4 \) action like (\ref{2.59})
with \( U(X)=\mbox {const.} \) is introduced. The \( D=4 \) version of identity
(\ref{2.53}), together with an appropriate transformation of \( X \) may be
used to transform that action so that the term involving \( R \) is reduced
to the Hilbert-Einstein form. At that time a controversy arose whether the latter
(the {}``Einstein-frame{}'') or the original one (the {}``Jordan frame{}'')
was the {}``correct{}'' one. As argued by Fierz \cite{Fierz:1956} the answer 
to that questions
depends on the definition of geodesics, to be used for the determination of
the global topology (cf.\ sect.\ \ref{se:2.1}). A geodesic depending on the
metric \( g \) in the Jordan frame is quite different from the one which feels
the metric of the conformally transformed \( \hat{g} \) in the Einstein-frame.
Of course, for a (globally) regular conformal transformation \( \Omega ^{2} \),
\( g_{\mu \nu }=\Omega ^{2}\, \hat{g}_{\mu \nu } \) it would be perfectly correct
to simultaneously transform \( g \) into the Jordan frame. But then the equation
of the geodesic, when expressed in terms of \( \hat{g} \) acquires an additional
dependence on \( \Omega (X) \), i.e.\ the test particle would feel a non-geodesic
external force exerted by the Jordan-field \( X \).

The confusion in \( D=2 \) probably also originated from the by now very familiar
situation in string theory \cite{Green:1987sp,Polchinski:1998rq}. Its conformal
invariance does not carry over automatically to the world-sheet, where it is
achieved 
by imposing the e.o.m.'s in target space (cf. sect. \ref{se:2.2.1.5}).
String theory yields dilaton gravity as its low energy limit also in higher
dimensions. In that context the Jordan frame usually now is called
the string frame and the old discussion referred to in the previous paragraph
has been resurrected in modern language \cite{Capozziello:1997xg,Dick:1998ke,
Casadio:1998wu,Faraoni:1999hp,Alvarez:2001qj}.

A simple example of a singular conformal transformation leading to a change
of (timelike) geodesic (in)completeness can be found in fig. 9.1 of \cite{waldgeneral}.
Another obvious case is provided by the Schwarzschild metric, eq. 
(\ref{eq:a35}) below.
A (singular) conformal transformation with \( \Om ^{2}=\xi ^{-1}=(1-2M/r)^{-1} \)
and a (singular) coordinate transformation \( \tilde{r}=\int ^{r}dy/\xi (y) \)
leads to Minkowski spacetime. This is, of course, a rather trivial consequence
of (patchwise) conformal flatness of \emph{any} \( 2D \) metric. 
It will be discussed
below why ADM mass (sect. \ref{admsec}) and Hawking radiation (sect.
\ref{se:3}) are, in general, different in conformally related theories.

\subsection{Equivalence to first-order formalism}

\label{se:2.2.2}

Cartan variables have been introduced in sect.\ \ref{se:2.1} in order to 
formulate
a very general class of \( D=2 \) first order gravity (FOG) theories by the
covariant Hamiltonian action 
\begin{equation}
\label{2.62}
L^{\rm (FOG)}=\int\limits _{\mathcal{M}_{2}}\, \left[ X_{a}(De)^{a}+Xd\omega +\epsilon \mathcal{V}(X^{a}X_{a},X)\right] \; ,
\end{equation}
 which seems to have been introduced first for the special case (\( \mathcal{V}=0 \))
in string theory \cite{Verlinde:1991rf}, then considered for a special model
in ref.\ \cite{Ikeda:1993aj} and finally generalized to the in \( D=2 \) most
general form (\ref{2.62}) for a theory of pure gravity in refs.\  \cite{Strobl:1994yk,Schaller:1994es}.
It depends on auxiliary fields \( X^{a} \) and \( X \) so that it is sufficient
to include only the first derivatives of the zweibeine (torsion) and of the
spin connection (curvature). The whole dynamical content is encoded in a (Lorentz-invariant)
potential \( \mathcal{V} \) multiplied by the volume form (\ref{2.44a}). In
the following very often light-cone coordinates (\ref{2.54}) and (\ref{2.56}) will
be used: \begin{equation}
\label{2.63}
X_{a}(De)^{a}=X^{+}(d-\omega )\, e^{-}+X^{-}(d+\omega )\, e^{+}
\end{equation}
We also recall (\ref{2.47}), the relation \( 2\ast d\omega =R \) between 
spin-connection and curvature scalar.

The component version of (\ref{2.62}) with (\ref{2.47}) follows from the identification
(cf.\ (\ref{2.50}),(\ref{2.52})) implying the Hodge-duals, 
\begin{equation}
\begin{split}
(d \pm \omega ) e^\pm &\Rightarrow \;  
\tilde{\epsilon}^{\mu\nu} d^2 x \frac{1}{2}T_{\mu\nu}^\pm = (e)\, 
\tau^\pm d^2 x \,,\\
d\omega &\Rightarrow\; \tilde{\epsilon}^{\mu\nu} 
\partial_\mu \omega_\nu d^2 x = (e) \frac{R}{2} d^2 x\,,\\
\epsilon &\Rightarrow  - \frac{1}{2} \epsilon_{ab} 
e_\mu^a e_\nu^b \tilde{\epsilon}^{\mu\nu} d^2 x = (e)\, 
d^2x\,, 
\label{2.65}
\end{split}
\end{equation}
 as \begin{multline}
L^{\rm (FOG)} = \int d^{2} x \left\{ \tilde{\epsilon}^{\mu\nu} 
[ X^{+} (\partial_{\mu }- \omega_{\mu }) e_{\nu}^{-} + 
X^{-} (\partial_{\mu }+ \omega_{\mu}) e_{\nu}^{+} \right.\\
\left. + X\, \partial_{\mu}\omega_{\nu}] + 
(e)\, \mathcal{V} (2X^{+} X^{-}, X )\, \right\}.
\label{2.66}
\end{multline}

The original intention of the formulation (\ref{2.62}) had been to express
a general $2D$ Lagrangian involving the only independent geometric
quantities (Ricci scalar \( R \), and torsion scalar \( T^{2}=\tau ^{a}\tau _{a}=2\tau ^{+}\tau ^{-} \), cf. (\ref{2.50}),(\ref{eq:RTgravity}))
\begin{equation}
\label{2.67}
L^{(R,\tau ^{2})}=\int \, d^{2}x\, \sqrt{-g}\, h(R,\tau ^{2})
\end{equation}
 in a simpler fashion. The variables \( X^{a} \) and \( X \) can be eliminated
by the algebraic e.o.m.-s from variation \( \delta X^{a},\delta X \)
in (\ref{2.62}) or (\ref{2.66}), \begin{equation}
\label{2.68}
\tau _{a}+\frac{\partial \mathcal{V}}{\partial X^{a}}=0\, ,\hspace {0.5cm}\frac{R}{2}+\frac{\partial \mathcal{V}}{\partial X}=0\, ,
\end{equation}
 provided the Hessian \( |\, \partial ^{2}\mathcal{V}/\partial X^{A}\partial X^{B}\, | \)
does not vanish \( (X^{A}=\{X,X^{a}\}) \). Evidently this is not always possible,
but also, inversely, not every action \( L(R,\tau ^{2}) \) permits a reformulation
as \( L^{\rm (FOG)} \) in (\ref{2.62})\footnote{%
For a mathematically more precise discussion of this point we refer to ref. \cite{Strobl:1999wv}.
}.

Fortunately the relation of (\ref{2.62}) to GDT (\ref{2.59})
and especially to models with a physical motivation (e.g.\ SRG) is more immediate
and subjected to weaker conditions. Then, instead, only \( X^{a} \) and the
torsion-dependent part of the spin connection are eliminated by e.o.m.-s
which are linear and algebraic and thus may be reinserted into the action\footnote{%
This equivalence has been published first in ref.~\cite{Katanaev:1996bh} for the KV-model 
\cite{Katanaev:1986wk,Katanaev:1990qm}. The proof below follows the 
formulation used in ref. \cite{Ertl:1998ib} for the even more general 
case of \( 2D \) dilaton supergravity (cf.\ also sect.\ \ref{se:2.4.2}).
}. From the definition for \( \ast T^{a} \) (\ref{2.50}) with \( \omega =\omega ^{a}e_{a} \)
in the local Lorentz basis \( e_{a} \), the identities \( e^{a}\wedge e^{b}=-\epsilon ^{ab}\cdot \epsilon  \)
and \( \ast \, \epsilon =1 \), one gets \begin{equation}
\label{2.69}
\ast T^{a}=\ast \, de^{a}-\omega ^{a}
\end{equation}
 or \begin{equation}
\label{2.70}
\omega =\omega ^{a}e_{a}=e_{a}\, \ast de^{a}-e_{a}\ast T^{a}=:\widetilde{\omega }-\ast T\; ,
\end{equation}
 where \( \widetilde{\omega } \) represents the torsion free part of the spin
connection.

The e.o.m.\ from variation of \( X^{a} \) in (\ref{2.62}) \begin{equation}
\label{2.71}
de^{a}+\epsilon ^{a}{}_{b}\, \omega \wedge e^{b}+\epsilon \, \frac{\partial \mathcal{V}}{\partial X^{a}}=0\; ,
\end{equation}
 after taking the Hodge dual, multiplication with \( e_{a} \) and comparison
with the identity (\ref{2.70}) yields the relation between \( \ast T \) and
\( \mathcal{V} \)\begin{equation}
\label{2.72}
\ast T=-e_{a}\, \frac{\partial \mathcal{V}}{\partial X_{a}}\; .
\end{equation}
 Reinserting this algebraic eq.\ into (\ref{2.62}) produces \begin{equation}
\label{2.73}
L_{1}^{\rm (FOG)}=\int\limits _{\mathcal{M}_{2}}\left[ X^{a}\epsilon _{ab}\, \frac{\partial \mathcal{V}}{\partial \, X^{c}}\; e^{c}\wedge e^{b}+Xd\widetilde{\omega }-dX\wedge e^{c}\, \frac{\partial \mathcal{V}}{\partial X^{c}}+\epsilon \mathcal{V}\, \right] \; ,
\end{equation}
 where the torsion dependent part of \( \omega  \) now has been eliminated,
but the dependence on \( X^{a} \) is retained. For potentials \( \partial \mathcal{V}/\partial X^{a}=0 \)
eq.\ (\ref{2.73}) already by the second and third eq.\ (\ref{2.65}) can be
identified directly as GDT (\ref{2.59}) with \( U=0,\mathcal{V}(X)=V(X) \).
When \( \partial \mathcal{V}/\partial X^{a}\neq 0 \) the e.o.m.\ from
\( \delta X^{a} \) in (\ref{2.73}) must be used, \begin{equation}
\label{2.74}
(dX\wedge e^{c}+X^{c}\epsilon )\; \frac{\partial ^{2}\mathcal{V}}{\partial X^{c}\, \partial X^{a}}=0\; ,
\end{equation}
 which for nonvanishing Hessian of \( \mathcal{V} \), now with respect to the
\( X^{a} \) alone, leads to\footnote{For potentials $\mathcal{V}$ of the form
(\ref{2.77}) eq. (\ref{2.75}) does not hold necessarily at points $X_0$
where $U(X_0)=0$.} 
\begin{equation}
\label{2.75}
X^{a}=\ast \left(e^{a}\wedge \, dX\right)=\frac{\tilde{\epsilon }^{\mu \nu }}{(e)}\, e^{a}_{\mu }\, (\partial _{\nu }X)\; .
\end{equation}

For easy comparison with the GDT (\ref{2.59}), before
using (\ref{2.75}) and (\ref{2.73}), the latter action is rewritten in component
form. After cancellation of two terms with \( \partial \mathcal{V}/\partial X^{a} \)
the final result is very simple \begin{equation}
\label{2.76}
L^{\rm (dil)}=\int d^{2}x\, (e)\, \left[ \frac{X\widetilde{R}}{2}+\mathcal{V}(-(\nabla X)^{2},X)\right] \, ,
\end{equation}
 where, according to (\ref{2.75}), the argument \( X^{a}X_{a} \) in \( \mathcal{V} \)
has been replaced by a second derivative term of \( X \), to be identified 
also here with the same dilaton field as in (\ref{2.59}). The curvature scalar
\( \widetilde{R}=\ast 2d\widetilde{\omega } \) refers to the torsionless part
of the spin connection in (\ref{2.70}). Thus it may be expressed equally well
directly by the \( 2D \) metric \( g_{\mu \nu } \).

For potentials quadratic in the torsion-related variable \( X^{a} \)\begin{equation}
\label{2.77}
\mathcal{V}=U(X)\; \frac{X^{a}X_{a}}{2}+V(X)
\end{equation}
 the action (\ref{2.76}) exactly coincides with (\ref{2.59}) in which torsion
had been zero from the beginning. As we have used
the algebraic (and even only linear) e.o.m.-s to reduce the configuration space
of \( L^{\rm (FOG)} \) to the one of \( L^{\rm (dil)} \), the two actions lead
to the same dynamics.
This equivalence can be verified easily as well by the study of the explicit
analytic solution (cf.\ sect.\ \ref{se:2.3.1}). We anticipate also that at
the quantum level the steps above can be simply reinterpreted as {}``integrating
out{}'' the torsion dependent part of \( \omega  \) and \( X^{a} \) \cite{Kummer:1997hy}, cf. footnote \ref{fn:60} on page \pageref{fn:60}.

Apart from covering torsionless dilaton theories (\ref{2.59}), the first order
formulation (\ref{2.62}) also permits the inclusion of \( 2D \) theories with
nonvanishing torsion. The choice \begin{equation}
\label{2.78}
\mathcal{V}_{KV}=\frac{\alpha }{2}X^{a}X_{a}+\frac{\beta }{2}\, X^{2}-\Lambda \; ,
\end{equation}
 after elimination of \( X^{a} \) and \( X \) according to (\ref{2.68}) produces
the KV model \cite{Katanaev:1986wk,Katanaev:1990qm} which is quadratic in curvature
and torsion and thus of the type of {}``Poincar\'{e} gauge{}'' theory
\cite{Hehl:1976kj,Hehl:1995ue}. By our equivalence relation it could also have
been written as the corresponding dilaton theory (\ref{2.59}), of course.

A final remark concerns the overall normalization of our action. By comparing
the term \( \propto R \) in (\ref{2.59}) and in SRG (\ref{2.58}), the factor
\( \mathcal{O}_{D-2}/(16\pi G_{N}\lambda ^{D-2}) \) is replaced by \( 1/2 \)
in (\ref{2.59}). We shall find it more convenient to stick to the latter normalization
so that e.g.\ for SRG (\ref{2.58}) \begin{equation}
U_{SRG}=-\frac{(D-3)}{(D-2)\, X}\,,\hspace{0.5cm}
V_{SRG}=-\frac{\lambda ^{2}}{2}\, (D-2)(D-3)\, X^{(D-4)/(D-2)}\; .
\label{2.79}
\end{equation}
 Of course, when introducing matter by spherical reduction (cf.\ sect. \ref{se:2.4.1})
the same overall normalization must be chosen.

\begin{figure}
\setlength{\extrarowheight}{6pt}
\begin{center}
\begin{tabular}{|l||>{$}c<{$}|>{$}c<{$}|c|} 
\hline
Model & V(X) & U(X) & Reference \\ \hline \hline
SRG ($D>3$) & -\frac{\lambda^2}{2}(D-2)(D-3) X^{\frac{(D-4)}{(D-2)}} & 
-\frac{(D-3)}{(D-2)X} & sect. \ref{se:2.2.1.1} \\ 
CGHS & -2\lambda^2 X & -\frac{1}{X} & sect. \ref{se:2.2.1.5} \\
JT & \La X  & 0 & sect. \ref{se:2.1.3} \\ 
KV & \frac{\be}{2}X^2-\La & \al & \cite{Katanaev:1986wk,Katanaev:1990qm} \\ 
\hline
\end{tabular} 
\end{center}
\setlength{\extrarowheight}{0pt}\index{twodimensional models}
\caption{A selection of dilaton theories}
\label{table1}
\end{figure}
The potentials for the most frequently used dilaton models are summarized in table \ref{table1}.

\subsection{Relation to Poisson-Sigma models}

\label{se:2.2.3}

Possibly the most important by-product of the approach to \( 2D \)-gravity
theories as presented in this report has been the realization that all models
of type (\ref{2.62}) are a special case of the more comprehensive concept
of Poisson-Sigma models (PSM) \cite{Ikeda:1994fh,Schaller:1994es,
Schaller:1994pm, Schaller:1994uj} with the action 
\begin{equation}
\label{2.80}
L^{\rm (PSM)}=\int\limits _{\mathcal{M}_{2}}\left[ dX^{I}\wedge A_{I}+\frac{1}{2}\mathcal{P}^{IJ}A_{J}\wedge A_{I}\right] \; ,
\end{equation}
 defined on a $2D$ base manifold \( \mathcal{M}_{2} \) with target
space \( \mathcal{N} \) with coordinates \( X^{I} \). Those coordinates as
well as the gauge fields \( A_{I} \) are functions of the coordinates \( x^{\mu } \)
on the base manifold (\( X^{I}(x),A_{I}(x) \)). The same symbols are used
to denote the mapping of \( \mathcal{M}_{2} \) to \( \mathcal{N} \). The \( dX^{I} \)
stand for the pullback of the target space differential \( dX^{I}=dx^{\mu }\partial _{\mu }X^{I} \)
and \( A_{I} \) are one-forms on \( \mathcal{M}_{2} \) with values in the
cotangent space of \( \mathcal{N} \).

The nontrivial (topological) content of a certain PSM is encoded in the Poisson
tensor \( \mathcal{P}^{IJ}=-\mathcal{P}^{JI} \) which only depends on the target
space coordinates. This tensor may be related to the Schouten-Nijenhuis bracket
\cite{Schouten:1954,Nijenhuis:1955}\begin{equation}
\label{2.81}
\{\; X^{I},X^{J}\; \}=\mathcal{P}^{IJ}
\end{equation}
 which is assumed to obey a vanishing bracket of \( \mathcal{P} \) with itself,
i.e.\ nothing else than a Jacobi identity which expresses the vanishing of the
Nijenhuis tensor \cite{Nijenhuis:1955} \begin{equation}
\label{2.82}
\mathcal{P}^{IL}\; \frac{\partial \mathcal{P}^{JK}}{\partial X^{L}}+\mbox {cycl}\, (I,J,K)=0\; .
\end{equation}
 Only for \( \mathcal{P}^{IJ} \) linear in \( X^{I} \) (in gravity theories
the Jackiw-Teitelboim model \cite{Barbashov:1979bm,D'Hoker:1982er,Teitelboim:1983ux,D'Hoker:1983is,D'Hoker:1983ef,Jackiw:1985je}),
eq.\ (\ref{2.82}) reduces to the Jacobi identity for the structure constants
of a Lie algebra and becomes independent of \( X \). In general the algebra
(\ref{2.81}) with (\ref{2.82}) covers a class of finite W-algebras \cite{deBoer:1996nu}.
Early versions of this nonlinear algebras from \( 2D \) gravity were discussed
as constraint algebra of the Hamiltonian in the context of the KV-model in \cite{Grosse:1992vc},
and with scalar and fermionic matter in \cite{Kummer:1992ef}. The interpretation
as a nonlinear gauge theory in a related approach goes back to \cite{Ikeda:1993aj, Ikeda:1994fh}.

Although we are dealing with bosonic fields in the present section our notation
anticipates already the graded PSM (gPSM) of supergravity in sect. \ref{se:2.4.3}.
Thus the index summation in (\ref{2.80}) is in agreement with the convention
used in supersymmetry and (just here and in sect.\ \ref{se:2.4.3}) we shall
also define instead of (\ref{2.20}) the exterior differentiation to act from
the right: \begin{equation}
\label{2.83}
d(\Omega _{p}\wedge \Omega _{q})=\Omega _{p}\wedge d\Omega _{q}+(-1)^{q}d\Omega _{p}\wedge \Omega _{q}
\end{equation}
 In the bosonic PSM for \( 2D \) gravity the action (\ref{2.80}) reduces to
(\ref{2.62}) with the identification \begin{equation}
X^{I}\rightarrow (X,X^{a})\;, \hspace{0.5cm}A_{I}\rightarrow (\om ,e_{a})\; .
\label{2.84}
\end{equation}
 The component \( \mathcal{P}^{aX} \) of \( \mathcal{P}^{IJ} \) is determined
by local Lorentz transformation for which (cf.\  (\ref{2.89}) below) \begin{equation}
\label{2.85}
\mathcal{P}^{aX}=X^{b}\epsilon _{b}{}^{a}
\end{equation}
is the generator. The components \begin{equation}
\label{2.86}
\mathcal{P}^{ab}=\mathcal{V}\, \epsilon ^{ab}
\end{equation}
 contain the potential \( \mathcal{V}(Y,X) \) which determines the specific
model ($Y=X^aX_a/2$).

With the present convention (\ref{2.83}), the e.o.m.-s from (\ref{2.80}) become
\begin{gather}
d X^{I} + \mathcal{P}^{IJ} A_{J} = 0 \label{2.87} \\
d A_{I} + \frac{1}{2}\, \left( \frac{\partial \mathcal{P}^{JK}}{
\partial X^{I}} \right)\; A_{K} \wedge A_{J} = 0\; .
\label{2.88}
\end{gather} The identities (\ref{2.82}) are the essential ingredient to show
the validity of the symmetries\footnote{%
Applying (\ref{2.89}) and (\ref{2.90}) to the commutator of infinitesimal transformations
the resulting one is again a symmetry only if the e.o.m.-s (\ref{2.87}) are used,
or if $\mathcal{P}^{IJ}$ is linear in \( X \) \cite{Strobl:1999wv}.
} \begin{align}
\delta X^{I} &= \mathcal{P}^{IJ}\, \epsilon_{J}\; ,\label{2.89} \\
\delta A_{I} &= - d \, \epsilon_{I} - 
\left( \frac{\partial \mathcal{P}^{JK}}{
\partial X^{I}} \right)\; \epsilon_{K} \, A_{J} 
\label{2.90}
\end{align} in terms of the local infinitesimal parameters \( \epsilon _{I}(x) \).
Eq.\ (\ref{2.90}) reveals the gauge field property of \( A_{I} \). Whereas
for \( 2D \) gravity with (\ref{2.84}), (\ref{2.85}), (\ref{2.86}) local
Lorentz-transformations (\( \epsilon _{I}\rightarrow \epsilon _{X} \)) can
be extracted easily from (\ref{2.89}) and (\ref{2.90}), diffeomorphisms (\ref{2.5})
are obtained by considering \( \epsilon _{I}=\xi ^{\mu }A_{\mu I} \) 
\cite{Strobl:1994yk}.
Evidently (\ref{2.80}) is invariant under target space diffeomorphisms too.
Only when those transformations are diffeomorphisms also globally, the topology
of \( \mathcal{M}_{2} \) remains unchanged. It should be noted that 
conformal transformations
of the world sheet metric can be expressed as target space diffeomorphisms.
Otherwise the problems discussed in sect.\ \ref{se:2.1.4} are relevant also at
the present, more general, level. Singular target space reparametrization (analogous
to the conformal transformations discussed there) could eliminate singularities
of the manifold \( \mathcal{M}_{2} \) if the identification (\ref{2.84}) of
the PSM variables is retained. Of course, an appropriate simultaneous (singular)
redefinition in the relation between \( A_{I} \) and the Cartan variables could
formally keep the topology of \( \mathcal{M}_{2} \) in terms of the new variables
intact, at the price of those singularities appearing in the relation between
\( A_{I} \) and \( (e_{a},\omega ) \).

In \( 2D \) gravity the Poisson tensor \( \mathcal{P}^{IJ} \) is not of full
rank, because the number of target space coordinates is odd. This also may happen
for general PSM-s and it implies the existence of {}``Casimir functions{}'',
whose commutator with \( X^{I} \) in the sense of (\ref{2.81}) vanishes. In
\( 2D \) gravity there is only one such function\footnote{%
For more details regarding generic PSM-s with more Casimir functions \( \mathcal{C} \)
we refer to ref. \cite{Strobl:1999wv}.
}\begin{equation}
\label{2.91}
\{X^{I},\mathcal{C}\}=\mathcal{P}^{IJ}\; \frac{\partial \mathcal{C}}{\partial X^{J}}=0\, .
\end{equation}
 The conservation of \( \mathcal{C} \) with respect to both coordinates of
the manifold \begin{equation}
\label{2.92}
d\mathcal{C}=dX^{I}\frac{\partial \mathcal{C}}{\partial X^{I}}=-\mathcal{P}^{IJ}\, A_{J}\, \frac{\partial \mathcal{C}}{\partial X^{I}}=0
\end{equation}
 follows from (\ref{2.91}) and the use of (\ref{2.87}) in (\ref{2.92}). Lorentz-invariance
requires \( \mathcal{C}=\mathcal{C}(Y,X) \) with \( Y=X^{a}X_{a}/2 \) and
thus according to (\ref{2.91}) \( \mathcal{C} \) must obey (cf.\ (\ref{2.85})
and (\ref{2.86})) \begin{equation}
\label{2.93}
\frac{\partial \mathcal{C}}{\partial X}-\mathcal{V}(Y,X)\; \frac{\partial \mathcal{C}}{\partial Y}=0\; .
\end{equation}
 This partial differential equation has a simple analytic solution for the physically
most interesting potentials of type (\ref{2.77}). It will be discussed in 
connection with the solution in closed form in sect. \ref{se:2.3.1}.

The rank of the Poisson tensor is not constant in general but may
change at special points in the target space or corresponding points 
on the world-sheet. A noteable example is a Killing-horizon. Thus,
the introduction of so-called Casimir-Darboux coordinates in which the Poisson 
tensor
\eq{
\mathcal{P}^{IJ}_{\textrm{CD}} = \left( \begin{array}{ccc} 
0 & 0 & 0 \\
0 & 0 & 1 \\
0 & -1 & 0 
\end{array} \right)
}{2.CD}
is constant only works patchwise. Such singular points may be modelled by
``Casimir-non-Darboux'' coordinates $Z^I$ 
\eq{
\mathcal{P}^{IJ}_{\textrm{CnD}} = \left( \begin{array}{ccc} 
0 & 0 & 0 \\
0 & 0 & Z^1 \\
0 & -Z^1 & 0 
\end{array} \right)\,.
}{2.CnD}
This allows the extension of patches over a point which is singular in 
Casimir-Darboux coordinates since for \(Z^1=0\) the rank changes 
from 2 to 0. In addition, however, such a coordinate system still may change 
the singularity structure of the original theory: e.g. a singularity like the 
one at $X=0$ in SRG is not visible in (\ref{2.CnD}); thus the transformation 
between these coordinate systems must be singular at $X=0$.

A different route to simplify the target space structure is symplectic 
extension \cite{Ertl:2001sj}. By adding an auxiliary target space coordinate 
one can elevate the Poisson structure to a symplectic structure. Again, 
this works only patchwise in general since the determinant of the 
Poisson-tensor can be singular. At a physical level, the symplectic extension 
resembles Kucha{\v{r}}'s geometrodynamics of the Schwarzschild BHs 
\cite{Kuchar:1994zk}: one introduces a canonically conjugate variable for the 
conserved quantity (in Kucha{\v{r}}'s scenario on the world-sheet boundary, in 
the symplectic extension in the bulk of target space).

For the application to gravity and supergravity theories in $D=2$ we
shall not need to know more about the PSM formulation. However, the field of
PSM-theories recently has attracted substantial interest in string theory \cite{Schomerus:1999ug, Seiberg:1999vs}
and, quite generally, in mathematical physics in connection with the Kontsevich
formula for the non-commutative star product\footnote{%
For the definition and physical applications of deformation quantization
the seminal papers \cite{Bayen:1978ha,Bayen:1978hb} may be consulted.
} \cite{Cattaneo:1999fm,Cattaneo:2001bp}.

The quantization of general PSM-s \cite{Hirshfeld:1999xm,Hirshfeld:2001cm} essentially
follows the approach which will be presented in sect. \ref{se:3.2} for the 
special case of dilaton gravity.


\clearpage

\section{General classical treatment}

\label{se:2.3} Simple counting of degrees of freedom shows that dilaton
gravity without matter fields in $2D$ has no propagating modes. 
Therefore, in terms of
suitable variables, the dynamics may be made essentially trivial. This suggests
(but in no way guarantees!) that all classical solutions can be found in a closed
form.

As pointed out already above, the fact that the solutions for dilaton theories
of type (\ref{2.59}) can be obtained in analytic form had been tested in many
specific cases \cite{Frolov:1992xx,Mann:1990gh,
Gegenberg:1993rg, Solodukhin:1993xs,  Louis-Martinez:1994cc, Gegenberg:1995pv},
always using the conformal gauge \eq{
ds^{2}=2e^{2\rho}dx^{+}dx^{-}
}{eq:conformalgauge} 
or the Schwarzschild gauge (cf. (\ref{eq:agensch}) below). With 
(\ref{eq:conformalgauge}) 
even for a theory as simple as (\ref{eq:JT}) the solution of a Liouville 
equation 
\begin{equation}
\label{eq:a1}
\square \rho =-\La e^{\rho }
\end{equation}
had been necessary. 

The advantages of the light cone gauge for Lorentz indices combined with a temporal
gauge for the Cartan variables \begin{equation}
\label{eq:a2}
e_{0}^{+}=0,\quad e_{0}^{-}=1,\quad \omega _{0}=0
\end{equation}
 was realized first \cite{Kummer:1992bg} in connection with the classical
solution of the KV-model \cite{Katanaev:1986wk}. In this gauge
the line element for any 2D gravity theory becomes \begin{equation}
\label{eq:a3}
(ds)^{2}=2e_{1}^{+}\, dx^{1}(dx^{0}+e_{1}^{-}dx^{1})
\end{equation}
 which in GR represents an Eddington-Finkelstein (EF) gauge
\cite{Eddington:1924,Finkelstein:1958}. In terms of the Killing field 
\( k^{\alpha }=(0,1) \), the existence of
which is a property of the solutions \( (\partial g_{\mu \nu }/\partial x^{1}=0) \)
and redefining the \( x^{0} \)-coordinate by \( d\bar{x}^{0}=e_{1}^{+}(x^{0})dx^{0} \),
the line element (\ref{eq:a3}) may be rewritten as \begin{equation}
\label{eq:a4}
(ds)^{2}=dx^{1}(2d\bar{x}^{0}+k^{2}dx^{1})
\end{equation}
 with the Killing norm \( k^{2}(\bar{x}^{0})=k^{\alpha }k_{\alpha } \) containing
all the information of the system (like \( \rho (x) \) in the conformal gauge
(\ref{eq:conformalgauge})). The key advantage of the ingoing (outgoing) EF
gauge as compared to the conformal or the Schwarzschild gauge is that it is
free from coordinate singularities on an ingoing (outgoing) horizon. The only
singularities of \( k^{2}(\bar{x}^{0}) \) correspond to singularities of the 
curvature; zeros \( k^{2}(\bar{x}^{0})=0 \) describe horizons. This gauge will 
turn out to be intimately related to the natural solution of the e.o.m.-s for 
all models in the first order formulation.

In the first subsection \ref{se:2.3.1} all classical solutions of GDTs 
without matter are determined in a very simple way, maintaining gauge
invariance. Among the specific gauge choices the EF gauge emerges as the most
natural one, also for the analysis of the global structure of these solutions.
The most important dilaton gravity models (cf. fig. \ref{table1}) belong to a 
two parameter sub-family of all possible theories.
This family of models is considered in more detail in the last subsection.

\subsection{All classical solutions}

\label{se:2.3.1}

In anticipation of what we shall need in sect. \ref{se:2.4.3} we derive the
e.o.m.s from an action (\ref{2.62}) supplemented as \( L=L^{(FOG)}+L^{(m)} \) 
by an, as yet, unspecified matter part \( L^{(m)} \). The quantities \eq{
W^{\pm }:= \delta L^{(m)}/\delta e^{\mp}, \quad W := \delta L^{(m)}/\delta X
}{eq:a9} contain the couplings to matter. A dependence of \( L^{(m)} \) on
the spin connection or the auxiliary fields \( X^{\pm } \) will be discarded
(cf.\ sect.\ \ref{se:2.4.3}). Variation of \( \delta \omega ,\delta e^{\mp },\delta X \),
and \( \delta X^{\mp } \), respectively, yields the e.o.m.-s 
\begin{eqnarray}
 &  & \hspace {3cm}dX+X^{-}e^{+}-X^{+}e^{-}=0\, ,\label{eq:a5} \\
 &  & \hspace {3cm}(d\pm \omega )X^{\pm }\mp \mathcal{V}e^{\pm }+W^{\pm }=0\, ,\label{eq:a6} \\
 &  & \hspace {3cm}d\omega +\epsilon \frac{\partial \mathcal{V}}{\partial X}+W=0\, ,\label{eq:a7} \\
 &  & \hspace {3cm}(d\pm \omega )e^{\pm }+\epsilon \, \frac{\partial \mathcal{V}}{\partial X^{\mp }}=0\, .\label{eq:a8} 
\end{eqnarray}
 The first equation (\ref{eq:a5}) can be used to eliminate the auxiliary fields
\( X^{\pm } \) in terms of \( e^{\pm } \) and \( dX \). The second 
pair (\ref{eq:a6}) is contained in the set of higher-dimensional Einstein equations
for the special case of dimensionally reduced gravity. Eq. (\ref{eq:a7})
yields the dilaton current \( W \) which is proportional to the trace of the
higher-dimensional energy momentum tensor for dimensionally reduced gravity,
and (\ref{eq:a8}) entails the torsion condition. 
If the potential
\( \mathcal{V} \) is independent of the auxiliary fields \( X^{\pm } \) the
condition for vanishing torsion (\ref{2.56}) is obtained. Of course, in addition
to (\ref{eq:a5})-(\ref{eq:a8}) the e.o.m.-s for matter \( \delta L^{(m)}/\delta \phi _{A}=0 \)
for generic matter fields \( \phi _{A} \) must be taken into account as well.

In the present section we are interested only in the direct solution of (\ref{eq:a5})-(\ref{eq:a8})
without fixing any gauge \cite{Kummer:1995qv}, in the absence of matter \( (W=W^{\pm }=0) \).
Linear combination of the two equations (\ref{eq:a6}), multiplied, respectively,
by \( X^{-} \) and \( X^{+} \) and using (\ref{eq:a5}) leads to \( (Y=X^{a}X_{a}/2=X^{+}X^{-}) \)\begin{equation}
\label{eq:a10}
d(X^{+}X^{-})+\mathcal{V}(Y,X)dX=0\; .
\end{equation}
 This indicates the existence of a conservation law for a function \( \mathcal{C}(Y,X)=C_{0}=\mbox {const.} \)
which is nothing else than the Casimir function of the Poisson-Sigma model of
sect.\ \ref{se:2.2.3}. In the application to physically motivated \( 2D \)
models, potentials of the form (\ref{2.77}) were found to be the most important ones.
We therefore concentrate on those. Multiplying (\ref{eq:a10}) by the integrating
factor \( \exp Q \) with \eq{
Q = \int^{X} U (y) dy, 
}{eq:aQ} one obtains the conservation law \begin{equation}
\label{eq:aC}
d\, \mathcal{C}=0
\end{equation}
 for the Casimir function \begin{equation}
\label{eq:a11}
\mathcal{C}=e^{Q}\, Y+w
\end{equation}
 with \begin{equation}
\label{eq:a12}
w(X)=\int\limits ^{X}\, e^{Q(y)}\, V(y)dy\, .
\end{equation}
 Of course, any function of \( \mathcal{C} \) is also absolutely conserved.
Therefore, for some specific model, among others, a suitable convention must
be used to fix the lower limit of integration in \( Q \) (influencing an overall
factor of \( \mathcal{C} \)) and the lower limit in (\ref{eq:a12}) (yielding
an additive overall contribution).

We assume \( X^{+}\neq 0 \) which will be realized (cf.\ (\ref{eq:a6}))
if \( V(X)=0 \) does not possess a nontrivial solution for \( X \). This is
true in SRG, but e.g.\  in the KV-model \cite{Katanaev:1986wk} such a {}``point-solution{}''
may appear for certain values of the parameters \cite{Schaller:1994np}. If
\( X^{+}\neq 0 \) the first component of eq.\  (\ref{eq:a6}) with a new one
form \( Z:=e^{+}/X^{+} \)
\begin{equation}
\label{eq:a13}
\omega =-\frac{dX^{+}}{X^{+}}+Z\mathcal{V}
\end{equation}
 determines the spin connection in terms of the other variables. In a similar
way eq.\ (\ref{eq:a5}) may be taken to define \( e^{-} \): \begin{equation}
\label{eq:a14}
e^{-}=\frac{dX}{X^{+}}+X^{-}Z
\end{equation}
 From (\ref{eq:a13}) and (\ref{eq:a14}) and eq.\  (\ref{eq:a8}) with the
upper sign, recalling that now \begin{equation}
\label{eq:a15}
\epsilon =-e^{-}e^{+}=-dX\,Z
\end{equation}
 for the potential (\ref{2.77}), the short relation \begin{equation}
\label{eq:a16}
dZ-dXZU=0
\end{equation}
 follows. The ansatz \( Z=\hat{Z}\exp Q \), with the same integrating factor
(\ref{eq:aQ})
as the one introduced above for \( \mathcal{C} \), reduces eq.\ (\ref{eq:a16})
to \( d\hat{Z}=0 \). Now by application of the Poincar\'{e} Lemma 
(cf.\ sect.\  \ref{se:2.1}) \( \hat{Z}=df \) 
is the only {}``integration{}'' necessary for the full solution\footnote{%
This type of solution has been given first in ref. \cite{Schaller:1994np}, 
starting from the Darboux coordinates for the KV-model \cite{Katanaev:1986wk}.
}: \begin{align}
e^{+} &= X^{+} e^{Q} df \label{eq:a17}\\
e^{-} &= \frac{dX}{X^{+}} + X^{-} e^{Q} df \label{eq:a18}\\
\omega &= - \frac{dX^{+}}{X^{+}} + \mathcal{V}\, e^{Q}\, df\label{eq:a19}\\
\mathcal{C} &= e^{Q}\, X^{+} X^{-} + w (X) = \mathcal{C}_{0} = \mbox{const.}
\label{eq:a20}
\end{align} Indeed, all the other equations are easily checked to be fulfilled
identically. Eq.\ (\ref{eq:a20}) can be used to express \( X^{-} \) in terms
\( X \) and \( X^{+} \), so that in addition to \( f \) beside the constant
\( \mathcal{C}_{0} \) we have the free functions \( X \) and \( X^{+} \).
Eqs. (\ref{eq:a5}-\ref{eq:a8}) are symmetric in the light cone coordinates.
Therefore, the whole derivation could have started as well from the assumption
\( X^{-}\neq 0 \).

It is straightforward, although eventually tedious in detail, to generalize
the solution (\ref{eq:a17})-(\ref{eq:a20}) to dilaton theories where in 
(\ref{2.59})
the factor of the Ricci scalar is a more general (non-invertible) function
\( Z(X) \). 

Comparing the number of arbitrary functions (\( f \), \( X \), \( X^{+} \)) in
the solution (\ref{eq:a17})-(\ref{eq:a20}), with the three continuous
gauge degrees of freedom, the theory is a topological one\footnote{%
In the sense that no continuous physical degrees of freedom are present \cite{Birmingham:1991ty}.
}, albeit of a different type from other topological theories like the Chern-Simons
theory \cite{Schwarz:1978cn,Schwarz:1979ae,Witten:1988ze,Witten:1988xj,Witten:1989hf}:
there is no discrete topological charge like the winding number associated to
the solutions. In agreement
with sect.\ \ref{se:2.2.3} the only variable which determines the different
solutions for a given action is the constant \( \mathcal{C}_{0}\in \mathbb {R} \).

The key role of \( \mathcal{C} \) is exhibited by the line element \begin{equation}
\label{eq:a21}
(ds)^{2}=2e^{+}\otimes e^{-}=e^{Q(X)}df\otimes [2dX+2(\mathcal{C}_{0}-w(X))\, df]
\end{equation}
 after elimination of \( X^{-} \) by (\ref{eq:a20}). Whenever a redefinition
of \( X \) by \eq{
d\widetilde{X} = dX \exp Q
}{eq:a21.5} is possible\footnote{%
If the redefinition is not possible for all \( X \) the computation of geodesics
should start from (\ref{eq:a21}).
} eq.\ (\ref{eq:a21}) becomes \begin{align}
(ds)^{2} &= 2df \otimes d\widetilde{X} + \xi(\widetilde{X}) df \otimes df\, , 
\label{eq:a22}\\
\xi(\widetilde{X}) &= 2e^{Q} (\mathcal{C}_{0} -w)_{X=X(\tilde{X})}\, ; 
\label{eq:a23}
\end{align}
i.e.\ the EF gauge is obtained when $f$ and $\widetilde{X}$ are taken directly 
as the coordinates. Then $\xi(\widetilde{X})$ coincides with the Killing norm 
$k^2$ (cf.\ (\ref{eq:a4})). As we shall see in the next subsection the 
``topological'' properties of $\xi (\widetilde{X})$, i.e.\ the sequence of 
singularities and zeros (horizons), and the behavior at the boundaries of the 
range for $\widetilde{X}$, completely determines the global 
structure of a solution. Eqs. (\ref{eq:a21}-\ref{eq:a23}), together with the definitions (\ref{eq:aQ}), (\ref{eq:a12}) represent the main result of this section. They are exact expressions for the geometric variables and thus also for the line element $(ds)^2$ valid for (almost) arbitrary dilaton gravity models without matter. 
It is now easy to specify other gauges by taking 
(\ref{eq:a21}) as the point of departure, that is 
\textit{after} having solved the e.o.m.-s 
(\ref{eq:a5})-(\ref{eq:a8}) in the simple manner 
demonstrated above, namely inserting 
$(x^0 = t, x^1=r, F'=\partial F/\partial r, \dot{F} = \partial F/\partial t$) 
\begin{equation}
\label{eq:a24}
d\widetilde{X} = \widetilde{X}' dr + \dot{\widetilde{X}} dt\, , 
\hspace{0.5cm}df = f'dr + \dot{f} dt\, ,
\end{equation}
into (\ref{eq:a22}) with (\ref{eq:a23}) with appropriate choices for 
$\tilde{X}$ and $f$. Of particular interest are diagonal gauges,
a class of gauges to which prominent choices (Schwarzschild and conformal 
gauge) belong. The absence of mixed 
terms $drdt$ in the metric can be guaranteed in a certain patch by the gauge 
conditions 
\begin{align}
\widetilde{X} &= \widetilde{X}(r)\, , \label{eq:a25}\\
\widetilde{X}'+ \xi f' &= 0\, , \label{eq:a26}
\end{align}
and $\dot{f} \neq 0$. The solution\footnote{$\bar{f}(t)$ is
arbitrary except $\dot{\bar{f}}\neq 0$.} for $f$ from  
(\ref{eq:a26}) 
\begin{align}
f &= - \int\limits^r dx \frac{1}{\xi\, (\tilde{X}(x))} 
\; \cdot \; \frac{d\tilde{X}(x)}{dx} + \bar{f} (t) = \nonumber\\
&= - \frac{\mathcal{K}(\widetilde{X})}{2} + \bar{f} (t) \; ,
\label{eq:a27}
\end{align}
contains the integral $\mathcal{K}(\widetilde{X})$ defined by 
\eq{
\mathcal{K} (r) = 2\, \int\limits_{r_0}^r \, dy 
\xi^{-1} (y).
}{eq:a50}

The diagonal line element
\eq{
(ds)^2=\xi\left[(\dot{f}dt)^2-(f'dr)^2\right],
}{eq:aconf}
for $\dot{f}=f'=1$ attains the form of the conformal gauge. Requiring
furthermore $\det g=-1$ as in Schwarzschild type gauges yields
\eq{
(ds)^2=\xi(dt)^2-\xi^{-1}(dr)^2.
}{eq:agensch}

As a concrete example we take SRG (\ref{2.79}) with 
$D=4$, where $Q_{SRG} = \int^X U_{SRG} (y) dy = -1/2\, \ln 
X$ with a natural choice $y=1$ for the lower limit of 
integration and $w_{SRG} = -2\lambda^2 \sqrt{X}$ with the 
lower limit $X=0$ (cf.(\ref{eq:aQ}), (\ref{eq:a12})). The conserved quantity (\ref{eq:a11}) becomes 
\begin{equation}
\label{eq:a33}
\mathcal{C}_{SRG} = \frac{X^+X^-}{\sqrt{X}} - 2\lambda^2 
\sqrt{X} = \mathcal{C}_0\, ,
\end{equation}
and the Killing norm (\ref{eq:a23}) reads
\begin{equation}
\label{eq:a34}
k^2_{SRG} = \xi_{SRG} = \frac{2\mathcal{C}_0}{\sqrt{X}} + 4\lambda^2\; .
\end{equation}
In terms of the new variable 
$\widetilde{X}=2\sqrt{X}$ the EF gauge (\ref{eq:a22}) follows 
with $\xi(\widetilde{X})=4\mathcal{C}_0/\widetilde{X}+4\la^2$. Further trivial
redefinitions ($r=\widetilde{X}/(2\la)$, $u=2\la f$, $dt=du+dr/\xi$) yield
the Schwarzschild metric \cite{Schwarzschild:1916uq}
\begin{equation}
\label{eq:a35}
(ds)^2_{sch} = \left( 1 - \frac{2M}{r} \right) (d t)^2 - 
\left( 1 - \frac{2M}{r} \right)^{-1}(dr)^2\,,
\end{equation}
where, as expected,  $\mathcal{C}_0$ is related to the mass $M$
\begin{equation}
\label{eq:a36}
M = - \frac{\mathcal{C}_0}{4\lambda^3}\; .
\end{equation}
From the steps leading to the line element (\ref{eq:a21}) or 
to one of its subsequent versions it is obvious that these steps can 
be retraced backwards as easily, say from a Killing norm 
$\xi (\widetilde{X})$ in (\ref{eq:a23}) towards an action. 
This procedure is not unique, because one function $\xi 
(\widetilde{X})$ is to be related to two other functions ($U$ 
and $V$).

For an associated model (\ref{2.59}) with $U=0$, from 
(\ref{eq:a12}) and (\ref{eq:a23}) the potential $V(X)$ simply results by 
differentiation $dw/dX$. However, as emphasized already 
above, these models essentially encode their topology in the parameters 
of the action determined by $V(X)$. In the Killing norm 
$\xi = 2 (\mathcal{C}_0 - w)$ the value of $\mathcal{C}_0$ 
by which the solutions differ in those models only influences the position 
of the zeros of $\xi$, the horizons. For instance,  the 
Reissner-Nordstr\"om solution of eq.\ (\ref{2.60}) follows 
from the potential anticipated in that equation. As will be 
seen in sect.\ \ref{se:2.4}, when interactions with matter are 
turned on, $\mathcal{C}$ of the present chapter is part of 
the conservation law involving matter contributions. Thus 
e.g.\ the influx of matter only changes $\mathcal{C}$ and 
hence in models with $U=0$ the 
position of the horizons, but does not change the ``strength'' 
of the singularities which is fixed here by the mass and the charge $Q$. 

Actually SRG belongs to the interesting class of models in 
which a given $\xi (\widetilde{X})$ is related to 
\textit{both} functions $U$ and $V$ in a very specific way, 
namely, such that 
$\mathcal{C}_0 = 0$ corresponds to a flat (Minkowski) 
manifold. According to (\ref{eq:a21}) this can be simply 
achieved by the condition
\begin{equation}
\label{eq:a37}
e^Q\, w = \alpha = \mbox{const.}\, ,
\end{equation}
because then the only dependence on $X$ resides in the term 
$\propto \mathcal{C}_0$ \cite{Katanaev:1996bh}. Thus all models of 
this class (``Minkowski ground state dilaton theories'') are 
characterized by the relation between $U$ and $V$ in (\ref{2.59}) 
following from (\ref{eq:a37}):
\begin{equation}
\label{eq:a38}
V(X) = \frac{\alpha}{2}\, \frac{d}{dX}\, \exp \left[ 
- 2 \int\limits^X U (y) dy \right] 
\end{equation}

A last related remark concerns the conformal transformation (\ref{2.53}) 
of a dilaton theory represented by the first order action (\ref{2.62}). 
It is always possible, starting from a 
theory with $\bar{V} (X)$ and vanishing $\bar{U} (X)$ to 
arrive at a model with nonvanishing $U(X)$ by the 
transformation 
\begin{align}
\label{eq:a39}
e^a &= \bar{e}^{a}\, e^{Q/2} \, \nonumber\\
X^a &= \bar{X}^a\, e^{-Q/2} \, \nonumber\\
\omega  &= \bar{\omega} + \frac{U}{2}\, \bar{X}^a \, 
\bar{e}_a\, \nonumber\\
\mathcal{V} &= e^{-Q}\, \bar{V} (X) + \frac{X^aX_a}{2}\, 
U(X) \, , 
\end{align}
where $Q$ is defined as in (\ref{eq:aQ}). In the language of the PSM 
formulation (sect. 2.3) this is an explicit example of a target space 
diffeomorphism\footnote{Cf. the discussion after (\ref{2.90}).}. However, as 
pointed out already 
several times, this mathematical transformation 
connects two models with solutions of, 
in general,  
completely different topology and/or properties regarding 
the role of the conserved quantity $\mathcal{C}_0$.

\subsection{Global structure}\label{se:2.3.2}

As emphasized in sect. \ref{se:2.1} the global properties of a 
solution for the geometric variables are obtained by 
following the path of some device on the manifold. The most 
important example is the geodesic (\ref{2.42}) which may 
penetrate horizons, but ends when singularities are encountered 
at finite affine parameter. When no geodesic can reach a boundary 
of the space-time for finite values of the affine parameter the space-time is
called geodesically complete, otherwise geodesically incomplete.  It should be
emphasized that the procedure presented below does not require the explicit or
implicit knowledge of Kruskal-like global coordinates.

For the analysis of the global structure it is convenient to use
outgoing or ingoing EF coordinates. In a simplified notation
\cite{Klosch:1996qv} the line element (\ref{eq:a22}) is written 
 for the outgoing case ($f \propto u, \widetilde{X} 
\propto r, k^2 (r) = \xi (r), \xi_{\infty}:=+1$)
\eq{
(ds)^2_{out} = du ( 2dr + \xi (r) du )\; .
}{eq:a39.5}
The ingoing EF gauge (still with $\xi_{\infty}=+1$)
\eq{
(ds)^2_{in}  = dv ( 2dr - \xi (r) dv )\; ,
}{eq:a40}
which will be used here in order to construct patch 
$\mathcal{A}$ of the conformal diagram for Schwarzschild space-time (see 
below), could have been obtained if one uses $Z:=e^-/X^-$ for $X^-\neq0$ in 
(\ref{eq:a13}-\ref{eq:a20}). Eq. (\ref{eq:a40}) is the most suitable starting point for our
subsequent arguments in the present example.

For the ingoing metric ($x^\alpha = \{ v,r \}$) 
\begin{equation}
g_{\mu\nu} = \left(
\begin{array}{cc}
-\xi(r) & 1 \\
1 & 0
\end{array}
\right)
\label{eq:a41}
\end{equation}
the geodesics ($ v = v(\tau),\, r = r(\tau)$) obey
\begin{gather}
\ddot{v} + \frac{\dot{v}^2}{2}\, \xi' = 0\; ,\label{eq:a43} \\
\label{eq:a42} \ddot{r} - \dot{r} \dot{v} \xi' + 
\frac{\dot{v}^2}{2}\xi'\xi = 0 \, .
\end{gather}
These are the e.o.m.-s of (\ref{2.41}) if the affine parameter 
$\tau$ is identified with $s$. The Killing field 
$\partial/\partial v$ implies a constant of motion 
($k^\alpha = (1,0)$) 
\begin{equation}
g_{\alpha\beta} k^\alpha \dot{x}^\beta = \dot{r} - 
\xi(r) \dot{v} = \sqrt{\vert \tilde{A}\vert } = \mbox{const.}
\label{eq:a44} 
\end{equation}
which could have been derived as well from (\ref{eq:a42}) and 
(\ref{eq:a43}) by taking a proper linear combination. 
From (\ref{eq:a44}) the affine parameter $\tau$ can be
identified with parameters describing the line-element (\ref{eq:a40})
\begin{equation}
\pm (d \tau)^2 = \frac{1}{\vert \tilde A \vert } \, 
(dr - \xi dv )^2 = (ds)^2\,, 
\label{eq:a45}
\end{equation}
where the two signs correspond to 
time-like, resp.\ space-like geodesics. 
The  first order differential equation 
\begin{equation}
\frac{dv}{dr} = \frac{1}{\xi (r)} \, 
\left[ 1 \mp  \left( 1 + \frac{\xi}{A}\, \right)^{-1/2} \, 
\right]\,,
\label{eq:a46}
\end{equation}
for the two signs in (\ref{eq:a46}) describes two types 
of geodesics $v^{(1)} (r)$ and $v^{(2)} (r)$ at each point 
$\{v,r\}$. To avoid confusion with the association of signs, in the new 
constant $A = \pm \vert\tilde{A} \vert$ the two signs from (\ref{eq:a45}) have 
been absorbed so that now $A>0$ and $A<0$ correspond to time-like, 
resp.\ space-like geodesics. Inserting (\ref{eq:a46}) into (\ref{eq:a45}) 
provides the relation 
\begin{equation}
s(r) = \pm \frac{1}{\sqrt{\vert A \vert}}\, 
\int\limits_{r_0}^r \, dy \,
\frac{1}{\sqrt{1 + \xi(y)/A}}.
\label{eq:a47}
\end{equation}
``Special'' geodesics $\tilde{A} = 0$ with 
\begin{align}
\label{eq:a51}
\frac{dv}{dr} &= \xi^{-1} (r)\,,\\
s &= \int\limits_{r_0}^r\, dy \; \vert \xi (y) \vert^{-1/2}\,,
\label{eq:a52}
\end{align}
and ``degenerate'' ones obeying $A = \xi (r)=\text{const.}$ must 
be considered separately. 

The advantage of the EF-gauge is visible e.g.\ in 
(\ref{eq:a46}). The geodesic with the upper sign in the 
square bracket clearly passes continuously ($C^\infty$) 
through a horizon where $\xi (r_h) = 0$ which, therefore, 
does not represent a boundary of the patch for the 
solution (provided $A\neq 0$).

For a first orientation of the global properties of a 
manifold it is sufficient to study the behavior of 
null-directions. Light-like directions are immediately 
read off from $(ds)^2 = 0$ in (\ref{eq:a40}):
\begin{align}
\label{eq:a48} 
(dv=0) \rightarrow v^{(1)} &= \mbox{const.} := v_0^{(1)}\\
\label{eq:a49} 
(dv\neq0) \rightarrow v^{(2)} &= \mathcal{K} (r) + \mbox{const.} := 
\mathcal{K} (r) +  v_0^{(2)}
\end{align}
with $\mathcal{K}$ defined in (\ref{eq:a50}).
In terms of the new variables
\begin{equation}
\tilde{v} = v^{(1)}\,, \hspace{0.5cm} \tilde{u} = v^{(2)} - \mathcal{K} (r)
\label{eq:a53} 
\end{equation}
those null-directions become the straight lines $\tilde{v} 
= \mbox{const.}\; \tilde{u} = \mbox{const.}$.  The line 
element in these (conformal) coordinates, of course, exhibits
(coordinate) singularities at the horizons. It should be stressed 
that conformal coordinates are being used here only in order to be in
agreement with standard diagrammatic representations.

Furthermore, it is convenient to map $(ds)^2$ by a conformal 
transformation onto a finite region by considering e.g. \cite{waldgeneral}
\begin{equation}
(d\tilde{s})^2 = \frac{2 d\tilde{u} d\tilde{v}\, \xi 
(r (\tilde{u}))}{(1 + \tilde{u}^2) ( 1 + \tilde{v}^2 )}
\label{eq:a55}
\end{equation}
where the powers of the two factors in the denominator are chosen 
appropriately. Light-like
geodesics are mapped onto light-like geodesics, i.e. the causal structure is 
not changed by this transformation. The conformal diagrams obtained in this 
way have been introduced by Carter and Penrose 
\cite{Carter:1969,Penrose:1969pc}. 

The trivial example is Minkowski space with $\xi = 1$. From (\ref{eq:a55}) 
with (\ref{eq:a50}) and (\ref{eq:a56}) both light-cone variables 
\begin{equation}
\tilde{u} = v - 2(r-r_0)\,, \hspace{0.5cm} \tilde{v} = v
\label{eq:a56}
\end{equation}
lie in $ -\infty \leq \{ \tilde{u}, \tilde{v} \} \leq + 
\infty$. By the compactification (\ref{eq:a55}), in the 
line element
\begin{equation}
(d \tilde{s})^2 = 2 dU dV
\label{eq:a57}
\end{equation}
the new variables $U = \arctan \tilde{u}$, $ V = \arctan 
\tilde{v}$ are restricted to the finite interval $-\pi/2 
\leq U,\, V \leq +\pi/2$. 

\subsubsection{Schwarzschild metric}

As a typical nontrivial example for the general procedure in curved 
space \cite{Klosch:1996qv} we take the Schwarzschild BH with 
\begin{equation}
\xi (r) = 1-\frac{2M}{r} \; .
\label{eq:a58}
\end{equation}
The second light-like coordinate, solving (\ref{eq:a49}) 
with (\ref{eq:a50})
\begin{equation}
v^{(2)} = v_0^{(2)} + 2r^\ast=v_0^{(2)} + 2r + 4M \ln 
\left\vert 1 - \frac{r}{2M} \right\vert \,,
\label{eq:a59}
\end{equation}
\begin{wrapfigure}{r}{160pt}
\begin{picture}(160,91)(0,10)
    \put(0,0){\mbox{\resizebox*{153pt}{101pt}{\includegraphics{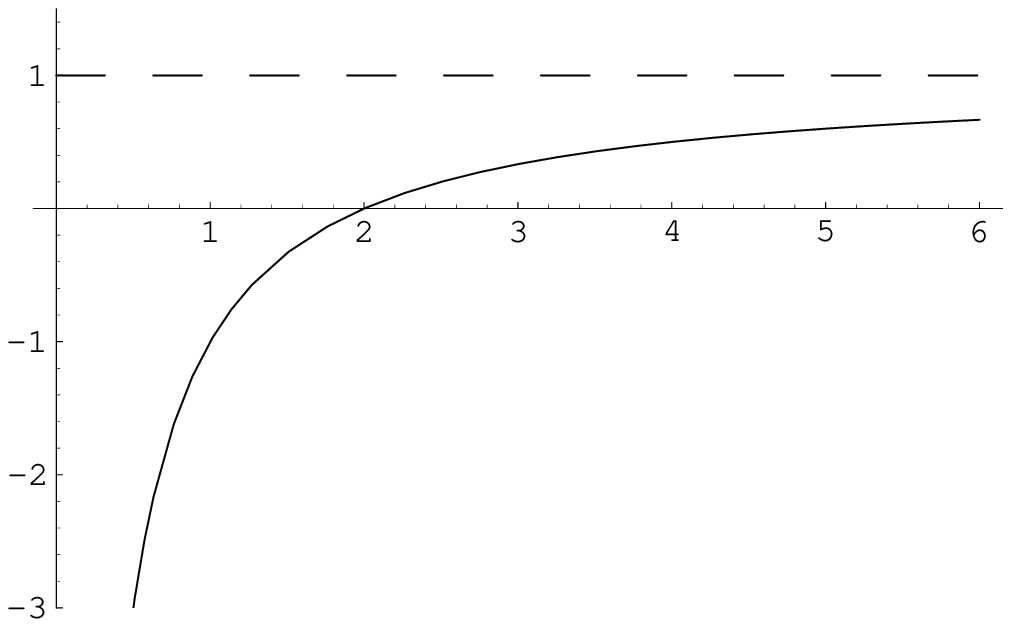}}}}
    \put(20,80){\makebox(0,0)[l]{A}}
    \put(0,110){\makebox(0,0)[l]{$\xi(r)$}}
    \put(100,50){\makebox(0,0)[l]{B}}
    \put(160,70){\makebox(0,0)[l]{$r$}}
\end{picture}
\caption[Killing norm for Schwarzschild metric]{}
\label{fig:2.2}
\begin{picture}(160,110)(0,10)
    \put(0,0){\mbox{\resizebox*{153pt}{101pt}{\includegraphics{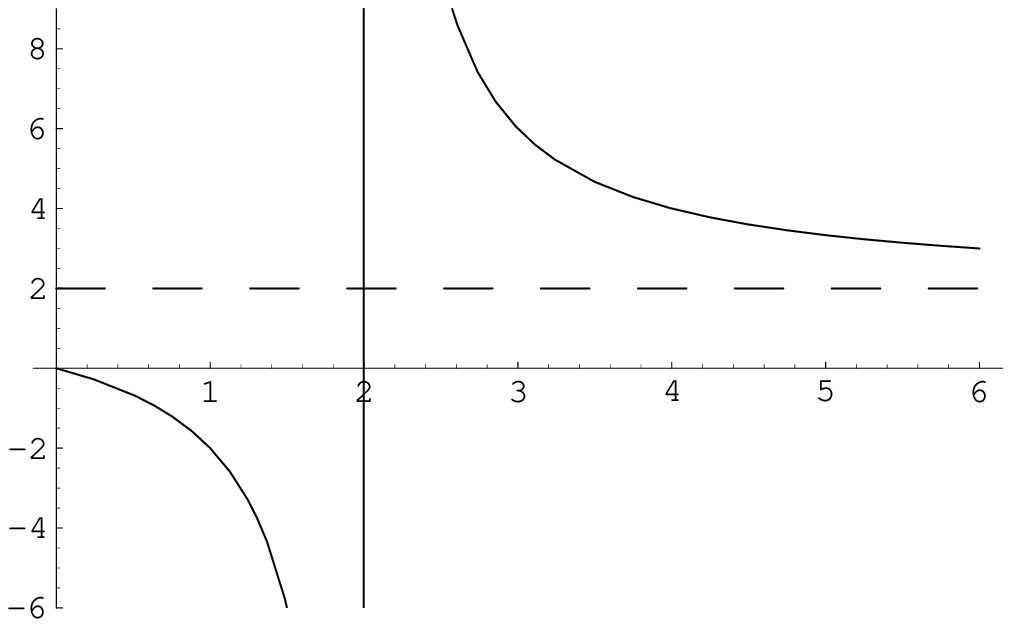}}}}
    \put(20,48){\makebox(0,0)[l]{A}}
    \put(0,110){\makebox(0,0)[l]{$dv^{(2)}/dr$}}
    \put(100,48){\makebox(0,0)[l]{B}}
    \put(160,44){\makebox(0,0)[l]{$r$}}
\end{picture}
\caption[Derivative of the second null direction]{}
\label{fig:2.3}
\begin{picture}(160,110)(0,10)
    \put(0,0){\mbox{\resizebox*{153pt}{101pt}{\includegraphics{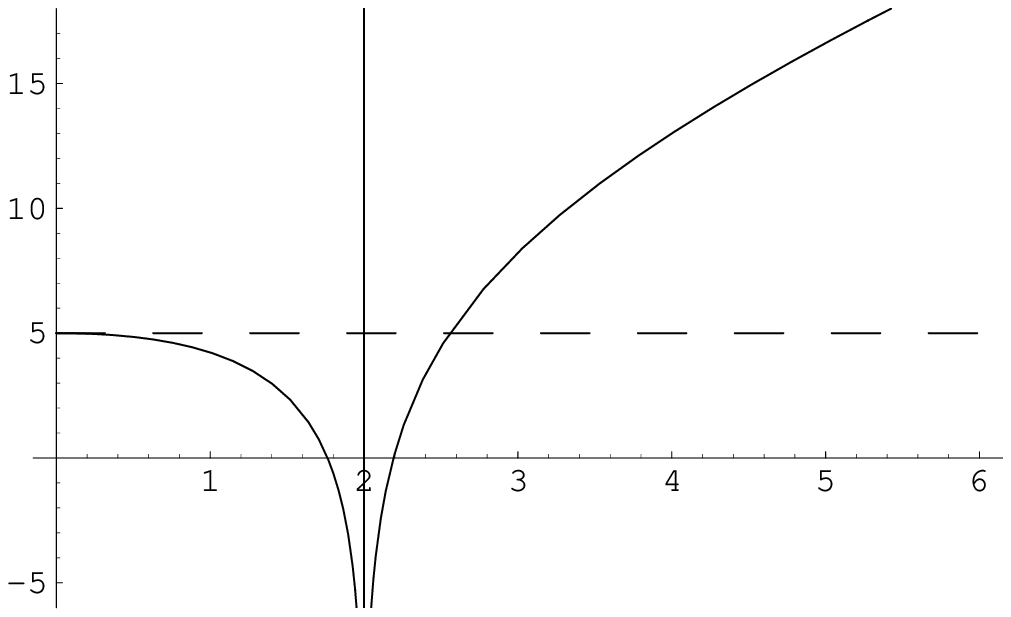}}}}
    \put(20,35){\makebox(0,0)[l]{A}}
    \put(0,110){\makebox(0,0)[l]{$v^{(2)}$}}
    \put(100,35){\makebox(0,0)[l]{B}}
    \put(160,28){\makebox(0,0)[l]{$r$}}
\end{picture}
\caption[Second null direction]{}
\label{fig:2.4}
\end{wrapfigure}
is intimately related to the ``Regge-Wheeler tortoise coordinate'' $r^\ast$, 
but, as we see below, the actual integration need not even be performed. It is 
sufficient to just regard the general features of the curves. 

\begin{figure}[t]
\centering
\epsfig{file=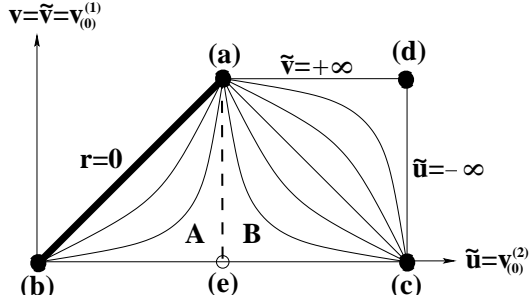,width=.5\linewidth}

\caption{Conformal coordinates with ``compression factor''}
\label{fig:2.5}
\end{figure}

The steps from Fig. \ref{fig:2.2} to Fig. \ref{fig:2.4} are obvious by 
inspection. The change to conformal (null) coordinates in Fig. \ref{fig:2.5} 
implies the introduction of $\tilde{u}$  as the horizontal axis. 
Thus the curves $v^{(2)} = \mbox{const.}$ in Fig. \ref{fig:2.4} are to be 
``straightened'' into vertical lines. Above the line $\tilde{u}=0$ 
this pushes the lines $r=\mbox{const.}$ in the regions $A$ 
and $B$ of Fig. \ref{fig:2.2} to \ref{fig:2.4} together so that they all 
terminate in the point $(a)$ in \ref{fig:2.5}. For negative $\tilde{u}$ those 
lines are pushed apart to end in the corners $(b)$ and $(c)$. The value $r=2M$
corresponds to the lines (b)-(c) and (a)-(e) with the exception of the
endpoints (a), (b) and (c). Similarly, the value $r=\infty$ corresponds to the 
lines (a)-(d) and (c)-(d), except for the endpoints (a) and (c).
The integration constant $v_0^{(2)}$, the endpoint of those 
curves for $r=0$ in Fig. \ref{fig:2.4}, always terminates at some finite 
value which is smaller than all $v^{(1)}\vert_{r=0} > v_0^{(2)}$. Therefore,
the left-hand boundary in Fig. \ref{fig:2.5} for $r=0$ 
experiences a ``cut off'', described by the line from $(a)$ 
to (b)\footnote{Whether this is really a straight 
line as drawn in Fig. \ref{fig:2.5} depends among others on the 
compression factor. The same is also true for the other 
boundaries at $r= \infty$, $\tilde{u} = -\infty$. However, the shape 
of those curves is irrelevant, as far as the topological 
properties are concerned which are determined by their mutual arrangement 
only.}. In the language of general relativity the nomenclature for the 
points (a), (c), (d) and (e) is, respectively, $i^+$, $i^-$, $i^0$ and the 
bifurcation-two-sphere. The lines (a)-(b), (a)-(d), (d)-(c) and
(a)-(e) are, respectively, the singularity, $\scri^+$, $\scri^-$ and the 
Killing horizon.

\begin{figure}
\centering
\epsfig{file=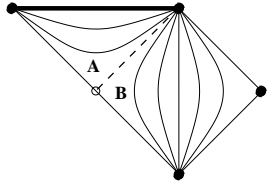,width=.25\linewidth}
\caption{Reorientation of Fig. \ref{fig:2.5}: patch $\mathcal{A}$}
\label{fig:2.6}
\end{figure}

We emphasize again that in the EF gauge the  whole patch of 
Fig. \ref{fig:2.5} is connected by continuous geodesics. A treatment in 
the conformal gauge \cite{Katanaev:1993fu}, although using simpler geodesics, 
suffers from the drawback that the connection between the 
regions $A$ and $B$ must be made by explicit continuation 
through the coordinate singularity at the horizon $r = 2M$. 
We now turn Fig. \ref{fig:2.5} by 45$^\circ$ (Fig. \ref{fig:2.6}) and call it 
patch $\mathcal{A}$. Clearly   $r \to \infty$ is complete in the 
sense of sect. \ref{se:2.1.1} because there the space becomes 
asymptotically flat (cf.\ (\ref{eq:a58})). 
The singularity at $r=0$ can be reached for finite affine 
parameter. At the edge $v=-\infty$ incompleteness is observed,
and (in conformal gauge) a coordinate singularity. 
Therefore, an extension must be possible. 

\begin{figure}
\centering
\epsfig{file=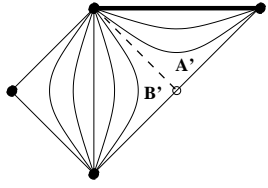,width=.25\linewidth}
\caption{Mirror image of Fig. \ref{fig:2.6}: patch $\mathcal{B}$}
\label{fig:2.7}
\end{figure}

Indeed, 
introducing coordinates $v^{\mathcal{B}}, r^{\mathcal{B}}$ in patch 
$\mathcal{B}$ by
\begin{equation}
\begin{split}
r^{\mathcal{B}} = r \\
v^{\mathcal{B}} = \mathcal{K}(r) - v
\end{split}
\label{eq:a61}
\end{equation}
with
\begin{equation}
\begin{split}
dr &= d r^{\mathcal{B}} \\
dv &= - d v^{\mathcal{B}} + 2 \xi^{-1}(r^{\mathcal{B}})\, dr^{\mathcal{B}}
\end{split}
\label{eq:a62}
\end{equation}
again transforms the line element (\ref{eq:a40}) into itself, but 
with the replacement $r \to r^{\mathcal{B}}, v \to v^{\mathcal{B}}$. 
Moreover, we obtain the same differential equations as the ones in the 
patch $\mathcal{A}$ except for the change of sign $v \to -v$ 
(cf.\ (\ref{eq:a61})). As a consequence patch $\mathcal{B}$ is 
given by Fig. \ref{fig:2.7}, the mirror image of Fig. \ref{fig:2.6}.

\begin{figure}
\centering
\epsfig{file=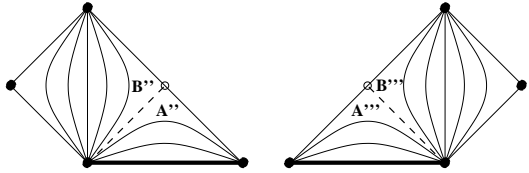,width=.5\linewidth}
\caption{Further flips: patches $\mathcal{C}$ and $\mathcal{D}$}
\label{fig:2.8}
\end{figure}

\begin{figure}
\centering
\epsfig{file=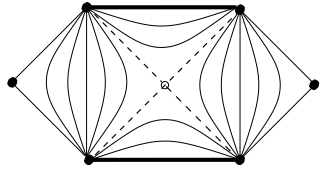,width=.3\linewidth}
\caption{CP diagram for the Schwarzschild solution}
\label{fig:2.9}
\end{figure}

Further patch solutions $\mathcal{C}$ and $\mathcal{D}$ can 
be obtained by simply changing both signs on the right-hand side of
(\ref{eq:a53})
resp.\ (\ref{eq:a61}), yielding the patches of Fig. \ref{fig:2.8}.
Now the key observation is that the lines $r=\mbox{const.}$ 
correspond to the same variable in the regions $A$ of $\mathcal{A}$ 
and $A'$ of $\mathcal{B}$. The same is true in $B'$ and 
$B''$ for $\mathcal{B}$ and $\mathcal{C}$, and for $A''$ and 
$A'''$ for $\mathcal{C}$ and $\mathcal{D}$. Superimposing 
those regions we arrive at the well-known Carter-Penrose (CP) diagram 
for the Schwarzschild solution (Fig. \ref{fig:2.9}).

\subsubsection{More general cases}

We have glossed over several delicate points in this procedure 
\cite{Klosch:1996fi,Klosch:1996qv}. As pointed out at the beginning of this 
section in a more complicated case a careful analysis of geodesics 
is necessary at external boundaries and, especially, at the 
corners of a diagram like Fig. \ref{fig:2.9}. One may encounter 
``completeness'' in this way in some corners, but also in the middle of a 
diagram, resembling Fig. \ref{fig:2.9}. Still, in all those cases the analysis 
does not need the full solution of the geodesic equations 
(\ref{eq:a43}), (\ref{eq:a42}). It suffices to check their 
properties in the appropriate limits only. 

Also the diagram alone may not be sufficient to read off 
some important ``physical'' properties. The line of 
reasoning, passing through the Figs.\ \ref{fig:2.2} - \ref{fig:2.6} shows that 
obviously all Killing norms $\xi(r)$ with one singularity, one (single) 
zero and $\xi_{\infty} = 1$ will lead to the same 
diagram Fig. \ref{fig:2.9}. However, e.g.\ the incomplete boundary at the 
singularity may behave differently. For the CGHS
model \cite{Callan:1992rs} in which the power $r^{-1}$ (or 
$r^{-(D-3)}$ for SRG from $D>4$) is replaced by an 
exponential $e^{-r}$, only time-like geodesics are 
incomplete at $r=0$. This means that light signals take 
``infinitely long time'' to reach the singularity (null completeness) , whereas 
massive objects do not. 

Another important point to be checked is whether by 
superimposing patches around some center as the bifurcation 2-sphere 
in Fig. \ref{fig:2.9} one 
really arrives at $B=B'''$ (uniqueness), or whether this can 
be obtained only by imposing certain further conditions. Otherwise 
not a planar picture like Fig. \ref{fig:2.9}, but an infinite 
continuation in the form of a ``spiral staircase'' extending 
above (and below) the drawing plane may emerge. 

\begin{figure}
\centering
\epsfig{file=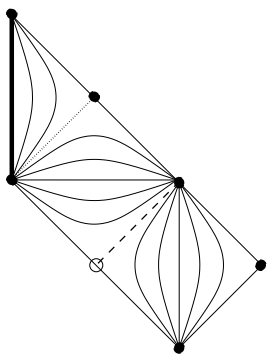,width=.25\linewidth}
\caption{Basic patch of  Reissner-Nordstr\"om metric}
\label{fig:2.10}
\end{figure}

\begin{figure}
\centering
\epsfig{file=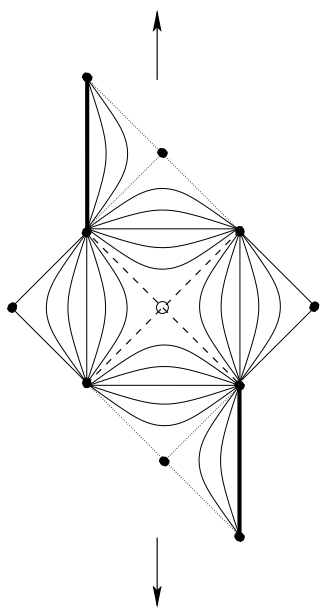,width=.3\linewidth}
\caption{Penrose diagram for Reissner-Nordstr\"om metric}
\label{fig:2.11}
\end{figure}

When $\xi$ exhibits two zeros as for the Reissner-Nordstr\"om metric
\begin{equation}
\xi_{RN} = 1-\frac{2M}{r} + \frac{Q^2}{r^2}
\label{eq:a63}
\end{equation}
the basic patch of Fig. \ref{fig:2.10}
with the superposition method described above, leads to the 
well-known \cite{Israel:1966,Klosch:1996bw} one-dimensional infinite periodic 
extension of Fig. \ref{fig:2.11}.

When even three zeros  are present in the Killing norm 
the global diagram becomes periodic in 
two directions (cf. e.g. Fig. 2.1-2.3 in \cite{Grumiller:2001ea}), 
i.e.\ covers the whole plane\footnote{Here we have even discarded the 
``uniqueness''-problem, referred to above.}.

\begin{figure}
\centering
\epsfig{file=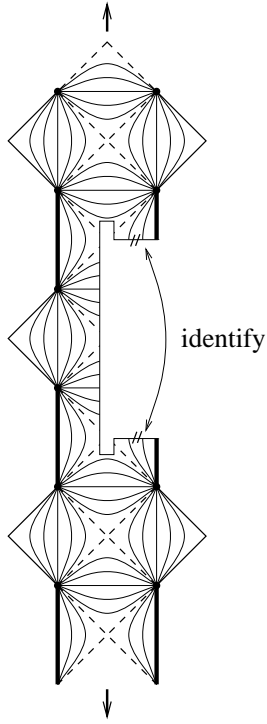,width=.25\linewidth}
\caption[A possible RN-kink]{A possible RN-kink (cf. \cite{Klosch:1998yh})}
\label{fig:2.11.5}
\end{figure}

Such one, two or more dimensional lattices exhibit discrete 
symmetries, which, in turn, may be used to compactify 
manifolds by identifying certain curves. If this 
identification occurs in a nontrivial manner, ``solitonic'' 
manifolds are produced \cite{Klosch:1998yh}, as illustrated by the example 
Fig. \ref{fig:2.11.5}. 




\subsection{Black hole in Minkowski, Rindler or de Sitter space}

\label{abfamily} 
As a further illustration for the application of the methods described in this 
section a family of dilaton gravities \cite{Katanaev:1997ni} is considered which includes the physically most interesting models describing a single BH in 
Minkowski
(cf. (\ref{2.79})), Rindler or de Sitter space. The potentials \( U \) and
\( V \) are assumed to be of a simple monomial form, \begin{equation}
\label{e23UV}
U(X)=-\frac{a}{X}\, ,\qquad V(X)=-\frac{B}{2}X^{a+b}.
\end{equation}
 Among the constants \( a \), \( b \) and \( B \) only \( a \) and \( b \)
distinguish between physically inequivalent models. \( B \) plays the same
role as \( \lambda ^{2} \) in (\ref{2.79}), defining an overall scale factor.

In the line element (\ref{eq:a21}) the functions \( Q \) and \( w \) read
(cf. (\ref{eq:aQ}) and (\ref{eq:a12})) \begin{equation}
\label{e23Qw}
e^{Q(X)}=X^{-a}\, ,\qquad w(X)=-\frac{B}{2(b+1)}X^{b+1}\, ,
\end{equation}
 so that \begin{equation}
\label{e23ds}
(ds)^{2}=X^{-a}df\otimes \left[ 2dX+2\left( \mathcal{C}_{0}+\frac{B}{2(b+1)}X^{b+1}\right) \, df\right] \, .
\end{equation}
The equation \( w=\mathcal{C}_{0} \) has at most one solution on the positive semi-axis. Hence the metric (\ref{e23ds}) exhibits at most one horizon. 

The most interesting models correspond to positive \( a \) for which the 
function
\( X^{-a} \) diverges at \( X=0 \). For $X>0$ in terms of the alternative 
definition of the dilaton field \( \Phi  \) (\ref{e23exp})
the dilaton action (\ref{2.59.5}) with the potentials (\ref{e23UV}) becomes
\begin{equation}
\label{e23eact}
L^{\rm (dil)}=\frac{1}{2}\int d^{2}x\sqrt{-g}e^{-2\Phi }\left[ R+4a(\nabla \Phi )^{2}-Be^{2(1-a-b)\Phi }\right] \, ,
\end{equation}
which may be more familiar to a string audience.

It is also instructive to calculate the scalar curvature: \begin{equation}
\label{e23R}
R=2a\mathcal{C}_{0}X^{a-2}+\frac{Bb}{b+1}(b+1-a)X^{a+b-1}\, .
\end{equation}
 In what follows only models with \( b\ne -1 \) will be considered (although
\( b=-1 \) can be analyzed too \cite{Katanaev:1997ni}). For the {}``ground
state{}'' solutions \( \mathcal{C}_{0}=0 \) only the second term in 
(\ref{e23R})
survives. If \( a=b+1 \) or \( b=0 \) the scalar curvature of the ground state
is zero. A more detailed analysis shows that the first case (\( a=b+1 \)) 
corresponds
to Minkowski space, and the second (\( b=0 \)) to Rindler space. The condition
\( a=b+1 \) for the Minkowski ground state models also follows from 
(\ref{eq:a38}).
If \( a=1-b \) the ground state has constant curvature and corresponds to 
(anti-) de Sitter space.

For the general solutions (\ref{e23ds}) with \( \mathcal{C}_{0}\ne 0 \) it
follows from (\ref{e23R}) that, depending on the values of \( a \) and \( b \),
they may show a curvature singularity at \( X=0 \), at \( X=\infty  \),
or at both values. In the special cases considered above there could be only
one singularity. Therefore, these models describe (in a somewhat generalized
sense) a single BH immersed in Minkowski, or Rindler, or de Sitter space.

\begin{figure}
\centering
\epsfig{file=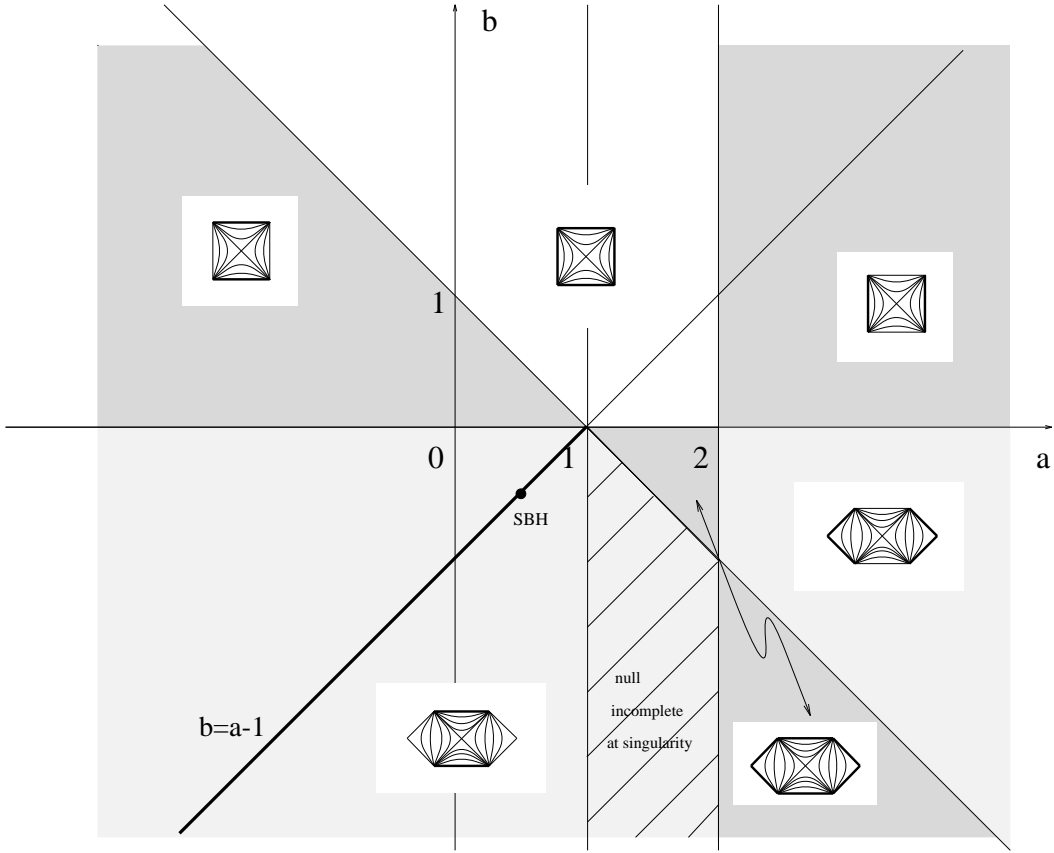,width=\linewidth}
\caption[``Phase'' diagram of CP diagrams]
{``Phase'' diagram showing the CP diagrams related to the 
($a$- and $b$-dependent) action (\ref{e23eact}). Bold lines in those diagrams
denote 
geodesically incomplete boundaries. Spherically reduced models lie on the 
half-line $b=a-1$, $b\leq0$, the endpoint of which corresponds to  
CGHS. The special case of SRG from $D=4$ is depicted by the point labelled 
by SBH.}
\label{fig:complete}
\end{figure}

Many interesting and important models belong to the two-parameter family of
this section. The SRG models (\ref{2.58}) for general dimension \( D \) lie
on the line \( a=b+1 \) between \( a=1/2 \) (this point corresponds to 
\( D=4 \))
and \( a=1 \). As \( D \) grows, these models approach the point \( a=1 \),
corresponding to the CGHS model \cite{Callan:1992rs}. The point \( a=0 \),
\( b=1 \) describes the Jackiw-Teitelboim model \cite{Barbashov:1979bm,
Teitelboim:1983ux,Jackiw:1985je}.
Lemos and Sa \cite{Lemos:1994py} gave the global solutions for \( b=1-a \),
Mignemi \cite{Mignemi:1994kz} considered \( a=1 \) and all values of \( b \).
The models of \cite{Fabbri:1996bz} correspond to \( b=0 \) and \( a\leq 1 \).
The general solution for the whole plane was obtained in ref. 
\cite{Katanaev:1997ni} and is summarized in Fig. \ref{fig:complete}.  

In order to calculate quantities like the ADM mass and the Hawking flux it is
essential to re-write the line element (\ref{e23ds}) for asymptotically 
Minkowski,
de Sitter and Rindler models in such a form that it becomes the standard one
in the asymptotic region. Here we give an explicit expression for the 
asymptotically
Minkowski solutions only (other cases can be found in ref. \cite{Liebl:1997ti})
where such a representation is possible for \( a\in (0,2) \). We repeat the
steps which led before from the metric (\ref{eq:a21}) to the Schwarzschild
black hole (\ref{eq:a35}). Namely, we first introduce the coordinates 
\( \widetilde{X}=r,f=u \)
(cf. (\ref{eq:a22})) \begin{equation}
\label{e23ur}
u=\sqrt{\frac{B}{a}}\, f\, {\textrm{sign}}(1-a)\, ,\qquad r=\sqrt{\frac{a}{B}}\frac{1}{|1-a|}X^{1-a}
\end{equation}
 and write the metric in EF form (\ref{eq:a23}) where \( \xi  \) now reads
\begin{equation}
\label{e23EF}
\xi \left( r\right) =2\mathcal{C}_{0}\left| 1-a\right| ^{\frac{a}{a-1}}r^{\frac{a}{a-1}}\left( \frac{B}{a}\right) ^{\frac{2-a}{2\left( a-1\right) }}+1
\end{equation}

Following the steps after (\ref{eq:a24}) one arrives at the generalized 
Schwarz\-schild
metric (\ref{eq:agensch}) with \( \xi  \) as in (\ref{e23EF}). 

For \( a\in (0,1) \) the asymptotic region corresponds to 
\( r\rightarrow \infty  \),
while for \( a\in (1,2) \) it is reached with the limit \( r\rightarrow 0 \).
In both cases \( \xi (r)\rightarrow 1 \), and the metric assumes the standard
Minkowski form \( g_{\mu \nu }\rightarrow {\textrm{diag}}(1,-1) \).

For the asymptotically Rindler and de Sitter solution with \( a \) belonging
to the same interval, \( a\in (0,2) \), the quantity \( \xi  \) becomes 
\begin{equation}
\label{e23rin}
\xi (r)=rB_{2}^{\frac{1}{2}}-Mr^{\frac{a}{a-1}}
\end{equation}
 for the Rindler solutions, and \begin{equation}
\label{e23des}
\xi (r)=r^{2}B_{2}-Mr^{\frac{a}{a-1}}
\end{equation}
 for the de Sitter solutions. Explicit expressions for the constants \( M \)
and \( B_{2} \) and for the variables \( t \) and \( r \) can be found in
ref. \cite{Liebl:1997ti}. The presence of typical linear (Rindler) and 
quadratic (de Sitter) terms in (\ref{e23rin}) and (\ref{e23des}) should be 
noted.

Although the CGHS model \( a=1 \), \( b=0 \) belongs to the family of the
asymptotically Minkowski models considered above, the equations (\ref{e23ur}) are singular at \( a=1 \). 
Appropriate coordinates for this case are \begin{equation}
\label{e23urs}
u=-\sqrt{B}f\, ,\qquad r=-\frac{1}{\sqrt{B}}\ln{X} \, .
\end{equation}
 The line element (\ref{eq:a23}) now contains
\begin{equation}
\label{e23ls}
\xi (r)=1+\frac{2\mathcal{C}_{0}e^{\sqrt{B}r}}{B}\, .
\end{equation}
 A somewhat non-standard feature of (\ref{e23ls}) is that the asymptotic region
is situated at \( r\rightarrow -\infty  \).


\clearpage

\section{Additional fields}\label{se:2.4}

In $2D$ there are neither gravitons nor photons, i.e. no propagating physical
modes exist \cite{Birmingham:1991ty}. This feature 
makes the inclusion of Yang-Mills fields in $2D$ dilaton gravity or an
extension to supergravity straightforward. 
Indeed, both generalizations can be treated again 
as a PSM (\ref{2.80}) with generalized $A_I$ and $X^I$. More locally 
conserved quantities (Casimir functions) may emerge and the 
integrability concept is extended.

Beside gauge fields also scalar and spinor fields may be added. If the latter
are derived from higher dimensions
through spherical reduction they are non-minimally coupled to
the dilaton.  The introduction of those fields in general destroys 
the integrability. Only in special cases exact solutions still can be
obtained. An example are chiral fermions \cite{Kummer:1992ef} or 
(anti-)selfdual scalars \cite{Pelzer:1998ea}. 

\subsection{Dilaton-Yang-Mills Theory}\label{se:2.4.1}

The interaction with additional one-form Yang-Mills fields $A^{\ul{a}}$ 
related to local gauge transformations 
belonging to a compact Lie group 
$\mathcal{G}$ is simply included in (\ref{2.62}) by introducing 
further auxiliary variables $Z^{\ul{a}}$ (additional target 
space coordinates in the PSM language) in the dilaton-Yang-Mills (DYM) action 
\begin{multline}
L^{\rm (DYM)}  = \int\limits_{\mathcal{M}_2} \, 
[\, X_a D e^a + X d\omega + Z^{\ul{a}} (\mathcal{D} A)^{\ul{a}} \\
+ \epsilon \, \mathcal{V} (X^a X_a, X, c_1 (Z), c_2 (Z), \; 
\ldots \; c_{n-1} (Z) ) \, ]\, .
\label{eq:a64}
\end{multline}
The gauge covariant derivative
\begin{equation}
(\mathcal{D} A)^{\ul{a}} = d A^{\ul{a}} + 
gf^{\ul{a}\ul{b}\ul{c}}\, A^{\ul{b}}  A^{\ul{c}}
\label{eq:a65}
\end{equation}
contains the structure constants $f^{\ul{a}\ul{b}\ul{c}}$ and the gauge
coupling $g$.
The potential $\mathcal{V}$, invariant under local 
Lorentz transformations and transformations $\mathcal{G}$ now also may 
depend on the Casimir invariants $c_i$ of the 
group $\mathcal{G}$. For instance in $\mathcal{G} = SU(N)$ there are $N-1$ 
independent invariant polynomials of degree $2,3,\, \ldots\,, N-1$ in 
terms of the components  $Z^{\ul{a}}$. 

The abelian case ($f=0$ in 
(\ref{eq:a65})) is especially simple. There $\mathcal{V}$ only 
depends on the single variable $Z$. Variation of 
(\ref{eq:a64}) with respect to $A$ directly yields $dZ=0$, 
i.e.\ $Z=Z_0=\mbox{const}$ is conserved, an additional 
Casimir function in the PSM-interpretation of 
(\ref{eq:a64}). Because $Z=Z_0$ is the result of solving a differential 
equation it cannot be simply reinserted into (\ref{eq:a64}). Variation 
of $Z$ yields
\begin{equation}
dA = -\epsilon\left.\frac{\partial \mathcal{V}}{\partial Z}\; 
\right\vert_{Z=Z_0} \; .
\label{eq:a66}
\end{equation}
The remaining e.o.m-s can be solved as for (\ref{2.62}) by 
(\ref{eq:a17}-\ref{eq:a20}) with just an additional dependence on the 
constant $Z=Z_0$ in the potential $V$ in (\ref{eq:a19}) and in 
$w(X,Z)$ of (\ref{eq:a20}) \cite{Kummer:1994ur,Louis-Martinez:1995rq}. 

For a nonabelian gauge group $\mathcal{G}$ the coupling between the 
gauge fields $A^{\ul{a}}$ and the geometric variables is  
somewhat more complicated, but as a PSM it can be treated 
again along the lines of sect. \ref{se:2.3.1}. 
For a potential of the type $\mathcal{V}(X^+X^-,X)+\alpha(X)Z^{\ul{a}}
Z^{\ul{a}}$ the solution of the geometric sector can be obtained like the one from of an ordinary GDT, because $Z^{\ul{a}}Z^{\ul{a}}$ is constant on-shell 
\cite{Klosch:1996fi,Klosch:1996qv,Strobl:1999wv}. Such a potential correctly 
reproduces the e.o.m.-s for ordinary D=2 dilaton-Yang-Mills theory:
\eq{
(\mathcal{D}A)^{\ul{a}} = - \epsilon \alpha(X) Z^{\ul{a}}\,, \hspace{0.5cm}
d(Z^{\ul{a}}Z^{\ul{a}}) = 0\,
}{eq:a666}
Some explicit solutions for the dilaton-Maxwell-Scalar system can be
found in ref. \cite{Park:1996zf}.  

The action (\ref{eq:a64}) does not contain the special case which emerges from spherical 
reduction of Einstein-Yang-Mills (EYM) in $D=4$. The
reason for this is obvious: While the Killing vectors $\xi_s$ associated with 
spherical symmetry act trivially on the metric, $\de_{\xi_s}g_{\mu\nu}=0$ with 
$\de_{\xi_s}$ being the Lie-derivative with respect to $\xi_s$, a
corresponding transformation of the gauge field $A_\mu \to A_\mu - 
\eps^s\de_{\xi_s} A_\mu$ can be compensated by a suitable gauge transformation,
$A_\mu \to A_\mu + \eps^s \mathcal{D}_\mu W_s$, such that $\de_{\xi_s}A_\mu=
\mathcal{D}_\mu W_s$ \cite{Forgacs:1980zs}. Already for $SU(2)$ the 
most general solution compatible with spherical symmetry, sometimes called 
``Witten's ansatz'' \cite{Witten:1977ck}, yields terms which are not described by (\ref{eq:a64}), namely an additional (minimally coupled) charged scalar 
field with dilaton-dependent mass term and quartic 
self-interaction\footnote{Cf. \cite{Volkov:1998cc} for a comprehensive review 
on non-trivial EYM solutions.}.

\subsection{Dilaton Supergravity}\label{se:2.4.2}

Already a long time ago the superfield approach has been applied 
in supersymmetric extensions of GDTs \cite{Howe:1979ia}. 
Superfields are expressed in terms of supercoordinates 
$z^M = (x^m, \theta^\mu)$, where  
the bosonic coordinates $x^m$ are supplemented\footnote{ 
Throughout this subsection we 
employ the generally accepted supersymmetry 
notation with Latin indices from 
the middle of the alphabet for the holonomic bosonic coordinates 
$x^n$ (denoted ``$x^\mu$'' in the rest of this Report). Greek 
indices are reserved for the fermionic coordinates $\theta^\mu$. 
A similar notation is 
used in tangential space for Lorentz vectors $X^A = 
(X^a,X^\alpha)$ with indices from the beginning of the 
alphabet. } by anticommuting (Grassmann) coordinates $\theta^\mu$. 

We assume the latter to be Majorana spinors, i.e.\ we restrict 
ourselves to $N=1$ superspace in $D=2$. 
Beside the $\mathbb{Z}_2$-grading property for coordinates\footnote{In the 
exponent of $(-1)$, $M$ or $N$ is zero for a bosonic component, 
for a fermionic one $M$ resp.\ $N$ are 1.},
\begin{equation}
z^Mz^N = (-1)^{MN} z^N z^M\, ,
\label{eq:s1}
\end{equation}
the derivatives with respect to $z$ are defined to act to 
the right $\rpartial_M = \rpartial/\partial z^M$. Only 
those derivatives will appear in the following, so the arrow will be dropped.

Any vector-field in superspace $V=V^M \partial_M$ is 
invariant under non-de\-gen\-er\-ate coordinate changes $z^M \to 
\bar{z}^M (z)$ 
\begin{equation}
V^M \partial_M = \bar{V}^M \; 
\frac{\partial z^L}{\partial \bar{z}^M}\; 
\frac{\partial \bar{z}^N}{\partial z^L}\; 
\frac{\partial}{\partial \bar{z}^N}\, ,
\label{eq:s2}
\end{equation}
which shows the advantage of the conventional summation of 
indices ``ten to four'' in supersymmetry, already introduced 
in the section on PSM-s (sect. \ref{se:2.2.3}). 

Now any formula of differential geometry in ordinary space of the PSM 
can be copied to superspace notation. The p-forms eq. (\ref{2.12}), 
of sect. \ref{se:2.1.1} turn into the same expressions written 
in terms of $dz^M$ instead of $dx^\mu$, adhering strictly to 
the summation of supersymmetry: 
\begin{equation}
  \Phi = \frac{1}{p!} dz^{M_p} \wedge \cdots \wedge dz^{M_1}
  \Phi_{M_1\cdots M_p}
  \label{eq:s3}
\end{equation}
Also the external differential is defined in agreement with sect. 
\ref{se:2.2.3} as
\begin{equation}
  d\Phi = \frac{1}{p!} dz^{M_p} \wedge \cdots \wedge dz^{M_1} \wedge
  dz^N \partial_N \Phi_{M_1\cdots M_p}
\label{eq:s4}
\end{equation}
which implies the Leibniz rule (\ref{2.83}) for superforms, already introduced 
in that section in anticipation of the graded PSM (gPSM) approach below. 
Clearly the (anti-)symmetry properties of the tensor 
$\Omega_{M_1\, \ldots \, M_p}$ now depend on the graded commutation 
properties (\ref{eq:s1}) of the indices. Instead of (\ref{2.11}), 
(\ref{2.12}) the one-forms of superzweibein and superconnection are 
$E_M{}^A$ and 
\begin{equation}
 \LC_{MA}\^B = \LC_M L_A\^B\, ,
\label{eq:s5}
\end{equation}
where in (\ref{eq:s5}) the simplification for two 
dimensions with 
\begin{equation}
 L_A\^B = \mtrx{\epsilon_a\^b}{0}{0}{-\frac12 (\gamma_\ast)\_\alpha\^\beta}
\label{eq:s6}
\end{equation}
already has been taken into account. The fermionic part in 
(\ref{eq:s6}), the generator of Lorentz transformations in 
spinor space, agrees with (\ref{eq:73}).  Covariant derivatives, 
defined by analogy to (\ref{2.21}) 
\begin{equation}
\begin{split}
\label{eq:s7}
  \nabla_M V^A &= \partial_M V^A + \LC_M V^B L_B\^A\,, \\
  \nabla_M V_A &= \partial_M V_A - \LC_M L_A\^B V_B\; .
\end{split}
\end{equation}
lead to the expressions for the components of supercurvature 
and supertorsion 
\begin{equation}
  R_{MNA}\^B = \left(\partial_M \LC_N - \partial_N \LC_M (-1)^{MN}\right)
  L_A\^B =: F_{MN} L_A\^B,
\label{eq:s8}
\end{equation}
\begin{equation}
T_{MN}\^A = \partial_M E_N\^A + \Omega_M E_N\^B \Omega_M 
L_B\^B - ( M \leftrightarrow N) (-1)^{MN}
\label{eq:s9}
\end{equation}
Again Bianchi identities, direct generalizations of (\ref{2.24}) 
and (\ref{2.26}), must hold (cf.\ \cite{Ertl:2001sj}, eqs.\ (\ref{2.55}), 
(\ref{2.66})). They restrict the component fields, contained in
$E_M\^A, \Omega_M$ when these expressions are expanded in 
terms of (a finite number of) ordinary fields, appearing as 
coefficients of powers $\Theta^\mu$ and $\Theta^2 = 
\Theta^\mu \Theta_\mu$. It turns out that to deal with  
supertorsion it is more convenient to use its anholonomic 
components: 
\begin{equation}
T_{AB}{}^C = (-1)^{A(B+N)} \, E_B{}^N E_A{}^M T_{MN}{}^C
\label{eq:s10}
\end{equation}
The literature on $2D$ supergravity (cf. e.g. \cite{Howe:1979ia,Brown:1979ma,
Martinec:1983um,Rocek:1986iz,Rivelles:1994xs,Leite:1995sa}) is strongly 
influenced by its close relation to string theory, where the bosonic torsion 
vanishes, $T_{ab}{}^c=0$. It uses the further constraints on $T_{AB}{}^C$
\begin{equation}
T_{\alpha\beta}{}^c = 2i\,(\gamma^c)_{\alpha\beta}, \qquad 
T_{\alpha\beta}{}^\gamma = 0\, , 
\label{eq:s11}
\end{equation}
the first of which is dictated by the requirement that in 
the limit of global transformations ordinary supersymmetry 
should be restored. The second one turns out to be a 
convenient choice, because then the Bianchi identities in a 
Wess-Zumino type gauge are fulfilled identically 
\cite{Ertl:1998ib}. 

In the application to $2D$ gravity including 
bosonic torsion it seems natural to retain (\ref{eq:s11}), 
but to simply drop the zero bosonic torsion condition. 
However, as a consequence of the Bianchi 
identities it turned out \cite{Ertl:1998ib} that the superfield components,  
obtained in this manner, did not permit the construction of 
$2D$ supergravity Lagrangians with nonvanishing bosonic 
torsion, after all. Only after replacing (\ref{eq:s11}) by 
the weaker set ($F_{\alpha\beta} = E_\alpha{}^M\, 
E_\beta{}^N\, F_{NM} (-1)^M$, cf. (\ref{eq:s8})) 
\begin{equation}
(\gamma_a)^{\beta\alpha} \, T_{\alpha\beta}{}^c = - 4i 
\delta_a{}^c, \quad T_{\alpha\beta}{}^\gamma = 0, \quad
(\gamma_a)^{\beta\alpha} F_{\alpha\beta} =0 
\label{eq:s12}
\end{equation}
a solution can be found \cite{Ertl:2001sj}. However, the 
mathematical complexity of this approach becomes considerable.

Instead, we turn to the generalization of the PSM, adding fermionic target 
space coordinates $\chi^\alpha$ and corresponding Rarita-Schwinger 1-form 
fields $\psi_\alpha$ to the degrees of freedom in (\ref{2.80}), 
(\ref{2.84}) as\footnote{Under the title ``free differential algebras'' this 
has been proposed for simple models in ref. \cite{Izquierdo:1998hg}, cf.\ also 
refs. \cite{Ikeda:1994dr,Ikeda:1994fh,Strobl:1999zz}.} 
\begin{equation}
\begin{split}
X^I & = ( X, X^a, \chi^\alpha )\, ,\\
A_I & = (\omega, e_a, \psi_\alpha)\; .
\label{eq:s13}
\end{split}
\end{equation}
Apart from that, the PSM action retains the form (eq. (\ref{2.80})). 

Both $\chi^\alpha$ and $\psi_\alpha$ denote Majorana fields, 
when, as in what follows, $N=1$ supergravity is considered\footnote{In higher
$N$ supergravity Majorana fields $\stackrel{(1)}{\chi^\al}\dots
\stackrel{(N)}{\chi^\al}$ and corresponding $\stackrel{(i)}{\psi}_\alpha$
are needed with an additional $SO(N)$ symmetry.}. 
The graded Poisson tensor $\mathcal{P}^{IJ} = (-1)^{IJ+1} 
\mathcal{P}^{JI}$ instead of (eq. (\ref{2.82})) 
is now assumed to fulfill a graded Jacobi identity 
\begin{equation}
  \Poisson^{IL} \rpartial_L \Poisson^{JK} + \gcycl(IJK) = 0 \, .
\label{eq:s15}
\end{equation}
Except for the range of fields to be summed over, the e.o.m-s 
are again (\ref{2.87}), (\ref{2.88}).  
The symmetries (\ref{2.89}), (\ref{2.90}) 
depend on infinitesimal local parameters $\epsilon_I = 
(\epsilon_\phi, \epsilon_a, \epsilon_\alpha)$.

The mixed components $\mathcal{P}^{\alpha X} $ are constructed by 
analogy to $\mathcal{P}^{aX}$ in (\ref{2.85}) 
with the appropriate 
generator $(-\gamma_\ast/2)$ of Lorentz transformations in $2D$ 
spinor space (cf.\ eq.\ (\ref{eq:s6})). 
Then $d\epsilon_\alpha$ in the second set of 
eq.\ (\ref{2.90}) acquires an additional term casting it into the 
covariant combination $(D\epsilon )_\alpha$, with covariant derivative 
(\ref{eq:74}). This is precisely the 
form required for the (dilaton deformed) supergravity transformation of the 
``gravitino'' $\psi_\al$. 
As the Poisson tensor $\mathcal{P}^{IJ}$ also here is not of full 
rank, Casimir functions $\mathcal{C} (Y,\phi,\chi^2)$ exist  
which, following the same line of argument as in 
sect.\ \ref{se:2.2.3} obey $d\mathcal{C} =0$. From Lorentz invariance 
a bosonic $\mathcal{C}$ in  supergravity is of the form 
\begin{equation}
\label{eq:s19}
  \mathcal{C} = \casimir + \half \chi^2 \casimir_2\, ,
\end{equation}
where $\casimir$ coincides with the quantity denoted by $\mathcal{C}$ in the 
pure bosonic case (\ref{2.92},\ref{eq:a20}). However, also 
fermionic Casimir functions may appear (see below).
 
The determination of all possible minimally extended 
supergravities reduces to the solution of the Jacobi 
identities (\ref{eq:s15}). In the general ansatz for 
$\mathcal{P}^{IJ}$ 
\begin{align}
\mathcal{P}^{ab} &=  V\, \epsilon^{ab}\, ,\label{eq:s20}\\
\mathcal{P}^{b\phi} &=  X^a\, \epsilon_a{}^b\, 
,\label{eq:s21}\\
\mathcal{P}^{\alpha\phi} &=  -\frac{1}{2} \chi^\beta 
(\gamma_\ast)_\beta{}^\alpha\, ,\label{eq:s22} \\
\mathcal{P}^{\alpha b} &=  \chi^\beta (F^b)_\beta{}^\alpha\, 
,\label{eq:s23} \\
\mathcal{P}^{\alpha\beta} &=  v^{\alpha\beta} + 
\frac{\chi^2}{2} \, v_2^{\alpha\beta} \, ,
\label{eq:s24}
\end{align}
the function 
\begin{equation}
\label{eq:s25}
V = \mathcal{V} (X,Y) + \frac{\chi^2}{2} \, v_2 (X,Y) 
\end{equation}
contains the original bosonic potential $\mathcal{V}$. As explained 
above, eqs.\ (\ref{eq:s21}) and 
(\ref{eq:s22}) are fixed completely by Lorentz invariance. 
This invariance also implies that the (symmetric) 
spinor-tensor $V^{\alpha\beta}$ in (\ref{eq:s24}) can be
expanded further into three scalar functions of $X$ and $Y$, multiplying 
the symmetric matrices $(\gamma_\ast)^{\alpha\beta}, \gamma_a^{\alpha\beta}, 
X^a (\gamma_\ast\gamma_a)^{\alpha\beta}$: 
\begin{equation}
\begin{split}
\label{eq:s26}
V^{\alpha\beta} &= \mathcal{U} \gamma_\ast{}^{\alpha\beta} + i\, 
\widetilde{\mathcal{U}} X^a \gamma_a{}^{\alpha\beta} + 
i\, \widehat{\mathcal{U}} X^a \epsilon_a{}^b 
\gamma_b{}^{\alpha\beta} =  \\
&= v^{\alpha\beta} + \frac{\chi^2}{2} v_2^{\alpha\beta}\; .
\end{split}
\end{equation}
Each function $\mathcal{U}$ again has a pure bosonic part 
and, as indicated in the second line of (\ref{eq:s26}), a 
term proportional to $\chi^2$. In a similar manner the 
spinor $F^b$ in (\ref{eq:s23}) can be expressed in terms 
proportional to $\delta_\beta{}^\alpha, 
(\gamma^a)_\beta{}^\alpha (\gamma_\ast)_\beta{}^\alpha$ which  
finally requires eight scalar functions of $X$ and $Y$, 
multiplied by appropriate factors constructed with the help 
of $X^a$ and $\epsilon^{ab}$ \cite{Ertl:2000si}. In (\ref{eq:s23}) the multiplying factor $\chi$  
precludes terms with factor $\chi^2$ in $F$. Thanks to 
(\ref{eq:s21}) and (\ref{eq:s22}) the Jacobi identities with 
one index  referring to $X$ are fulfilled automatically. The 
remaining ones can be solved \textit{algebraically}, 
provided a quite specific sequence of steps is followed (for 
details see \cite{Ertl:2000si}). Three main cases are 
determined by the rank of the $2\times2$ spinor matrix 
$v^{\alpha\beta}$ in (\ref{eq:s26}). 

For full rank ($\det\, v^{\alpha\beta} \neq 0$) 
the solution is found to depend on five scalar functions of 
$X,Y$ and the derivatives thereof, if the bosonic potential 
$\mathcal{V}$ in (\ref{eq:s25}) is assumed to be given. 

If the fermionic rank is reduced, beside the bosonic 
Casimir function (\ref{eq:s19}) one or two fermionic ones exists. 
They are of the generic form 
\begin{equation}
\label{eq:s27}
\mathcal{C}^{(\pm)} = 
\chi^\pm \; \left\vert\, 
\frac{X^{--}}{X^{++}}\,\right\vert^{\pm 1/4}\; 
c_{(\pm)}\, (X,Y)
\end{equation}
and owe their Lorentz invariance to the interplay of the abelian boost 
transformations $\exp (\pm \beta)$ of the light cone 
coordinates $X^{\pm\pm}$ related to $X^a$, and $\exp (\pm 
\beta/2)$ of the chiral spinor components $\chi^{\pm}$ of $\chi^\al$. 
For fermionic rank 1 the general solution contains four arbitrary functions 
beside $\mathcal{V}$ and one additional Casimir function of type 
(\ref{eq:s27}). For rank zero of the fermionic extension   
in $\mathcal{P}^{IJ}$ beside (\ref{eq:s19}) 
both fermionic Casimir functions (\ref{eq:s27}) are conserved and 
three functions remain arbitrary for a given bosonic 
potential. 

In order to avoid solving differential equations by imposing 
the Jacobi identities (\ref{eq:s15}) also for reduced 
fermionic rank, it is important to make intensive use of the 
information on the Casimir functions. 

The arbitrariness of the solution of the Jacobi identities can 
be understood as well by studying  
reparametrizations of the target space, spanned by the $X^I$ 
in the gPSM. Those reparametrizations may generate  new 
models. Therefore, they can be useful to create a more 
general gPSM from a simpler one, although this 
approach is 
difficult to handle if $\mathcal{V}$ in (\ref{eq:s25}) is 
assumed to be the given 
starting point. However, the 
subset of those reparametrizations may be analyzed which 
leaves a given bosonic theory unchanged. Again the same number 
of arbitrary functions emerges for the different cases 
described in the paragraphs above. 

As an illustration we quote eq.\ (4.252) from 
\cite{Ertl:2000si} which represents one (of many) supergravity 
actions which possess a bosonic potential (\ref{2.77}) 
quadratic in torsion\footnote{We comply with our present 
notation by the replacements $\phi \to X, Z\to U(X)$ in ref.
\cite{Ertl:2000si}. Furthermore an arbitrary constant is now fixed 
as $\tilde{u}_0 =1$.\label{fn:phi}}: 
\begin{equation}
\begin{split}
  \label{eq:s28}
  \Action^{\rm (QBT)} &= \int_\BMf X d\omega + X^a De_a +
  \chi^\alpha D\psi_\alpha + \epsilon \left( V + \half X^a X_a U +
    \half\chi^2 v_2 \right) \\
  & + \frac{U}{4} X^a (\chi\gamma^3\gamma_a\gamma^b e_b \psi) -
  \frac{i\, V}{2u} (\chi \gamma^a e_a \psi) \\
 &  - \frac{i }{2} X^a (\psi \gamma_a \psi) - \frac{1}{2}
  \left( u + \frac{U}{8} \chi^2 \right) (\psi \gamma^3 \psi),\\
  v_2 & = -\frac{1}{2u} \left( V U + V' + \frac{V^2}{u^2}
  \right).
\end{split}
\end{equation}
In this formula $U(X)$ is the quantity defined in (\ref{2.77}). The 
prepotential $u$ is related to $U(X)$ and $V(X)$ by
\begin{equation}
u^2 (X) = - 2\, e^{-Q(X)}\, w(X)\, ,
\label{eq:s29}
\end{equation}
where $Q$ and $w$ have been defined in (\ref{eq:aQ}) and (\ref{eq:a12}).

The supergravity transformations of $e_a$ and $\psi_\alpha$  with small fermionic parameter $\epsilon$ for this action (\ref{eq:s28})(cf.\ eqs.\ 
(4.255) - (4.259) of \cite{Ertl:2000si} and footnote 
\ref{fn:phi}) are of the form
\eq{
\delta\, e_a = i\, (\epsilon \gamma_a \psi)+ \cdots,\hspace{0.5cm}
\delta\, \psi_\alpha = - (D\, \epsilon)_\alpha + \cdots\;.
}{eq:s29.5} 
Thus, they contain the essential terms, but, not shown here, also others, 
because of the deformation by the dilaton field. SRG is the special case 
(\ref{2.79}) for $U$ and $V$, but as a supergravity extension (\ref{eq:s28}) 
is not unique.

A generic property of the fermionic extensions obtained in 
this analysis is the appearance of obstructions, which is a typical
feature of supersymmetric theories (cf. e.g. \cite{Duff:1986hr}). The 
first type of those consists in singular functions of the 
bosonic variables $X$ and $Y$, multiplying the fermionic 
parts of a supergravity action, when no such singularities 
are present in the bosonic part. But even in the absence of 
such additional singularities, a relation like 
(\ref{eq:s29}) between the original 
potential and some prepotential $u$, dictated by the 
corresponding supergravity theory, either leads to a 
restriction of the range of $X$ and/or $Y$ as given by 
the original bosonic one, or even altogether prevents any 
extension of the latter. Remarkably, a known $2D$ supergravity 
model like the one of Howe \cite{Howe:1979ia}  which 
originally had been  constructed 
with the full machinery of the superfield technique, is one example
which escapes such obstructions. There, in our language, the PSM potential 
$\mathcal{V} =  -2 \lambda^2 
X^3$ permits an expansion in terms of the prepotential $u 
(X)$ through $\mathcal{V} = -du^2/d X$ because $Q=0$. An example where 
obstructions seem to be inevitable is the KV-model \cite{Katanaev:1986wk}
with quadratic bosonic torsion. 

On the other hand, the supergravity extension of SRG 
following from the action (\ref{eq:s28}) is free from such 
problems. However, it is not the only possible extension of 
the bosonic theory. Indeed, 
the hope that a link could be found between the 
possibility of reducing the arbitrariness of extensions 
referred to above, and of the absence of such obstructions, did 
not materialize. Several counter examples could be given 
including different singular and nonsingular extensions of SRG.

Another very important point concerns the ``triviality'' of supergravity
extensions, proved earlier by Strobl \cite{Strobl:1999zz}. It was based upon 
the observation that locally a formulation of the dynamics in 
terms of Darboux  coordinates allows to elevate the infinitesimal 
transformations (\ref{2.89},\ref{2.90}) (on-shell) to finite ones. 
Then the  latter may be used to gauge the fermionic fields in $2D$ 
supergravity to zero. 
Providing now the explicit form of those Darboux coordinates 
in the explicit solution of a generic model,  
additional support has been given to the original argument of 
ref. \cite{Strobl:1999zz}. 
However, the appearance of the obstructions and the ensuing 
singular factors in the transition to the Darboux 
coordinates may introduce a new aspect. When those new 
singularities appear at isolated points without restriction 
of the range for the original bosonic field variable, they 
may be interpreted and discarded much like coordinate singularities. 
Another way to circumvent this problem in the presence of 
restrictions to the range and thus to 
retain triviality is to 
allow a continuation of our (real) theory to complex 
variables.  This triviality disappears anyhow, when 
interactions with additional matter fields are introduced, 
obeying the same symmetry as given by the gPSM-theory. A proposal in this
direction can be found in ref.\ \cite{Izquierdo:1998hg}. 

In order to eliminate the arbitrariness of superdilaton 
extensions the only viable argument seems to start from a supergravity theory 
in higher dimensions (e.g.\ $D=4$) and to reduce it (spherically or 
toroidally) to a $D=2$ effective theory. It turns out that the Killing spinors 
needed in that case must be Dirac spinors, requiring the 
generalization of the work of ref. \cite{Ertl:2000si} to (at least) 
$N=2$, where, however, the same technique of gPSM-s can  
be applied. 

As in the bosonic case an action like (\ref{eq:s28}) or, better, directly its 
gPSM form (\ref{2.80}) can be converted into a dilaton theory by 
elimination of the torsion dependent part of the spin 
connection and of $X^a$. Also for supergravity the equation for $X^a$ is 
independent of the potential $\mathcal{V}$, eq.\ (\ref{2.75}) being 
replaced by 
\begin{equation}
\label{eq:s30a}
X^a = - e^a_m \epsilon^{mn} [ ( \partial_n X ) + \frac{1}{2} 
(\chi \gamma_\ast \psi_n ) ]\; . 
\end{equation}
The corresponding dilaton action in the gPSM notation 
becomes 
\begin{equation}
L = \int\limits_{\mathcal{M}} [ 
X d \tilde{\omega} + \chi^\alpha (D \psi)_\alpha + 
\frac{1}{2} \mathcal{P}^{AB}\, \vert_{X^a}\, 
e_B\, e_A\, ]\, ,
\label{eq:s30b}
\end{equation}
where $\tilde\omega$ is the torsionless part of the 
curvature as in (\ref{2.73}) and $\vert_{X^a}$ means that the 
components of the Poisson tensor (in the anholonomic basis) 
are to be taken with $X^a$ given by (\ref{eq:s30a}).  When 
(\ref{eq:s30b}) 
is written in components (cf.\ eq. (4.246) of \cite{Ertl:2000si}) 
with the bosonic potential (\ref{2.77}) for the Howe model ($V = 
- \frac{1}{2} X^3, U=0$), it can be checked that 
the superdilaton theory, obtained in this way differs from 
the direct superextension of the bosonic theory \cite{Park:1993sd}. 
The reason is that in our approach $X$ is directly promoted 
to be the bosonic component of the superfield, whereas in ref. 
\cite{Park:1993sd} the dilaton represents the bosonic part of yet
another scalar superfield.

\subsection{Dilaton gravity with matter}\label{se:2.4.3}

\subsubsection{Scalar and fermionic matter, quintessence}

For the inclusion of scalar matter as in sect. \ref{se:2.2.1} we 
start with the example of spherical reduction of Einstein 
theory. When massless scalar fields are coupled minimally 
in $D$ dimensions ($( \nabla_{(D)} \Phi )^2 = g^{MN} \partial_M \Phi\, 
\partial_N \Phi$\,; $M,N = 0,1, \ldots \, D-1$)
\begin{equation}
L^{(m,D)}_{(\phi)} = \frac{1}{2}\, 
\int d^D x \, \sqrt{-g_{(D)}} \; ( \nabla_{(D)} \Phi )^2\, ,
\label{eq:a67}
\end{equation}
for scalar fields $\phi=\phi(x^0,x^1)$ the $2D$ reduced action 
in terms of the components of $g^{MN}$ as derived from the line-element 
(\ref{2.57}), becomes
\begin{align}
\label{eq:a68}
L^{(m)}_{(\phi)}  &= \frac{\mathcal{O}_{D-2}}{\, 
\lambda^{D-2}}\; 
\int d^2 x \sqrt{-g}\, F(X)\, (\nabla \phi)^2\, ,\\
\label{eq:a69}
F_{SRG} (X) &= \frac{X}{2}\; .
\end{align}
Such an interaction is an example of nonminimal coupling 
($F(X) \neq \rm const$.) in the reduced case. Admitting a general 
function $F(X)$, a dilaton field dependence different from 
SRG (eq.\ (\ref{eq:a69})) can be covered as well. An especially 
simple theory follows for $F={\rm const.}$, minimal coupling at the 
$D=2$ level\footnote{This is not to be confused with $F\propto X$, 
corresponding to minimal coupling in the {\em original} dimension in the case 
of SRG. We will refer to that case as non-minimal coupling.}. Below we shall
absorb the relative factor $1/2$ between the dilaton action (\ref{2.58})
and the first order action (\ref{2.62}) -- which of course must also be 
adjusted properly in the matter action -- into the coupling function $F(X)$.

Dropping as in our convention for the geometrical part of the FOG action in 
(\ref{2.62}) the prefactor $2\mathcal{O}_{D-2}/\lambda^{D-2}$, the action 
(\ref{eq:a68}) can be written also as 
\begin{equation}
L^{(m)}_{(\phi)} = \frac{1}{2} \; \int\limits_{\mathcal{M}_2}\;
F\, d \phi  \ast d \phi = \frac{1}{2}\; \int F\, (d\phi e^a) 
\ast (d\phi e^b)\, \eta_{ab}\; .
\label{eq:a70}
\end{equation}

In order to avoid the delicate subject of Killing spinors, necessary for the 
(spherical) reduction of fermions\footnote{Such a reduction yields a dilaton
dependent ``mass'' term and coupling of spinors to the auxiliary fields 
$X^\pm$.} (cf. e.g. \cite{penrosespinors} for $D=4$) we shall only deal with 
fermions introduced directly in $D=2$. The diffeomorphism 
covariant generalization (cf. (\ref{eq:74})) of the Dirac action 
\begin{equation}
L^{(m)}_{(\psi)} = \frac{i}{4}\; 
\int d^2 x \, \sqrt{-g}\; e^\mu_a \, ( 
\bar{\psi} \gamma^a 
\overset{\leftrightarrow}{D}_\mu^{(\psi)}\, \psi )
\label{eq:a71}
\end{equation}
as in Minkowski space must contain a two-sided derivative 
$a \overset{\leftrightarrow}{D} b = a( D b) - (D a) b$ in order to yield a 
real action. 
The cancellation of $\om$ in that derivative is a peculiar feature of $D=2$, 
i.e.\ the simplification 
$\overset{\leftrightarrow}{D}_\mu = \overset{\leftrightarrow}{\partial}_\mu$ 
is possible there. Thus both interactions 
(\ref{eq:a70}) and (\ref{eq:a71}) do not depend on the spin connection, as 
anticipated already in deriving the (classical) e.o.m.-s 
with matter, eqs.\ (\ref{eq:a9})-(\ref{eq:a8}).

An additional geometrical degree of freedom may also appear 
already at the $D=4$ level. Recently in connection with the 
observation of supernovae at high values of the 
redshift \cite{Riess:2001gk,Riess:2000,Riess:1998cb,Perlmutter:1998np} the 
validity of the Hilbert-Einstein theory has 
been put into doubt \cite{Bahcall:1999xn}. 
The simplest theoretical description requires the 
introduction of a (still very small) cosmological constant 
$\Lambda$ (de Sitter theory). 
Therefore, extensions of the Einstein theory towards the old 
Jordan-Brans-Dicke (JBD) theory \cite{Fierz:1956,Jordan:1959eg,Brans:1961sx}, 
have been revived, where already at $D=4$ an additional scalar field 
$\Phi$ (Jordan field, ``quintessence'') is assumed to exist 
\cite{Wetterich:1988fm,Wang:1998gt,Carroll:1998zi,Zlatev:1998tr,Ratra:1988rm}. 
Then, already in $D=4$ an action like (\ref{2.59}) is postulated with 
$X\to\Phi$ and appropriate assumptions for functions 
$U_{(4)}(\Phi), V_{(4)}(\Phi)$ so that the ``effective cosmological constant'' 
is driven to its present (small) value\footnote{Recently dilaton gravity in $4D$ has been discussed in a framework where even the dilaton can be understood in geometrical terms \cite{Graf:2002tw}.} 
\begin{equation}
L^{\rm (Q)} = \int d^4x \, \sqrt{-g_{(4)}}\; 
\left[ R_{(4)} \, \Phi + U_{(4)}(\Phi)\, 
(\nabla_{(4)} \Phi)^2 \, + V_{(4)} (\Phi)\; 
\right] \; .
\label{eq:a75}
\end{equation}
After spherical reduction of (\ref{eq:a75}), even 
\textit{without} including a genuine matter interaction a 
$2D$ theory emerges where the (in $D=4$ geometric) variable 
$\Phi$ turns into something like an additional 
scalar field (beside the genuine $D=2$ dilaton field $X$). These
``two-dilaton theories'' have been studied in more detail in
\cite{Grumiller:2000wt}. The most interesting feature of such theories is that
one dilaton field plays the role of a ``geometric'' dilaton field in $D=2$ and 
the other one behaves like matter, providing continuous physical degrees of 
freedom. We shall discuss the classical and quantum properties of dilaton 
gravity with (\ref{eq:a68}), (\ref{eq:a69}) in sect. \ref{se:3.2}.

\subsubsection{Exact solutions -- conservation law for geometry and 
matter}\label{se:2.4.4}

In the presence of interactions with additional fields which -- in contrast to 
gauge fields (cf.\ subsection \ref{se:2.4.1}) or supergravity 
(subsection \ref{se:2.4.2}) -- cannot be incorporated into the PSM approach, 
the possibility to find analytic solutions 
is restricted\footnote{It is possible, however, to adjust the dilaton
potentials in such a way that some exact solutions with matter can be
obtained \cite{Filippov:1996ye,Filippov:1997db}.
}. Nevertheless, the interest in such solutions had been raised, 
especially by the work on the CGHS model (cf. sect. \ref{se:2.2.1.5}). 
It possesses a global structure very much like the one of the 
genuine Schwarzschild BH (Fig.\ \ref{fig:2.9}). Also application of the 
(singular) dilaton field dependent conformal transformation (\ref{2.53})
with dilaton dependent conformal factor
\eq{
\rho_{CGHS} = -\frac{1}{2} \int^X U_{CGHS}(X')dX' = 
\frac{1}{2}\ln{X}
}{eq:conformalfactor}
is found to cancel the torsion term (cf. sect. \ref{se:2.1.4}). The 
transformed potentials read
\eq{
\tilde{U}(X) = 0, \hspace{0.5cm} \tilde{V}(X) = V_{CGHS}(X) 
e^{Q_{CGHS}(X)} =  \rm const. 
}{eq:trafopotential}
Then also the action for the scalar 
field (\ref{eq:a70}) becomes the one in a flat background ($g_{\mu\nu} 
\to \eta_{\mu\nu}$). In this model $F$ is taken to be constant 
(minimal coupling). After (trivially) solving for the (free) 
scalar or fermionic fields, the inverse conformal transformation 
is applied. This method has also been extended to include 
one-loop quantum effects in the semi-classical approach, by 
describing that lowest order quantum effect through the Polyakov 
effective action \cite{Polyakov:1981rd}. Adding to this action  another piece, 
adapted suitably so that the exact solubility is maintained, more 
semi-classical solutions have been studied \cite{Russo:1992ht,Bose:1995pz,
Cruz:1996zt,Fabbri:1998hs,Kim:1999wa,Zaslavsky:1999zh,Zaslavsky:1998ca}.  An
approximate analysis of the solutions is possible, of course,
for a larger class of models. For instance, it has been demonstrated 
\cite{Kazakov:1994ha} that adding the Polyakov term to SRG shifts and 
attenuates the BH singularity. 

In the exact solution without matter, the important step has been 
to use one of the eqs.\ (\ref{eq:a6}) to express $\omega$ as 
in (\ref{eq:a13}). If the theory only depends on one type of chiral fields  
either $W^-$ or $W^+$ in (\ref{eq:a6}) vanishes. Then the same elimination of 
$\om$ can be used. Separating e.g.\ Dirac fermions into chiral components as 
\begin{equation}
\Psi = \sqrt[4]{2}\, \left(
\begin{array}{c}
\chi_R\\
\chi_L
\end{array} \right)\, ,\quad
\bar{\Psi} = \sqrt[4]{2}\; 
\left(\chi_R^\dag, \chi_L^\dag\right)\, ,
\label{eq:e1}
\end{equation}
the interaction (\ref{eq:a71}) may be written as
\begin{align}
\label{eq:e2}
L^{(m)}_{(\Psi)} &= - \int\limits_{\mathcal{M}_2}\,
( e^+  J^- + e^-  J^+ )\, ,\\
J^{-,+} = J^{R,L} &= i\, \left[ 
\chi^\dag_{R,L}\, (d\chi_{R,L}) - (d\chi^\dag_{R,L})\, 
\chi_{R,L} \; \right]\; .
\label{eq:e3}
\end{align}
Also ``amplitudes'' $k$ and ``phases'' $\varphi$ can be introduced, 
\begin{equation}
\chi_{R,L} = \frac{1}{\sqrt{2}}\, 
k_{R,L}\, e^{i\varphi_{R,L}}\, ,
\label{eq:e4}
\end{equation}
in terms of which the chiral currents (\ref{eq:e3})  become
\begin{equation}
J^{+,-} = - k^2_{L,R}\, d\varphi_{L,R}\; .
\label{eq:e5}
\end{equation}
Since the theory only depends upon one type of chiral fields, either 
$J^+$ or $J^-$ in (\ref{eq:e2}) is zero \cite{Kummer:1992ef,Pelzer:1998ea}. 
Still an equation like (\ref{eq:a13}) without matter contribution 
holds, and the further steps of the solution for the 
geometric variables are exactly as in the matterless situation, 
except for the matter contribution in the conservation law.  
The general case from (\ref{eq:a6}) and (\ref{eq:a5})  may be 
derived from
\begin{equation}
d(X^+ X^-) + \mathcal{V} (X,Y)\, dX + X^- W^+  + X^+ W^- = 0\; .
\label{eq:e6}
\end{equation}
For chiral fermions only one of the two last terms 
remains. As the consequences of (\ref{eq:e6}) are of more 
general importance we will come back to that relation when the 
situation without restrictions on matter will be discussed 
below.

The matter equation for chiral fermions $k_L = 0$  in 
(\ref{eq:e4}) by variation of $\varphi_R$ and $k^2_R$ 
\begin{eqnarray}
&&e^+  d\, \varphi_R =  0\label{eq:e6a} \,,\\
&&d\,( k^2_R\, e^+ ) = 0\label{eq:e6b}
\end{eqnarray}
are solved easily, because $e^+$ is the previous solution 
(\ref{eq:a17}). Eq.\ (\ref{eq:e6a}) implies that $\varphi_R = 
\varphi_R \,(f)$ where $f$ is the arbitrary function introduced in 
(\ref{eq:a17}). Inserting the latter into (\ref{eq:e6b}) 
determines the amplitude:
\begin{equation}
k_R^2  = \frac{e^{-Q(X)}}{X^+} \; g(f)
\label{eq:e6c}
\end{equation}
An analogous procedure works for chiral scalars. Variation with respect to $e^\mp$ of the action (\ref{eq:a70}) for $F = \mbox{const.}$ 
in light-like coordinates ($\phi^\pm = \ast (d\phi  e^\pm)$) yields
\begin{equation}
W^\pm = -F\left[\phi^\pm\, d\phi\mp e^\pm\phi^+\phi^-\right]\, ,
\label{eq:e7}
\end{equation}
so that for (anti-)selfdual scalars with either $\phi^+ = 0$ 
(selfdual: $\ast d\phi = d\phi$) or $\phi^- = 0$ (anti-selfdual: $\ast d\phi 
= -d\phi$) again one of the eqs.\ (\ref{eq:a6}) is independent of the 
matter contribution. The subsequent steps to obtain the exact 
solution proceed as for chiral fermions. 

The e.o.m. for minimally coupled scalars can be written as 
\begin{equation}
d\ast d\phi = d (\phi^- e^+) = d (\phi^+ e^- ) = 0
\label{eq:e8}
\end{equation}
where the last two forms are to be used, respectively, in the 
(anti-)selfdual cases. Thus (anti-)selfdual matter 
identically solves (\ref{eq:e8}). Furthermore, e.g.\ for 
selfdual $\phi$ the condition $e^+  d\phi = 0$ makes the 
lines of $\phi = \mbox{const.}$ light-like. With the same solution 
(\ref{eq:a17}) for $e^+$ as without matter, the latter relation 
leads to $d\phi  df = 0$, i.e.\ $\phi=\phi(f)$, or vice versa. As 
$f$ represented a null coordinate in the EF line element, the 
peculiar light-like nature of $\phi$ is confirmed. The 
similarity between \mbox{(anti-)}\-selfdual scalars and chiral fermions is 
not surprising in view of the well-known close relation between 
scalars and fermions in $D=2$ \cite{Witten:1984ar}. 

Other nontrivial examples of exactly soluble systems are static 
solutions\footnote{An extensive discussion of these solutions can be found in
refs. \cite{Katanaev:2000kc,Grumiller:2001ea}.} \cite{Fisher:1948yn} or 
continuously self-similar solutions \cite{Roberts:1989sk} of SRG with a 
massless (non-minimally coupled) scalar field. 

Minimally coupled scalars with SRG for the geometric sector can be solved in a 
perturbative manner \cite{Mikovic:1997xm} or in the static 
limit \cite{Gergely:1998ba,Gergely:1999im}.

With the exception of aforementioned special cases, the CGHS model, the models
of refs. \cite{Filippov:1996ye,Filippov:1997db} or some 
other simple potentials like e.g.\ constant $V$ (i.e.\ Rindler metric) and of 
teleparallelism (pure torsion, $V=0$) \cite{Tieber:1997} no exact solution 
with general (scalar or fermionic) matter seems to be known\footnote{
Recently cosmological solutions in the JT model have been obtained 
\cite{Cadoni:2002pu,Kremer:2002dp}.}.
Indeed, one of the main open problems in classical dilaton theory with matter 
is an analytic (as opposed to numerical) description of non-trivial systems
showing the feature of critical collapse\footnote{Here ref. 
\cite{Choptuik:1993jv} for the seminal work of Choptuik should be quoted. It 
had been triggered by previous analytic studies of Christodoulou 
\cite{Christodoulou:1986du,Christodoulou:1986zr,Christodoulou:1987vu,
Christodoulou:1987vv}. Recent reviews are refs.
\cite{Gundlach:1998wm,Gundlach:1999cu}.}.


\clearpage

\section{Energy considerations}
It is important to clarify the relation of certain quantities appearing in 
generic \( D=2 \) gravity theories (like the absolutely conserved quantity 
\( \mathcal{C}^{\left( g\right) } \)) with respect to corresponding concepts 
well-known from \( D=4 \) Einstein gravity.



\label{se:5.1}

\subsection{ADM mass and quasilocal energy}

\label{admsec} Because of diffeomorphism invariance the Hamiltonian density
vanishes on the surface of the constraints in all gravity theories including
all $2D$ dilaton models\footnote{%
Cf. also the general Hamiltonian analysis in sect. \ref{se:3.2} below.
}. However, a boundary term must be included in the Hamiltonian which can be
used to define a global or {}``quasilocal{}'' energy. This is the essence
of the Arnowitt--Deser--Misner procedure \cite{Arnowitt:1962}. As
discussed in detail by Faddeev \cite{Faddeev:1982id} in the context of 
$4D$ Einstein gravity, this boundary is the one at (spatial) 
infinity, and
the value of the gravitational energy is very sensitive to the choice of
asymptotic conditions and it is \textit{not} invariant under the change of 
coordinates at infinity. This reflects the fact that the energy is related to
an observer connected with the asymptotic coordinate system who measures this
energy. In generic $2D$ dilaton gravity the situation becomes even 
more complicated
since a natural asymptotic coordinate system does not exist for some of the
models\footnote{%
In the absence of asymptotic flatness as in the KV-model 
\cite{Katanaev:1986wk,Katanaev:1990qm} other possibilities than the 
ADM-mass were discussed in ref. \cite{Kummer:1995qv}.}. This lack of 
asymptotic diffeomorphism invariance (and of conformal 
invariance) has led to much confusion in the literature.

To facilitate comparison with other works on the ADM approach we consider here
the second order dilaton action (\ref{e23eact}) with the exponential 
parameterization
\( X=e^{-2\Phi } \) for the dilaton field. Since only the first term \( e^{-2\Phi }R \)
is essential for the calculation of the ADM mass, this does not imply any 
restrictions on the potentials \( U \) and \( V \).

The diagonal gauge for the metric 
(\( N=\sqrt{\xi }\dot{f},\Lambda =\sqrt{\xi }f' \) in (\ref{eq:aconf}))
\begin{equation}
\label{emdiag}
(ds)^{2}=N^{2}(dt)^{2}-\Lambda ^{2}(dr)^{2}\, .
\end{equation}
is sufficient for our purposes since the lapse \( N \) is a Lagrange multiplier
for the Hamiltonian constraint in generally covariant theories. A more complete
canonical analysis can be found in refs.
\cite{Kuchar:1994zk,Louko:1995tv,Lau:1996fr,Kummer:1997si,Liebl:1997ti}.

The next step is to supplement the volume action (\ref{e23eact}) by a suitable
boundary term. The role of the latter is to convert second order derivatives
to first order ones. More exactly, such a term must provide standard
variation equations for the boundary data, the induced metric on the boundary.
In particular, this means that the first variation of the total action must not
contain boundary contributions with \textit{normal} derivative for the 
variation of the lapse \( N \). Since only the curvature term contains second 
derivatives, it also must not depend on the potentials \( U \) and 
\( V \) in (\ref{2.59}). Therefore, we may adopt for all models the expression 
for the SRG model. It can be obtained by spherical reduction of the standard
extrinsic curvature term in four dimensions. This yields \begin{equation}
\label{embterm}
L^{b}=\int\limits _{\partial \mathcal{M}}Ndt\, e^{-2\Phi }K\, ,
\end{equation}
 where we assume that \( \partial \mathcal{M} \) corresponds to
a constant value of \( r \). \( N\, dt \) is the surface element on the boundary,
\( e^{-2\Phi } \) is produced by the spherical reduction. \( K \) is the extrinsic
curvature, \begin{equation}
\label{emK}
K=-\frac{1}{N}\partial _{n}N=\mp \frac{1}{\Lambda }\partial _{r}(\log N)\, ,
\end{equation}
 where \( \partial _{n} \) denotes the derivative with respect to an outward
pointing unit normal. The upper (lower) sign in (\ref{emK}) should be taken on
the {}``right{}'' ({}``left{}'') component of the boundary.

For linear variations around a static background in the diagonal gauge (\ref{eq:aconf})
the curvature follows from the simplified formula \begin{equation}
\label{emR}
\sqrt{-g}R=2\partial _{r}\frac{1}{\Lambda }\partial _{r}N\, .
\end{equation}

Now the lapse \( N \) is varied in the total action \begin{equation}
\label{emvar}
\delta (L^{dil}+L^{b})=\int\limits _{\mathcal{M}}(\delta N)[{\textrm{e}.\textrm{o}.\textrm{m}}]d^{2}x+\int\limits _{\partial \mathcal{M}}dt(\delta N)2e^{-2\Phi }\partial _{n}\Phi \, .
\end{equation}
  
The second integral generates the so-called quasilocal energy associated
with the observer at the boundary: 
\begin{equation}
\label{emEql}
E_{ql}=2e^{-2\Phi }\partial _{n}\Phi =-\partial _{n}X\, 
\end{equation}
 To obtain the Hamiltonian (or the ADM mass) (\ref{emEql}) 
must be multiplied by \( N \): \begin{equation}
\label{emH}
H=E_{ADM}=-N\partial _{n}X\, 
\end{equation}
Evidently, solutions with constant dilaton (see, e.g., \cite{Kim:1998wy}) 
will lead to zero ADM mass.

If one moves the boundary to the asymptotic region for a generic \( D=2 \) 
dilaton theory, the right hand side of (\ref{emH})
diverges. In order to arrive at a finite value of a {}``generalized{}'' ADM 
mass a procedure like the Gibbons-Hawking subtraction 
\cite{Gibbons:1977ue,Brown:1993br} is needed. A rather natural idea 
\cite{Abbott:1982ff,Hawking:1996fd} is to subtract from the total action an 
action functional calculated for some reference space-time with the same 
induced metric on the spatial boundary (\( \Phi  \)
or \( X \) for SRG). Note that normal derivatives of the boundary data and
the normal metric (\( \Lambda  \)) may be different for the reference space-time.
Effectively, this means to subtract from the quasi-local {}``physical{}''
energy the one of some reference ({}``empty space{}'') configuration, denoted 
by a subscript \( 0 \): \begin{equation}
\label{emEr}
E_{ql}^{reg}=-\partial _{n}X+[\partial _{n}X]_{0}\, .
\end{equation}
To obtain the ADM mass measured in a physical space-time (\ref{emEr}) should be
multiplied by the lapse function \( N \) corresponding also to the physical
space-time: \begin{equation}
\label{emMr}
M_{ADM}^{reg}=N\left( -\partial _{n}X+[\partial _{n}X]_{0}\right) \, .
\end{equation}

Eq. (\ref{emMr}) may be evaluated directly for dilaton theories admitting 
solutions which are flat everywhere, i.e. where \( U \) and \( V \) are 
related by (\ref{eq:a38}).
SRG is a special case of this class, where it is natural to take the reference
frame to be the Minkowski space solution with \( \mathcal{C}_{0}=0 \). Instead
of studying the most general situation we concentrate on the subclass of models
(\ref{e23eact}). Identifying values of the dilaton field on the boundary for
physical and reference space-times is equivalent to identifying the coordinate
\( r \) of the boundary. This yields the total ADM mass \begin{equation}
\label{emM1}
M_{ADM}^{reg}=\pm \lim _{r\rightarrow {\mathcal{I}}}\sqrt{\xi (r)}\left( -\sqrt{\xi (r)}+\sqrt{\xi _{0}(r)}\right) \partial _{r}X\, ,
\end{equation}
 where the upper sign ($+$) should be taken if the asymptotic region 
\( \mathcal{I} \) corresponds to \( r\rightarrow \infty  \), and the lower one 
($-$) if \( \mathcal{I} \) corresponds to \( r\rightarrow 0 \). By substituting
e.g. the expressions (\ref{e23EF}) and (\ref{e23ur}) for \( \xi (r) \) and
\( r(X) \) and taking into account the position of the asymptotic region one
obtains \begin{equation}
\label{emM2}
M_{ADM}^{reg}=-\mathcal{C}_{0}\sqrt{\frac{a}{B}}\, .
\end{equation}
 It is instructive to check whether the mass of the Schwarzschild black hole
fits correctly into this procedure. For $D=4$ the values \( a=\frac{1}{2} \) 
and \( B=2\lambda ^{2} \) follow from (\ref{2.79}) and (\ref{e23UV}). 
Multiplying (\ref{emM2}) by the coefficient which has been omitted while 
passing from (\ref{2.58}) to (\ref{2.59}) indeed yields 
\begin{equation}
\label{emM3}
M_{ADM}^{reg}=-\frac{\mathcal{C}_{0}}{4G_{N}\lambda ^{3}}\, .
\end{equation}
 This value exactly coincides with (\ref{eq:a36}) where units with
\( G_{N}=1 \) have been used.

For the CGHS model (\( a=1 \) in (\ref{emM2})) the different coordinate system
(\ref{e23urs}) has to be used. Substituting also (\ref{e23ls}) in (\ref{emM1})
and taking the lower sign there - since the asymptotic region corresponds to
the lower limit of \( r \) - the ADM mass becomes \begin{equation}
\label{emM4}
M_{ADM}^{reg}=-\mathcal{C}_{0}B^{-\frac{1}{2}}\, ,
\end{equation}
 which somewhat surprisingly coincides with the naive limit \( a\rightarrow 1 \)
in (\ref{emM2}). This value is also consistent with the calculations existing
in the literature. For example, Witten's result \cite{Witten:1991yr} is recovered
if we take \( B=8/k' \), replace \( \mathcal{C}_{0} \) by a constant shift
of the dilaton \( \Phi  \) and take into account the overall factor of \( 1/2 \)
assumed in our action.

It should be noted that in all examples considered above positive mass BHs
correspond to negative values of \( \mathcal{C}_{0} \) whereas positive values 
of the latter describe naked singularities.

For Minkowski ground state theories (cf. (\ref{eq:a38})) of which
SRG is a special case, that subtraction procedure
appears very natural. For more complicated models it might be preferable to
subtract the total energy of a reference space from the total energy of the
physical space \cite{Liebl:1997ti}: \begin{equation}
\label{emMnew}
\tilde{M}_{ADM}^{reg}=-[N\partial _{n}X]+[N\partial _{n}X]_{0}\, .
\end{equation}
This formula, of course, cannot reproduce the correct mass e.g. in the case
of the Schwarzschild BH, but may be useful is a different context. 
There are considerable variations in the details of such a subtraction 
being used by different authors and in different models. Sometimes, 
it appeared more appropriate to subtract an extremal BH solution 
instead of the Minkowski space \cite{Berkovits:2001tg}. As noted in 
\cite{Kazakov:2001pj} the subtraction procedures of \cite{Witten:1991yr,
Gibbons:1992rh,Nappi:1992as,Mann:1993yv,Liebl:1997ti}, applied to the 
so-called exact string BH \cite{Dijkgraaf:1992ba}, lead to different 
results. Treating this model is especially tricky since the corresponding 
action is not known. 

It should always be kept in mind that a strong
dependence on the asymptotic conditions and on the subtraction procedure has
a clear physical origin: energy depends on the observer who measures it. This
is true for both reference and physical space-times.

Definitions of the ADM mass for higher dimensional dilaton gravities have been
considered in refs. \cite{Chan:1996sx,Iofa:1995xk}. 
For asymptotically Rindler and de Sitter models the ADM mass 
has been calculated in ref. \cite{Liebl:1997ti} (see also \cite{Kim:1999un}). 
This concept can be also introduced in the presence of 
radiation \cite{Kim:1996jt} and of a shock wave of matter fields 
\cite{Cruz:1999yq}. An extension of the Hamiltonian analysis to the case
of charged BH-s is  possible too\cite{Medved:1998ks}.

Obviously according to this procedure the ADM mass is \textit{not} conformally
invariant. This means that it will change, in general, if a conformal 
transformation, for example, removes the kinetic term for the dilaton. The
reason was clearly stated in ref. \cite{Chan:1996sx,Chan:1997ns}. 
Even though it is possible to make the unregularized energy conformally 
invariant for a selected class of models, the subtraction term will inevitably 
destroy that invariance since an {}``empty{}'' reference space is 
mapped into a non-trivial configuration. In other words, physical and 
reference observers are being transformed differently.

A final remark concerns approaches where instead of the ADM mass the
conserved quantity $\mathcal{C}^{(g)}$ has been related directly to a 
quasilocal energy expression. The notion of quasilocal energy has been 
investigated thoroughly in the context of General Relativity by Brown and 
York \cite{Brown:1993br,Brown:2000dz} and in the context of $2D$ dilaton 
gravity by Kummer and Lau \cite{Kummer:1997si}. As shown in ref. 
\cite{Kummer:1995qv} a relation to $\mathcal{C}^{(g)}$ is 
possible following the arguments leading to Wald's energy density 
\cite{Wald:1993nt,Iyer:1994ys}. Approaches which require no explicit 
subtraction have been suggested as well \cite{Kummer:1995qv, Mann:1993yv}.

\subsection{Conservation laws}\label{se:5.2}

Even in the most general dilaton theory including matter interactions, when no 
exact solution is known, the conservation law which may be derived from 
(\ref{eq:e6}) contains important information. Again potentials with quadratic 
torsion (\ref{2.77}) only are considered because there the integrating factor 
$\exp{Q}$ can be determined easily. Attaching that factor as in 
(\ref{eq:a11}), (\ref{eq:a12}) yields
\begin{equation}
d \mathcal{C}^{(g)} + W^{(m)} = 0\, ,
\label{eq:e9}
\end{equation}
where $\mathcal{C}^{(g)}$ is the quantity defined in 
(\ref{eq:a11}) for the geometric variable and (cf.\ 
(\ref{eq:a6}), (\ref{eq:a9}))
\begin{equation}
W^{(m)} = e^Q\, (X^+ W^- + X^- W^+ ) \; .
\label{eq:e10}
\end{equation}
Clearly $W^{(m)}$ from (\ref{eq:e9}) must obey the integrability 
condition $d W^{(m)} = 0$, a relation which in turn must be 
expressible in terms of the e.o.m.-s. 
With $W^{(m)} = d \mathcal{C}^{(m)}$ eq. (\ref{eq:e9}) simply 
becomes\footnote{An early version of a 
conservation law of this type \cite{Mann:1993yv} was not general enough to 
cover interacting scalars and fermions as introduced in the 
present chapter.}
\begin{equation}
d \mathcal{C}^{(tot)} = d (\mathcal{C}^{(g)} + \mathcal{C}^{(m)} ) = 0
\label{eq:e11}
\end{equation}
and $\mathcal{C}^{(tot)} = \mathcal{C}_0 = \mbox{const.}$ is an 
absolutely (i.e. in both coordinates) conserved quantity \cite{Kummer:1995qv}. 
Obviously in the presence of further gauge fields the present 
argument can be generalized easily following the steps of 
sect.\ \ref{se:2.4.1}.

Using the integrability condition for the 
components $W_\mu$ in  $W^{(m)} = W_\mu\, dx^\mu$, eq.\ 
(\ref{eq:e9}) can be integrated in two equivalent ways ($x^0 = t, 
x^1 = r$, $\mathcal{C}_0$ is an integration constant). 
\begin{align}
\mathcal{C}^{(g)} (t,r; t_0, r_0) &= 
- \int\limits_{t_0}^t \, dt'\; W_0 (t',r) - 
\int\limits_{r_0}^r\, dr' \, W_1 (t_0,r') + \mathcal{C}_0 = 
\nonumber\\
&=  - \int\limits_{r_0}^r\; dr'\, W_1 (t,r') - 
\int\limits_{t_0}^t \, dt'\, W_0 (t',r_0) + \mathcal{C}_0
\label{eq:e12}
\end{align}
for any first order gravity action (\ref{2.62}) or its equivalent 
dilaton form (\ref{2.59}). 

SRG may serve as a concrete example 
\cite{Grumiller:1999rz}. In the ``diagonal'' gauge widely used in spherical BH 
simulations \cite{Choptuik:1993jv,Gundlach:1998wm,Gundlach:1999cu} with 
$g_{00}=\alpha^2(t,r)$, $g_{11}=-a^2=-(1-2m/r)^{-1}$, $g_{01}=g_{10}=0$, the 
zweibein must be fixed as $e_0^+ = e_0^- = \alpha/\sqrt{2}, \;
e_1^+ = -e_1^- = a/\sqrt{2}$ and the dilaton field by 
 $X=\lambda^2 r^2/4$. Then the quantity $m (t,r)$ (sometimes called mass 
aspect function) is proportional to 
$\mathcal{C}^{(g)}$ with the proportionality factor for SRG 
given by (\ref{eq:a36}). As shown in \cite{Grumiller:1999rz} e.g.\ the first 
version of (\ref{eq:e12}) turns into
\begin{multline}
m(t,r; t_0,r_0) = \int\limits_{t_0}^t\, dt'\;
\frac{4\pi r^2}{a^2}\; (\partial'_0 \phi) (\partial_1 \phi) + \\
+ \int\limits_{r_0}^r \; dr' \; 
2\pi {r'}^2\, \left(\; \frac{(\partial_0 \phi)^2}{\alpha^2} + 
\frac{(\partial'_1 \phi)}{a^2}\; \right)_{t=t_0} + m_0\; .
\label{eq:e13}
\end{multline}

Defining the ADM-mass as $m_{ADM} = m(t_0,\infty,t_0,r_0)$ in the limit 
$r \to \infty$ in (\ref{eq:e13}) and using asymptotically free ingoing and 
outgoing spherical scalar waves 
$\phi \sim [ f_+ (t-r) + f_- (t+r) ] / r\sqrt{4\pi}$ 
yields the effective time-dependent mass of the (eventual) BH
\begin{equation}
m^{\rm (eff)}_{BH} (t) = m (t,\infty; t_0, 0) = m_{ADM} + 
\int\limits_{t_0}^t \; dt' \; [(f'_-)^2 - (f'_+)^2 ]\; .
\label{eq:e14}
\end{equation}
The matter contribution has the 
intuitive interpretation as the total incoming, resp. outgoing flux to 
infinity of matter at a certain time $t$, starting from $t=t_0$. 
In fact, when such fluxes exist, it is necessary to use $m_{BH}^{\rm (eff)}$ 
as a measure of BH formation rather than $m_{ADM}$ alone 
\cite{Gundlach:1998wm}\footnote{As pointed out in ref. 
\cite{Grumiller:1999rz} the first order formulation also seems to be much more convenient in gauges of the Sachs-Bondi type, where no coordinate singularity is created at the horizon. Then the introduction of the extrinsic curvature as an additional variable \cite{Marsa:1996fa} can be avoided altogether.}.
It is remarkable that, in contrast to the situation in $D=4$, in 
$D=2$ something like a standard energy conservation law can be 
formulated in this manner.

Clearly, the importance of that conservation (\ref{eq:e11}) and 
(\ref{eq:e12}) is not restricted to SRG where the ADM-mass can be 
defined from an asymptotically flat region. For generic $2D$ 
dilaton theories (\ref{2.59}) or (\ref{2.62}) there is a close 
relation of $\mathcal{C}^{(tot)}$ to the concept of ``quasilocal 
energy'' \cite{Brown:1993br,Kummer:1995qv,Liebl:1997ti,Kummer:1997si}
which has been dealt with also in the previous subsection. 


\subsection{Symmetries}

The final topic of this subsection is the question of symmetries, 
to be attached to (\ref{eq:e11}). It should be emphasized that 
these symmetries are quite different from the (gauge-like) ones 
incorporated automatically in the PSM approach, because the 
latter is valid for the geometric part of the action alone. When 
matter is absent the N{\"o}ther symmetry of $\mathcal{C}^{(tot)} = 
\mathcal{C}^{(g)}$ is realized by a translation in the 
Killing direction \cite{Kummer:1994ur}. In the 
presence of matter the integrability condition $dW^{(m)} = 0$ for 
(\ref{eq:e9}) can be interpreted as a conservation law of 
\textit{another} one form current $W^{(m)}$ which is related to a 
symmetry transformation with \textit{another} type of parameters. 
Both ingredients are necessary for the peculiar ``two-stage'' 
N{\"o}ther symmetry which has been encountered here \cite{Kummer:1998yg}. It 
seems to be yet another special feature of a generic $D=2$ 
theory.

In order to simplify the discussion of this unusual symmetry, 
the mechanism is explained in the frame of a toy-model 
\cite{Kummer:1994ur} which, nevertheless, 
contains all essential features:
\begin{equation}
\hat{L} = \int_{\mathcal{M}_2} (X dw + Kw  d \phi )\; .
\label{eq:e15}
\end{equation}
The first (``geometric'')  term in (\ref{eq:e15}) can be considered 
as a simplification of the Lagrangian (\ref{2.62}) whereas the 
second (``matter'')  term resembles the fermion interaction, as 
written in (\ref{eq:e2}) with (\ref{eq:e3}), and a current 
expressed in terms of amplitude (1-form $w$), phase (0-form $\phi$), and 
Lagrange multipliers (0-forms $X,K$). In the 
e.o.m-s to be derived from (\ref{eq:e15})
\begin{gather}
dX + Kd\phi = 0 \, , \label{eq:e16}\\
dw = 0 \, , \label{eq:e17}\\
w  d \phi = 0 \, , \label{eq:e18}\\
d (K w) = 0 \, , \label{eq:e19}
\end{gather}
eq.\ (\ref{eq:e16}) represents the analogue of the conservation 
law in the form (\ref{eq:e6}) with $W^{(m)} = K d\phi$. The 
integrability condition $d W^{(m)} = 0$ becomes $dK  d \phi 
= 0$. This implies $K = K (\phi)$ so that $K d\phi = 
d\,(\int_{y_0}^\phi\, K (y)\, dy )$, and
\begin{equation}
d \mathcal{C} = d \left( 
X + \int\limits_{y_0}^\phi \, K (y) \, dy \right) = 0
\label{eq:e20}
\end{equation}
is the counterpart of (\ref{eq:e11}). Thus $\mathcal{C} = 
\mathcal{C}_0 = \mbox{const.}$ characterizes the solutions of 
this theory. From (\ref{eq:e17}) and (\ref{eq:e18}) similarly $w 
= w (\phi)$ can be concluded, so that (\ref{eq:e19}) is fulfilled 
identically. 

In the ``matterless'' case ($K=0$) the ``geometric'' symmetry 
transformations are constant translations $\delta w = \delta 
\gamma = \mbox{const.}$  The integrability condition $d(K d \phi) = 0$ allows 
an expansion in terms of e.o.m.-s 
(\ref{eq:e17})-(\ref{eq:e19})  which correspond, 
respectively, to  the variations 
$\delta \hat{L}/\delta X, \delta \hat{L}/\delta K, 
\delta \hat{L}/\delta \phi$:
\begin{equation}
d\left(K  d\phi\right) = \left(
\frac{\delta \hat{L}}{\delta\phi} - K\, \frac{\delta \hat{L}}{\delta 
X} \, \right) \; \frac{\partial_0 \phi}{w_0} + 
\frac{\delta \hat{L}}{\delta K}\, \frac{\partial_0 K}{w_0}
\label{eq:e21}
\end{equation}
The apparent dependence on the specific coordinate $x^0$ is 
spurious (e.g. $\partial_0 \phi / w_0 = \partial_1\phi / w_1$ 
from (\ref{eq:e18})). Thus (\ref{eq:e21}) permits the 
introduction of a ``matter'' symmetry with global parameter $\delta\rho$
\begin{equation}
\delta \phi = \frac{\partial_0 \phi}{w_0} \, \delta \rho, \quad
\delta X = - K \frac{\partial_0 \phi}{w_0}\, \delta \rho, \quad
\delta K = \frac{\partial_0 K}{w_0}\, \delta \rho
\label{eq:e22}
\end{equation}
or an equivalent one with $\partial_0 \to \partial_1,\; w_0 \to w_1$. 
It can be checked that the Lagrangian 
$\hat{\mathcal{L}}$ in $\hat{L} = \int \hat{\mathcal{L}}$ of 
(\ref{eq:e15}) indeed  transforms as a total divergence: 
\begin{equation}
\delta \hat{\mathcal{L}} = d \left( K d\phi - 
K \, \frac{\partial_0 \phi}{w_0}\, w \, \right)\, \de \rho
\label{eq:e23}
\end{equation}
The related conserved N{\"o}ther one form current becomes $J = 
Kd\phi$, or  $\ast J^\mu = \epsilon^{\mu\nu} K 
\partial_\nu \phi$ in components for the Hodge dual 
of $J$. Hence the conservation law for 
the complete expression (\ref{eq:e20}) is related to a 
simultaneous transformation of the action $\hat{L}$ with respect 
to both the symmetry parameters $\delta \gamma$ and $\delta\rho$ the second of
whom belongs to a different (one-form) current $W^{(m)}=Kd\phi$.

It is straightforward to apply the procedure, as outlined in this 
simple example, to a general theory with matter interactions in 
$D=2$. The resulting formulas are quite lengthy (cf.\ 
\cite{Kummer:1998yg}) and, therefore, will not be reproduced here.

\clearpage

\section{Hawking radiation }\label{se:3}

One of the main motivations for studying low dimensional gravity theories is
the hope to get insight into the dynamics of a BH, its quantum radiation and
eventual evaporation \cite{Hawking:1975sw}. Therefore, it is important to make
sure that especially the effect of Hawking radiation still exists in two-dimensional
theories and to study its basic properties like the temperature-mass relation.

It should be kept in mind, though, that this effect, discovered more than a quarter
of a century ago, is a fixed background phenomenon. No quantum gravity is involved;
only the matter field action is taken into account in the one-loop approximation.
The vacuum polarization is described by the energy momentum tensor, induced
by this quantum effect, \begin{equation}
\label{ehT}
T_{\mu \nu }=\frac{2}{\sqrt{-g}}\frac{\delta W}{\delta g^{\mu \nu }}\, ,
\end{equation}
 where \( W \) is the one-loop effective action for the matter fields on a
classical background manifold with metric \( g^{\mu \nu } \). For minimal coupling
of scalars in $2D$ \( W \) in (\ref{ehT}) is the famous Polyakov
action \cite{Polyakov:1981rd}. In a suitable coordinate system the Hawking
flux is given by the light-cone component \( T_{--} \) calculated in the asymptotic
region. 

To this end various methods have been developed \cite{Frolov:1998wf}. Most
of them can be applied in $2D$. Variation of \( W \) as in (\ref{ehT})
allows the direct determination of \( T_{\mu \nu}  \). Alternatively, the thermal
particle distribution may be reproduced by comparing different vacuum states
from the Bogoliubov coefficients \cite{Christodoulakis:2000yr,Vagenas:2000am}.

In this review we follow the approach of Christensen and Fulling \cite{Christensen:1977jc}
based upon the conformal anomaly. Like the comparison of thermal distributions
it should not be sensitive to the dimensionality of space-time. Here the computation
for minimally coupled scalars is very simple, and a closed expression for the
energy momentum tensor may be given for any dilaton gravity model. For non-minimal
coupling, the situation is much more complicated. Several problems still remain
unsolved, although the result for the flux from \( D=4 \) can be reproduced
correctly. A detailed and elementary discussion of the non-minimal case can
be found in ref. \cite{Kummer:1999zy}, where it was shown that the use of 
the fully integrated effective action could be avoided altogether.

\subsection{Minimally coupled scalars}

The simplest example is a minimally coupled scalar field with action (\ref{eq:a68})
and $F\mathcal{O}_{D-2}/\la^{D-2}=1/2$: \begin{equation}
\label{ehmact}
L_{(\phi )}^{min}=\frac{1}{2}\int d^{2}x\sqrt{-g}g^{\mu \nu }(\partial _{\mu }\phi )(\partial _{\nu }\phi )\, .
\end{equation}
 If \( \phi  \) is taken to be an on-shell classical field
the energy-momentum tensor satisfies the usual conservation
equation \begin{equation}
\label{ehmcon}
\nabla ^{\mu }T_{\mu \nu }=0\, .
\end{equation}
 The same relation holds for 1-loop quantum corrections with a trivial 
background field $\phi=0$ where in (\ref{ehmcon})
the effective action \( W \) by (\ref{ehT}) appears. Eq. (\ref{ehmcon}) is
most conveniently analyzed in the conformal gauge (\ref{eq:conformalgauge}).
We change variables \( dz=dr/\xi (r) \) in the generalized Schwarzschild gauge
(\ref{eq:agensch}) to obtain (\ref{eq:aconf}) in the form \begin{equation}
\label{ehmc1}
(ds)^{2}=\xi (r)((dt)^{2}-(dz)^{2})\, .
\end{equation}
 In light cone coordinates \( x^{\pm }=(t\pm z)/\sqrt{2} \) the line element
(\ref{ehmc1}) will be expressed as \begin{equation}
\label{ehmc2}
(ds)^{2}=2e^{2\rho }dx^{+}dx^{-}\, ,\qquad \rho =\frac{1}{2}\log (\xi )\, .
\end{equation}

For the asymptotically Minkowski models $0<a<1$ considered in (\ref{e23EF})
it is convenient to write \( \xi  \) as \begin{equation}
\label{ehml}
\xi (r)=1-\left( \frac{r_{h}}{r}\right) ^{\frac{a}{1-a}}\, ,
\end{equation}
 where \begin{equation}
\label{ehmrh}
r_{h}=(-2\mathcal{C}_{0})^{\frac{1-a}{a}}\frac{1}{|1-a|}\left( \frac{B}{a}\right) ^{\frac{a-2}{2a}}
\end{equation}
 is the value of \( r \) at the horizon. The explicit form (\ref{ehmrh}) will
not be needed until the very end of this calculation.

There are only two non-zero components of the Levi-Civit\'{a} connection: \( \Gamma _{++}{}^{+}=2\partial _{+}\rho  \)
and \( \Gamma _{--}{}^{-}=2\partial _{-}\rho  \). The minus component of \( \nu  \)
in (\ref{ehmcon}) yields \begin{equation}
\label{ehmc-}
\partial _{+}T_{--}+\partial _{-}T_{+-}-2(\partial _{-}\rho )T_{+-}=0\, .
\end{equation}
 On static backgrounds, which depend on the variable \( r \) alone, the relations
\begin{equation}
\label{ehmpd}
\partial _{+}=-\partial _{-}=\frac{1}{\sqrt{2}}\partial _{z}=\frac{1}{\sqrt{2}}\xi (r)\partial _{r}\, ,
\end{equation}
 between partial derivatives hold. Therefore, (\ref{ehmc-}) becomes a simple
first order ordinary differential equation \begin{equation}
\label{ehmod}
(\partial _{z}-2(\partial _{z}\rho ))T_{+-}=\partial _{z}T_{--}\, .
\end{equation}
The flux component \( T_{--} \) can be found easily from the trace
\cite{Christensen:1977jc}\eq{
T^{\mu}_{\mu }= 2 e^{-2\rho} T_{+-}\,.
}{ehmTrace}

As the classical trace of $T_{\mu\nu}$ for a massless field is zero in \( D=2 \),
the whole contribution to \( T_{\mu }^{\mu } \) arises from the conformal (or
Weyl) anomaly (cf. \cite{Duff:1994wm} for a historical review). With minimally
coupled scalars in $2D$ the calculations are especially simple, but
they permit to illustrate several important points. As a first step, in the 
action (\ref{ehmact}) an integration by parts is performed and it is 
continued to the Euclidean domain,
\begin{equation}
\label{ehme}
L_{E}=\frac{1}{2}\int d^{2}x\sqrt{g}\phi A\phi \, ,
\end{equation}
 where \( A=-\Delta =-g^{\mu \nu }\nabla _{\mu }\nabla _{\nu } \) is the Laplace
operator on the curved background.

The path integral measure is defined by the relation \begin{equation}
\label{ehmm}
1=\int (D\phi )\exp \left( -\int d^{2}x\sqrt{g}\phi ^{2}\right) \,,
\end{equation}
so that the procedure maintains diffeomorphism invariance and thus preserves the
conservation equation (\ref{ehmcon}). It is also possible to trade part of
the diffeomorphism invariance for Weyl invariance \cite{Karakhanian:1994gs,Jackiw:1995qh,Amelino-Camelia:1995nk,La:1995mc, Cruz:1996zt},
but this option will not be considered here.

The partition function for the field \( \phi  \) reads \begin{equation}
\label{ehmZ}
Z=\int (D\phi )\exp \left( -\int d^{2}x\sqrt{g}\phi A\phi \right) =(\det A)^{-\frac{1}{2}}\, .
\end{equation}
 where the determinant is divergent. The zeta function
regularization \cite{Dowker:1976tf,Hawking:1977ja}\begin{equation}
\label{ehmW}
W=-\ln Z=-\frac{1}{2}\zeta' _{A}(0),\qquad \zeta _{A}(s)={\textrm{Tr}}(A^{-s})\, ,
\end{equation}
 is very convenient in the present context. Prime denotes differentiation with
respect to \( s \). Strictly speaking, to keep the argument of the zeta function
in (\ref{ehmW}) dimensionless, one has to multiply it by \( \mu ^{2s} \) where
\( \mu  \) is a parameter with mass dimension one. Then the effective
action \( W \) will be shifted by \( -\frac{1}{2}\zeta _{A}(0)\ln{\mu^2} \)
which represents the usual renormalization ambiguity. This term, however, does
not contribute to the anomaly.

The following analysis will be valid for an arbitrary conformally covariant
operator which means that under an infinitesimal conformal transformation  
\( \delta g_{\mu \nu }=2g_{\mu \nu } \de\rho(x)\) of
the metric (\ref{ehmc2}) the operator \( A \) changes as \begin{equation}
\label{ehmdA}
\delta A=-2(\de\rho(x)) A\, .
\end{equation}
 Because of this property, the variation of the zeta function is simply \begin{equation}
\label{ehmdz}
\delta \zeta _{A}(s)=-s{\textrm{Tr}}((\delta A)A^{-1-s})=2s{\textrm{Tr}}((\delta \rho )A^{-s})\, ,
\end{equation}
i.e. the operator \( A^{-s} \) is restored with its original power. The corresponding
change of the effective action is expressed in terms of a generalized ({}``smeared{}'')
zeta function: \begin{equation}
\label{ehmdW}
\delta W=-\zeta (0|\delta \rho,A)\, ,\qquad \zeta (s|\delta \rho,A):={\textrm{Tr}}((\delta \rho )A^{-s})\, .
\end{equation}
At vanishing argument \( s\rightarrow 0 \) eq. (\ref{ehmdW}) can be evaluated
easily by heat kernel methods. For the operator \( A=-\Delta  \) the result
is\footnote{%
See Appendix B for the details.
} \begin{equation}
\label{ehmz0}
\zeta (0|\delta \rho,-\Delta )=\frac{1}{24\pi }\int d^{2}x\sqrt{g}R\de\rho\, .
\end{equation}
On the other hand, the definition of the energy-momentum tensor (\ref{ehT})
yields \begin{equation}
\label{ehmdWT}
\delta W=\frac{1}{2}\int d^{2}x\sqrt{g}T_{\mu \nu }\delta g^{\mu \nu }=-\int d^{2}x\sqrt{g}T_{\mu }^{\mu }\de\rho\, .
\end{equation}
 Comparing (\ref{ehmdW}) with (\ref{ehmdWT}) and (\ref{ehmz0}) the well-known
expression for the trace anomaly \begin{equation}
\label{ehmca}
T_{\mu }^{\mu }=\frac{1}{24\pi }R
\end{equation}
 follows, which remains unchanged after continuation back to Minkowski signature.

In conformal gauge the Ricci scalar becomes\footnote{%
This expression may be obtained most easily from the identity (\ref{2.53})
with $g_{\mu\nu}$ from the line element (\ref{ehmc2}), $\hat{R}=0$, 
$\partial_r=\partial_z$.
}\begin{equation}
\label{ehmRc}
R=2e^{-2\rho }\partial _{z}^{2}\rho \, .
\end{equation}
 In light cone coordinates (\ref{ehmTrace}) yields \begin{equation}
\label{ehmTpm}
T_{+-}=\frac{1}{24\pi }\partial _{z}^{2}\rho \, .
\end{equation}
 With this input, the conservation equation (\ref{ehmod}) is solved easily,
\begin{equation}
\label{ehmTmm}
T_{--}=\frac{1}{24\pi }\left[ \partial _{z}^{2}\rho -(\partial _{z}\rho )^{2}\right] +t_{-}\, ,
\end{equation}
 where \( t_{-} \) is the integration constant.

Different choices of \( t_{-} \) correspond to different {}``quantum vacua{}''
\cite{Boulware:1975dm,Unruh:1976db,Israel:1976ur,Gibbons:1976es}. There is
nothing specific for $2D$ models in this respect. We assume that
the Killing horizon is non-degenerate, i.e. \( \xi (r) \) has a simple zero
at \( r=r_{h} \) as for \( \xi (r) \) in (\ref{ehml}). To ensure regularity
of the energy-momentum tensor at the horizon in global (Kruskal) coordinates
one has to require that \( T_{--} \) exhibits a second order zero at 
\( r=r_{h} \). There is only one integration constant \( t_{-} \) available. 
Therefore, fixing it by the requirement 
\begin{equation}
\label{ehmUvac}
\left. T_{--}\right|_{h}=0\,,
\end{equation}
it must be checked later on whether (\ref{ehmUvac}) indeed produces a second
order zero.

In terms of the function \( \xi  \) the energy momentum tensor (\ref{ehmTmm})
can be expressed as (cf. (\ref{ehmc2}) and (\ref{ehmpd})) \begin{equation}
\label{ehmTl}
T_{--}=\frac{1}{96\pi }\left[ 2\xi \xi'' -(\xi' )^{2}\right] +t_{-}\, .
\end{equation}
With $t_-$ determined from (\ref{ehmUvac}) it is an easy exercise to show that 
for the asymptotically Minkowski models
(\ref{ehml}) the Hawking flux in the asymptotic region becomes \begin{equation}
\label{ehmHf}
\left. T_{--}\right| _{as}=\frac{a^{2}}{96\pi \, (a-1)^{2}r_{h}^{2}}\, .
\end{equation}
 This flux defines the Hawking temperature \( T_{H} \) of the BH. In $2D$
the Stefan-Boltzmann law contains \( T_{H}^{2} \): \begin{equation}
\label{ehmTH}
\left. T_{--}\right| _{as}=\frac{\pi }{6}T_{H}^{2}\, .
\end{equation}
 Comparing (\ref{ehmHf}) and (\ref{ehmTH}) the value of \( T_{H} \) agrees
with the one derived from surface gravity \( T_{H}=\frac{1}{4\pi }\left. \xi' \right| _{r_{h}} \).
These equations together with (\ref{ehmrh}) and (\ref{emM2}) fix the dependence
of the Hawking temperature on the ADM mass for this class of models: \begin{equation}
\label{ehmTM}
T_{H}\propto (M_{ADM})^{\frac{a-1}{a}}\, .
\end{equation}
 The well known inverse mass law for the Schwarzschild BH (\( a=1/2 \)) is
reproduced. Eq. (\ref{ehmTM}) reveals an intriguing property 
\cite{Liebl:1997ti} of the class of $2D$ models discussed in sect. 
\ref{abfamily}: depending on the parameter \( a \)
the Hawking temperature may be proportional to a negative, but also 
a positive power of the BH mass.

It is easy to check that near the horizon indeed \begin{equation}
\label{ehmTh}
\left. T_{--}\right| _{r\rightarrow r_{h}}\sim (r-r_{h})^{2}
\end{equation}
 for all values of \( a \), i.e. the requirement of a continuous flux in Kruskal
coordinates is fulfilled.

Again, the CGHS model must be considered separately. By substituting (\ref{e23ls})
in (\ref{ehmTl}) the Hawking flux \begin{equation}
\label{ehmTs}
\left. T_{--}\right| _{as}=\frac{B}{96\pi }
\end{equation}
 is obtained, consistent with the earlier calculation \cite{Callan:1992rs}.
It is important to note that in the CGHS model Hawking radiation does not depend
on the ADM mass.

Hawking radiation can be studied as well for asymptotically Rindler and 
de Sitter models. Explicit expressions can be found in ref. 
\cite{Liebl:1997ti}. $T_{--}$ for {}``exotic{}'' configurations with constant 
dilaton has been calculated in refs. \cite{Kim:1998wy,Kim:1999un}.
It is very sensitive with respect to asymptotic conditions on the metric. 
Physically this means that one has to fix length and time scales used in its 
measurement. Clearly, different scales yield different results, as may be seen 
by comparing refs. \cite{Fabbri:1996bz} and \cite{Liebl:1997ti} where 
asymptotically Rindler spaces were studied. By choosing an accelerated 
reference system Hawking radiation may be converted into Unruh radiation 
\cite{Cadoni:1995mi}.

It should be stressed that Hawking radiation behaves quite differently in 
conformally related models as witnessed by the results of ref.
\cite{Cadoni:1996dd} vs. ref. \cite{Liebl:1997ti}. Conformal transformations 
change, in general, also the asymptotic behavior of the metric and of the path 
integral measure. Indeed the very existence of the conformal anomaly means 
conformal non-invariance of the theory.

The case of minimally coupled spinor fields interacting (again minimally) with
an abelian gauge field can be also analyzed along the same lines. One has to
add a contribution of the chiral anomaly to 
the Polyakov action \cite{Nojiri:1992st,Ori:2001xc}\footnote{
The case of neutral matter on the background of a charged BH
is even simpler. One has to modify only the metric in the Polyakov
action \cite{Diba:2002hb}. The expression (\ref{ehmTl}) still holds in terms
of a different $\xi$.}.
Another generalization \cite{Christodoulakis:2001ps} consists in considering
the Casimir force due to a minimally coupled scalar field between two surfaces on a
CGHS background.

\subsection{Non-minimally coupled scalars}

The scalar field action (\ref{eq:a68}),(\ref{eq:a69})
contains a non-minimal coupling to the dilaton from spherical reduction. 
On dimensional and symmetry
grounds for a GDT in the path integral measure also a general function \( \Psi  \) of the dilaton \( \Phi  \)
may be introduced, \begin{equation}
\label{ehnm}
1=\int (D\phi )\exp \left( -\int d^{2}x\sqrt{g}e^{-2\Psi }\phi ^{2}\right) \, ,
\end{equation}
 instead of the standard mode normalization condition following from \( D \)
dimensional spherical reduction \( (\Psi =\Phi ) \), using the exponential
parameterization of the dilaton (\ref{e23exp}).
Then the rescaled field \( \varphi =e^{-\Psi }\phi  \) still possesses the
standard dilaton independent path integral measure in $2D$ (\ref{ehmm}).
In terms of this new field the action (\ref{eq:a68}) reads \begin{eqnarray}
 &  & L^{(nm)}=\frac{1}{2}\int d^{2}x\sqrt{g}\varphi A^{(nm)}\varphi \, ,\label{ehnac} \\
 &  & A^{(nm)}=-e^{2(\Psi -\Phi )}g^{\mu \nu }(\nabla _{\mu }\nabla _{\nu }+2(\Psi _{,\mu }-\Phi _{,\mu })\partial _{\nu }\nonumber \\
 &  & \qquad \qquad \qquad \qquad +\Psi _{,\mu \nu }+\Psi _{,\mu }\Psi _{,\nu }-2\Psi _{,\mu }\Phi _{,\nu }),\label{ehnA} 
\end{eqnarray}
 where an integration by parts has been performed and an 
irrelevant overall factor
in the action has been dropped \( (\Psi _{,\mu }=\nabla _{\mu }\Psi ) \).

The first calculation of the conformal anomaly for non-minimally coupled scalar
fields with the spherically reduced path integral measure (\( \Psi =\Phi  \))
has been presented by Mukhanov, Wipf and Zelnikov \cite{Mukhanov:1994ax} who
were also the first to address the problem of Hawking radiation for spherically
reduced matter. Their result was confirmed later 
\cite{Chiba:1997ex,Ichinose:1998gq} and extended to arbitrary measure 
\cite{Kummer:1997jr}\footnote{%
The literature on this subject is quite large (cf. e.g. \cite{Bousso:1997cg, Nojiri:1997hx,Mikovic:1998xq,Nojiri:1998yg,Ichinose:1998kh,Nojiri:2000ja}).
}.

As in the minimally coupled case the zeta function regularization (\ref{ehmW})
may be employed. The operator \( A^{(nm)} \) being conformally covariant, the equations
(\ref{ehmdA}) -- (\ref{ehmdW}) and (\ref{ehmdWT}) (after the replacement
\( A\rightarrow A^{(nm)} \)) are still valid. The conformal anomaly is derived
from \( \zeta (0|\delta \rho,A^{(nm)}) \). Again the general formulae from
Appendix \ref{appB} can be used, because \( A^{(nm)} \) may be expressed in
the standard form (\ref{ebA}) 
\begin{equation}
\label{ehnAs}
A^{(nm)}=-(\hat{g}^{\mu \nu }\hat{\nabla }_{\mu }\hat{\nabla }_{\nu }+\hat{E})\, ,
\end{equation}
 with the {}``effective{}'' metric \( \hat{g}^{\mu \nu }=e^{2(\Psi -\Phi )}g^{\mu \nu } \)
and the covariant derivative \begin{equation}
\label{ehno}
\hat{\nabla }_{\mu }=\partial _{\mu }+\hat{\Gamma }_{\mu }+\hat{\omega }_{\mu }\, ,\qquad \hat{\omega }_{\mu }=\Psi _{,\mu }-\Phi _{,\mu }\, ,
\end{equation}
 where \( \hat{\Gamma } \) is the Christoffel connection for the metric \( \hat{g} \).
Here the potential \( \hat{E} \) reads \begin{equation}
\label{ehnEh}
\hat{E}=\hat{g}^{\mu \nu }(-\Phi _{,\mu }\Phi _{,\nu }+\Phi _{,\mu \nu })\, .
\end{equation}
According to Appendix \ref{appB}, (eq. (\ref{eb2gen}) with (\ref{ebalp1})
and (\ref{ebalp2})), after returning to Minkowski space one obtains for the
smeared $\zeta$-function (\ref{ehmdW})
\begin{equation}
\label{ehnz}
\zeta (0|\delta \rho,A^{(nm)})=\frac{1}{24\pi }\int d^{2}x\sqrt{-\hat{g}}
(\hat{R}+6\hat{E})\de\rho\, ,
\end{equation}
 where \( \hat{R} \) is the scalar curvature determined from \( \hat{g} \),
so that (\ref{ehnEh}) and (\ref{ehnz}) yield the trace anomaly \cite{Kummer:1997jr}
\begin{equation}
\label{ehnT}
T_{\mu }^{\mu }=\frac{1}{24\pi }(R-6(\nabla \Phi )^{2}+4\nabla ^{2}\Phi +2\nabla ^{2}\Psi )\,. 
\end{equation}

For the spherically reduced measure \( \Psi =\Phi  \) this expression agrees
with refs. \cite{Mukhanov:1994ax,Chiba:1997ex,Ichinose:1998gq}. Different expressions
for the conformal anomaly with various choices of the measure were reported
too \cite{Bousso:1997cg,Nojiri:1997hx,Mikovic:1998xq,Nojiri:1998sr}. It is often
important
to keep track of total derivatives (or zero modes in the compact case) in
computations of $T^\mu_\mu$. A careful analysis of this type has been 
performed by Dowker \cite{Dowker:1998bm} (cf. also 
\cite{Kummer:1998sp,Kummer:1999zy}) who confirmed the result (\ref{ehnT}) for
SRG.

When scalars are coupled nonminimally to a dilaton field the conservation law
for the one-loop energy-momentum tensor has to be modified, \begin{equation}
\label{ehncon}
\nabla ^{\mu }T_{\mu \nu }=-(\partial _{\nu }\Phi )\frac{1}{\sqrt{-g}}\frac{\delta W}{\delta \Phi }\, ,
\end{equation}
 as can be seen by applying the usual assumption of diffeomorphism invariance
to a \( \phi  \) dependent matter action. In the absence of classical matter 
fluxes the matter action can be replaced again by the one-loop effective action of
non-minimally coupled scalars \cite{Hofmann:2002}.

In contrast to conformal transformations, a shift of the dilaton field \( \Phi \rightarrow \Phi +\delta \Phi  \)
does not act on the operator \( A^{(nm)} \) in a \textit{covariant} way\footnote{%
From now on we assume that the function \( \Psi  \) in the measure is fixed.
The most relevant choice (SRG) is \( \Psi =\Phi  \).
}. Therefore, the variation of \( \left({A^{(nm)}}\right)^{-s} \) in the zeta 
function does
not yield a power of \( A^{(nm)} \) anymore. As a consequence the variation
of the effective action cannot be expressed in terms of known heat kernel coefficients.

A way to overcome this difficulty has been suggested in \cite{Kummer:1998dc}.
By keeping the same classical action, but by changing the hermiticity requirements
of relevant operators, \( A^{(nm)} \) has been transformed into a product of
two operators of Dirac type each of which transforms homogeneously under the
shifts of the dilaton field. This allowed us to calculate the energy-momentum
tensor and the effective action in a closed analytical form. Although this procedure
changes the original spectral problem, the results exhibit several attractive
features which have to be present in spherically reduced theories. 
For example, the Hawking
temperature coincides with its geometrical expression through surface gravity.
Thus, in hindsight
one may conjecture \cite{Kummer:1999zy} that this procedure somehow takes into
account the {}``dimensional reduction anomaly{}'' (see below). Another model
where the energy-momentum tensor can be calculated exactly has been proposed
recently \cite{Frolov:2000jh}.

In this connection it should be remarked that in $2D$ a generic differential
operator can be represented in {}``dilaton{}'' form\footnote{%
In $2D$ locally any operator of Laplace type can be represented as
a product of an operator of Dirac type and its conjugate \cite{Vassilevich:2001at}.
In contrast to the present case, however, these Dirac operators will not necessarily
transform homogeneously under the shift of the dilaton.
} \begin{equation}
\label{ehndp}
\bar{A}=-\left( e^{\Psi }\nabla^{\mu }e^{-\Phi }\right) \left( e^{-\Phi }\nabla _{\mu }e^{\Psi }\right) :=L^{\mu }L_{\mu }^{\dag }\, .
\end{equation}
 This parameterization proved very convenient in resumming the perturbative expansion
of the effective action \cite{Gusev:1999cv}. It also allows to prove some symmetry
relations between functional determinants even in higher dimensions or if \( \Psi  \)
and \( \Phi  \) are matrix-valued fields \cite{Vassilevich:2000kt}. Roughly
speaking, these symmetry relations allow to interchange \( L_{\mu } \) and
\( L_{\mu }^{\dag } \) inside the determinant which is a rather non-trivial
operation because of the summation over \( \mu  \) in (\ref{ehndp}).

In order to study the quantum back reaction upon the classical BH, solutions of
the field equations obtained from an action containing both classical and one-loop
parts are needed. It is not possible to solve such equations in general, even
if the quantum effects are represented by the simplest Polyakov action. However,
for particular dilaton theories exact solutions can be obtained \cite{Russo:1992ht,Russo:1992ax,Bose:1995pz,Cruz:1996zt,Fabbri:1998hs, Kim:1999wa,Zaslavsky:1999zh,Zaslavsky:1998ca},
although in some of these papers the {}``quantum{}'' part was rather introduced
by hand than derived.

An effective action for non-minimally coupled fields was presented in 
\cite{Kummer:1998dc}. Admittedly, its derivation lacked complete rigor. 
Many authors \cite{Bousso:1998wi,Bousso:1998bn,
Bousso:1999ms,Buric:2001xw,Buric:1999it, Buric:1999wf,Buric:2000cj,
Medved:1999nt,Medved:2000fk,Medved:2001pk,Medved:2001sk,Barbachoux:2002tt} 
employ the {}``conformal{}'' action 
\begin{equation}
\label{ehnWc}
W^{({\textrm{conf}})}=\frac{1}{96\pi }\int d^{2}x\sqrt{-g}\left[ R\frac{1}{\Box }R-12(\nabla \Phi )^{2}\frac{1}{\Box }R+12R\Phi \right] 
\end{equation}
 which correctly reproduces the conformal anomaly (\ref{ehnT}) for \( \Psi =\Phi  \)
but neglects (an undetermined) conformally invariant part\footnote{%
Some physical motivations why the conformally non-invariant part may dominate 
at a certain energy scale can be found in ref. \cite{Novozhilov:1989sn}.
}. The first term under the integral in (\ref{ehnWc}) yields the Polyakov 
action. It is interesting to note that although (\ref{ehnWc}) differs from
the full effective action obtained in \cite{Kummer:1998dc}, many physical 
predictions are identical.

Even if it is assumed that \( W^{({\textrm{conf}})} \)
provides a correct description of one-loop effects for non-minimally coupled
matter, many problems remain open. The first one is how to deal with the
non-local
terms in (\ref{ehnWc}). Direct variation of this equation with
respect to the metric leads to very complicated expressions \cite{Lombardo:1998iw}. 
It was proposed in ref. \cite{Buric:1998xv} to convert (\ref{ehnWc}) into a 
local
action by introducing two auxiliary fields, \( f_{1}=(1/\Box )R \) and 
\( f_{2}=(1/\Box )(\nabla \Phi )^{2} \). Various versions of this method were 
frequently used since (e.g. \cite{Buric:2001xw,Buric:1999it,Buric:1999wf,
Buric:2000cj,Balbinot:1998yh,Balbinot:2000at}). As the new action
in terms of \( f_{1} \) and \( f_{2} \)
is local it is quite straightforward to vary it with respect to the 
metric in order to arrive at the energy
momentum tensor. However, since \( f_{1} \) and \( f_{2} \) are to be found
from \( R \) and \( (\nabla \Phi )^{2} \) by solving second order differential
equations, the energy momentum tensor obtained in this way will in general depend
on four integration constants. This is an indication that such an extended action
does not necessarily yield the same physics as the original one. For the latter
a single \textit{first order} equation (\ref{ehncon}) must be solved in the 
conformal gauge. Indeed, many physical predictions, as, e.g., the 
BH {}``anti-evaporation{}''
\cite{Bousso:1998wi} may depend on the way the action (\ref{ehnWc}) is treated.

It has been noted \cite{Balbinot:1999ir,Balbinot:1998yh,Balbinot:2000at}
that \( T_{--} \) for non-minimal coupling at the horizon behaves as \( (r-r_{h})^{2}\ln (r-r_{h}) \)
instead of (\ref{ehmTh}) for the minimal coupling. This means that in Kruskal
coordinates the energy momentum tensor exhibits a singularity, although a rather
weak (logarithmic, integrable) one \cite{Kummer:1999zy}. In ref.
\cite{Balbinot:2000iy} this singularity
has been attributed to a breakdown of the WKB approximation.

In the case of non-minimally coupled scalars derived from SRG, action and path 
integral measure (the mode normalization condition) coincide with the ones for
the \( s \)-wave
parts of the corresponding quantities in four dimensions. Does this guarantee
that the $2D$ Hawking flux will be just the \( s \)-wave part of
the Hawking flux in four dimensions? The answer is negative, because
renormalization and dimensional reduction do not commute.
Indeed, even if each individual angular momentum contribution to the energy-momentum
tensor or to the effective action were finite, the sum over the angular momenta
will, in general, diverge. In fact, the effective action \( W \) can be written
in the zeta function regularization as \begin{equation}
\label{ehnWreg}
W^{\textrm{reg}}=-\frac{1}{2}\Gamma (s)\sum _{l}\sum _{n_{l}}\lambda _{l,n_{l}}^{-s}=\sum _{l}W^{\textrm{reg}}_{l}\, ,
\end{equation}
 where the {}``partial wave{}'' effective action is \begin{equation}
\label{ehWpren}
W^{\textrm{reg}}_{l}=-\frac{1}{2}\Gamma (s)\sum _{n_{l}}\lambda _{l,n_{l}}^{-s}\, .
\end{equation}
 Here \( \lambda _{l,n_{l}} \) are eigenvalues of the kinetic operator in four
dimensions corresponding to the angular momentum \( l \). To remove divergences
in \( W^{\textrm{reg}}_{l} \) as \( s\rightarrow 0 \) one must subtract the
pole terms: \begin{equation}
\label{ehWlsub}
\delta W_{l}=\frac{1}{2s}\zeta _{l}(0)+\dots \, ,\qquad \zeta _{l}(s)=\sum _{n_{l}}\lambda _{l,n_{l}}^{-s}\, .
\end{equation}
 Here dots denote finite renormalization terms. After that one obtains the familiar
expression (\ref{ehmW}) for each \( l \) with an appropriate operator \( A \).
However, the sum \begin{equation}
\label{ehWsum}
W=\sum _{l}\left( W^{\textrm{reg}}_{l}+\delta W_{l}\right) 
\end{equation}
 will diverge. Thus, a subtraction term which is needed to make (\ref{ehnWreg})
finite has nothing to do with the sum over \( l \) of the individual pole terms
(\ref{ehWlsub}). This latter sum simply does not exist! This means that the
four-dimensional theory requires more counterterms and counter terms of a 
different type than
the spherically reduced one. This problem was noted long ago \cite{Wipf:1986mr}
in calculations of tunnel determinants. In the context of SRG the 
non-commutativity
of renormalization and dimensional reduction has been called {}``dimensional
reduction anomaly{}'' \cite{Frolov:1999an}; it has been the subject of 
extensive studies over recent
years \cite{Sutton:2000gm,Balbinot:2000iy,Cognola:2000xp}. 

We conclude this section by noting that for massive matter fields
the situation is simpler than for massless ones. One can apply e.g.
the high frequency approximation \cite{Balbinot:2002bz} to estimate
the energy-momentum tensor. The zero mass limit in such calculations is, of course,
singular. The massive case is also less interesting because the
Hawking flux is suppressed by the mass.


\clearpage

\section{Nonperturbative path integral quantization}\label{se:3.2}

As pointed out in the previous section, dilaton gravity is a convenient
laboratory for studying semi-classical effects like Hawking radiation. For
quantum effects the simplicity of $2D$ theories becomes even more important,
since quantum gravity is beset with well-known conceptual problems,
which are essentially independent of the considered dimension (cf. e.g.
\cite{Carlip:2001wq}).
Thus, a study in a framework were the purely technical challenges are not as 
demanding as in higher-dimensional theories is desirable. 

GDT in $2D$ with matter as a theoretical laboratory for quantum gravity has 
several advantages as compared to other models:
\blist
\item As outlined in the introduction, it encompasses many different theories 
proposed in the literature, including models with strong physical motivation,
like SRG.
\item It exhibits continuous physical degrees of freedom and thus provides
physical scattering processes with a non-trivial $S$-matrix, as opposed to
pure SRG or GDT without matter.
\item It is still simple enough to allow a non-perturbative treatment in the 
geometric sector. 
\elist
The main advantage of the first point is that the same techniques can be
used uniformly for a large class of theories.
The second point will be elaborated in detail in the next section, where the
$S$-matrix for $s$-wave gravitational scattering will be calculated.
The third point is conceptually and technically very important: the split of 
geometric variables into background plus fluctuations in perturbation theory 
is something which can be avoided here. From the viewpoint of GR this is 
very attractive.

After integrating out geometry exactly a non-local and non-polynomial 
action is obtained, depending solely on the matter fields and external 
sources. When perturbation theory is introduced at this point geometry can be 
reconstructed self-consistently to each given order. In particular, the proper 
back-reaction from matter is included automatically.

From a technical point of view the use of Cartan variables in a first order
formulation has been crucial. The ensuing constraint algebra also with matter shares the essential features with the one in the PSM model (cf. sect. \ref{se:2.2.3}) which governs the matterless case: It becomes a finite W-algebra for minimally coupled matter and a Lie-algebra for the
JT model \cite{Barbashov:1979bm,D'Hoker:1982er,Teitelboim:1983ux,
D'Hoker:1983is,D'Hoker:1983ef,Jackiw:1985je}. Moreover it still closes with 
$\de$-functions rather than derivatives of them. Within the BRST
quantization procedure the ``temporal'' gauge has turned out to be 
extremely useful. It will lead to an effective metric in Sachs-Bondi form. 
As seen above (cf. sect. \ref{se:2.3.1}) that gauge appeared to be already the most 
natural one in the absence of matter interactions. Even when the latter are
present the classical action in that gauge remains linear in the canonical
coordinates of the geometrical sector. Consequently, by integration
three functional delta functions are generated which are used to perform 
an exact path integration over the corresponding
canonical momenta. If no matter fields are present in this way 
an exact generating functional for Green functions is obtained. The 
effective field theory with non-local interactions for the case with matter 
fields will be considered in detail in sect.\,\ref{sec:VBH}.

\subsection{Constraint algebra}

A prerequisite for the proper formulation of a path integral is 
the Hamiltonian analysis. The key advantage of the formulation 
(\ref{2.62}) for the geometric part of the action is its 
``Hamiltonian'' form. The component version (\ref{2.66}), together with 
the one for scalar fields, nonminimally coupled by $F(X) \neq const$. to the 
dilaton field (cf. (\ref{eq:a70})), 
\begin{equation}
{\mathcal L}^{(m)}= (e) F(X) \left[ \frac{1}{2}\eta_{ab}
\left(\tilde{\epsilon}^{\mu\nu}e_\nu^a \partial_\mu\phi\right)
\left(\tilde{\epsilon}^{\ka\la}e_\ka^b\partial_\la\phi\right) - f(\phi) \right]
\label{eq:4.1}
\end{equation}
in terms of the canonical coordinates $\phi$ and 
\begin{equation}
q_i=(\om_1,e_1^-,e_1^+), \hspace{0.5cm}\bar{q}_i=(\om_0,e_0^-,e_0^+)
\label{eq:4.2}
\end{equation}
allow the identification of the respective canonical 
momenta\footnote{Strictly speaking the relation between $p_i$ 
and the variables $X,X^\pm$ yields primary second class constraints. 
However, the canonical procedure using Dirac brackets in the 
present case justifies the shortcut implied by (\ref{eq:4.3}). We use the 
standard nomenclature of Hamiltonian analysis 
(cf. e.g. \cite{Gitman:1990,Henneaux:1992,Dirac:1996}). Note that $L^{\rm FOG}=-\int d^2x{\mathcal L}^{(g)}$, the minus sign being a consequence of our notation which relates the first order action (2.17) to minus the second order action (1.1).}
from the total Lagrangian $\mathcal{L}=\mathcal{L}^{(g)}+\mathcal{L}^{(m)}$ 
($L = \int d^2 x \mathcal{L}, \partial_0 q_i = \dot{q}_i\;$ etc.)
\begin{align}
p_i &= \frac{\partial \mathcal{L}}{\partial \dot{q_i}} = (X,X^+,X^-)\,, 
\label{eq:4.3} \\
\pi &= \frac{\partial \mathcal{L}}{\partial \dot{\phi}}\,, \label{eq:4.4} \\
\bar{p}_i &= \frac{\partial \mathcal{L}}{\partial \dot{\bar{q}_i}} = 0 \,.
\label{eq:4.5}
\end{align}
Eqs.\ (\ref{eq:4.5}) are three primary constraints. The canonical Hamiltonian density 
\begin{equation}
{\mathcal{H}}_c = p_i\dot{q}_i+\pi\dot{\phi}-{\mathcal{L}},
\label{eq:4.6}
\end{equation}
after elimination of $\dot{\phi}$ and $\dot{q}_i$ becomes
\begin{equation}
{\mathcal{H}}_c = -\bar{q}_iG_i,
\label{eq:4.7}
\end{equation}
with the secondary first class constraints
\begin{equation}
G_i(q,p,\phi,\pi) := G_i^{(g)}(q,p)+G_i^{(m)}(q,p,\phi,\pi), 
\label{eq:4.8}
\end{equation}
the geometric part of which is given by ($\mathcal{V} = \mathcal{V} 
(p_2 p_3, p_1)$ as defined in (\ref{2.77}))
\begin{align}
G_1^{(g)} &= \partial_1 p_1 + p_3q_3 - p_2q_2, \label{eq:4.9} \\
G_2^{(g)} &= \partial_1 p_2 + q_1p_2 - q_3 \mathcal{V}, \label{eq:4.10} \\
G_3^{(g)} &= \partial_1 p_3 - q_1p_3 + q_2 \mathcal{V}, \label{eq:4.11} 
\end{align}
and its matter part reads
\begin{align}
G_1^{(m)} &= 0 \label{eq:4.12}, \\
G_2^{(m)} &= \frac{F(p_1)}{4q_2} \left[ \left( \partial_1 \phi \right) - 
\frac{\pi}{F(p_1)} \right]^2 - F(p_1)q_3f(\phi), \label{eq:4.13} \\
G_3^{(m)} &= - \frac{F(p_1)}{4q_3} \left[ \left( \partial_1 \phi \right) + 
\frac{\pi}{F(p_1)} \right]^2 + F(p_1)q_2f(\phi). \label{eq:4.14} 
\end{align}

By means of the Poisson bracket (${p_i\,}' = p_i (x')$ etc.)
\eq{
\left\{q_i, p_j'\right\} = \de_{ij} \de(x^1-{x^1}')
}{eq:4.15}
the stability of the first class primary constraints (\ref{eq:4.5}) identifies 
the $G_i$ of (\ref{eq:4.8}) as first class secondary 
constraints. There are no ternary 
constraints as can be seen from the Poisson algebra of the $G_i$
\eq{
\left\{G_i, G_j' \right\} = C_{ijk} G_k \de(x-x')\,,
}{eq:4.16}
with ($C_{ijk}=-C_{jik}$; all non-listed $C_{ijk}$-components vanish)
\newline \parbox{3cm}{\begin{eqnarray*}
&& C_{122} = -1, \\
&& C_{133} = 1,
\end{eqnarray*}} \hfill
\parbox{7cm}{\begin{eqnarray*}
&& C_{231} = -\frac{\partial{\mathcal{V}}}{\partial p_1}+\frac{F'(p_1)}
{(e)F(p_1)}{\mathcal{L}}^{(m)}, \\
&& C_{232} = -\frac{\partial {\mathcal{V}}}{\partial p_2}, \\
&& C_{233} = -\frac{\partial {\mathcal{V}}}{\partial p_3}.
\end{eqnarray*}} \hfill
\parbox{1cm}{\begin{equation} \label{eq:4.18} \end{equation}} \hfill
\newline
In the matterless case and for minimal coupling ($F' = 0$ in (\ref{eq:4.18})) 
the structure functions $C_{ijk}$ depend on the momenta only. For the 
JT model (\ref{eq:JT}) with 
$\mathcal{V} = \Lambda p_1$ again the Lie-algebra of $SO(1,2)$ is reproduced 
(cf.\ the observation after eq. (\ref{2.82})). Already without matter the 
symmetry generated by the $G_i=G_i^{(g)}$ corresponds to a nonlinear 
(finite $W$-) algebra $\mathcal{A}^{(g)}$. Including also the 
(mutually commuting) momenta $p_i$ in $\bar{\mathcal{A}}^{(g)} = 
\{ \mathcal{A}^{(g)}, p_i\}$ that algebra closes and the Casimir invariant 
$\mathcal{C}^{(g)}$ of the PSM appears as one of the two elements of 
the center \cite{Grosse:1992vc}, the second of which can be expressed as 
$\partial_1\mathcal{C}^{(g)}$.

It is remarkable that the commutators (\ref{eq:4.16}) resemble, though, 
the ones of an ordinary gauge theory or the Ashtekar approach to gravity
\cite{Ashtekar:1986yd,Ashtekar:1987gu}  in the sense that no 
space-derivatives of the delta functions appear{\footnote{It is also 
possible to switch to other sets of constraints, e.g.\ 
including $\partial_1 \mathcal{C}^{(g)}$ as one of them, thus abelianizing
them in the matterless case. This, however, works only in a given patch, since
the transformation involved breaks down at a horizon (cf. e.g. 
\cite{Grumiller:2001ea}).}. The usual Hamiltonian constraints $H$ and the diffeomorphism constraints $H_1$ in an analysis of the ADM-type \cite{Arnowitt:1962} always lead to such derivatives (cf. e.g. \cite{Thiemann:2001yy}). 
Indeed, $H$ and $H_1$ can be reproduced by suitable linear combination of the 
$G_i$ \cite{Katanaev:1994qf,Kummer:1994ur}.

We emphasize that upon quantization we have 
no ordering problems in the present 
formalism\footnote{We are grateful to P. van Nieuwenhuizen for discussions on 
that point.}. Due to the linear appearance of coordinates in the geometric 
parts $G_i^{(g)}$ of the constraints any hermitian version of them 
is automatically Weyl ordered\footnote{To be more explicit: classical terms of 
the form $pq$ have a unique hermitian representation, namely $(qp+pq)/2$. This
is also their Weyl ordered version.}. Moreover, the Hamiltonian is hermitian 
if the constraints exhibit that property (since the Hamiltonian essentially is 
just a sum over them).  This property carries over to $G_i$ since $C_{ijk}$ 
for minimal coupling depend only on the momenta and for 
non-minimal coupling the only addition consists of the matter Lagrangian.
The commutator between structure functions and constraints vanishes: for 
minimal coupling this is a trivial consequence of the PSM structure of the 
Hamiltonian and for 
non-minimal coupling the only non-trivial term (present in $C_{231}$) vanishes 
as well since it commutes with $G_1$. Moreover, the commutator of two 
(Weyl ordered) constraints again yields the classical expressions 
(\ref{eq:4.18}) in Weyl ordered form. Therefore, the Poisson algebra 
(\ref{eq:4.16}) can be elevated without problems to a commutator 
algebra for quantum operators.

As no ternary constraints exist and as all constraints are first 
class the extended phase space with (anticommuting) ghost fields 
can be constructed easily, following the approach of Batalin, Vilkovisky and 
Fradkin \cite{Fradkin:1975cq,Batalin:1977pb,Fradkin:1978xi}. 
One first determines the BRST 
charge $\Omega$ which fulfills  $\Omega^2 = \frac{1}{2} \{ \Omega, \Omega \}= 
0$. Treating $\bar{q}_i$ as canonical variable one obtains
two quadruplets of con\-straint/ca\-non\-ic\-al coordinate/ghost/ghost 
momentum
\begin{equation}
(\bar{p}_i,\bar{q}_i,b_i,p_i^b),\hspace{0.5cm}(G_i,-,c_i,p_i^c)\,,
\label{eq:4.19}
\end{equation}
with canonical (graded) brackets
\eq{
\left\{c_i,{p_i^c}'\right\} = - \de_{ij} \de(x^1-{x^1}') =
\left\{b_i,{p_i^b}'\right\}\,.
}{eq:4.19.5}
In (\ref{eq:4.19}) no ``coordinate'' conjugate to the secondary 
constraints $G_i$ appears, although one could try to construct 
some quantities which fulfill canonical Poisson bracket relations 
with them. These quantities are not needed for the BRST procedure 
which according to \cite{Grumiller:2001ea} yields
\begin{equation}
\label{eq:4.20}
\Omega = \int \left(c_i G_i + \frac{1}{2}c_i c_j C_{ijk} p^c_k + b_i \bar{p}_i
\right)d^2x\,.
\end{equation}
 Since the structure functions (\ref{eq:4.18}) are field dependent it is 
 non-trivial that the homological perturbation series stops at 
 $\mbox{rank} = 1$. In general one would expect the presence of higher 
 order ghost terms (``ghost self interactions''). However, it can be verified
 easily that $\Omega$ as defined in (\ref{eq:4.20}) is nilpotent by itself. 
For the matterless case this is a simple consequence of the 
 Poisson structure: The Jacobi identity (\ref{2.82}) for the Poisson 
 tensor implies that the homological perturbation series already 
 stops at the Yang-Mills level \cite{Schaller:1994es}. It turns out that 
the inclusion of (dynamical) scalars does not change this feature. 
 
 The quantity (\ref{eq:4.20}) generates BRST transformations with 
 anticommuting constant parameter $\delta \lambda$ by 
 $\delta \mathcal{H}_c = \{ \delta\la \Omega, \mathcal{H}_c\} = 0$. It 
 not only leaves the canonical Hamiltonian density
$\mathcal{H}_c$ invariant, but also the extended 
 Hamiltonian density
\begin{equation}
\mathcal{H}_{ext} = \mathcal{H}_c + \{ \psi, \Omega\}
\label{eq:4.23}
\end{equation}
in which $\mathcal{H}_c$ has been supplemented by a (BRST exact) term with the 
gauge-fixing fermion $\psi$. In our case also 
$\mathcal{H}_c=\left\{p^c_i\bar{q}_i,\Omega\right\}$ is exact, a well-known 
feature of reparametrization invariant theories. A useful class of gauge 
fixing fermions is given by \cite{Haider:1994cw}
\begin{equation}
\psi = p_i^b \chi_i
\label{eq:4.24}
\end{equation}
where $\chi_i$ are some gauge fixing functions. The class of 
temporal gauges (\ref{eq:a2}) 
\begin{equation}
\bar{q}_i = a_i,\quad a_i = (0,1,0)
\label{eq:4.25}
\end{equation}
has turned out to be very convenient for the exact path integration 
of the geometric part of the action \cite{Kummer:1992rt,Haider:1994cw,
Kummer:1997hy,Kummer:1998zs}. It can be incorporated in (\ref{eq:4.24}) by the 
choice
\begin{equation}
\chi_i = \frac{1}{\epsilon} (\bar{q}_i - a_i)
\label{eq:4.26}
\end{equation}
with $\epsilon$ being a positive constant. Then (\ref{eq:4.23}) 
reduces to
\begin{equation}
\mathcal{H}_{ext} = \frac{1}{\epsilon} p_i^b b_i - 
\frac{1}{\epsilon} ( \bar{q}_i - a_i) \bar{p}_i - \bar{q}_i G_i - \bar{q}_ic_j 
\mathcal{C}_{ijk} p_k^c + p_i^c b_i \; .
\label{eq:4.27}
\end{equation}
It is necessary to perform the limit $\epsilon \to 0$ to impose (\ref{eq:4.25})
in the path integral. This can be achieved by a redefinition of the canonical 
momenta
\begin{equation}
\bar{p}_i \to \hat{\bar{p}}_i=\epsilon \bar{p}_i, \quad p_i^b \to \hat{p}_i^b = 
\epsilon p_i^b\, ,
\label{eq:4.28}
\end{equation}
which has unit super-Jacobian in the path integral measure. Taking 
$\epsilon \to 0$ afterwards, in terms
of the new momenta yields a well-defined extended Hamiltonian.

\subsection{Path integral quantization}

After that step the path integral in extended 
phase space becomes
\begin{align}
W &= \int\, (\mathcal{D} q_i )\, (\mathcal{D} p_i )\,  
(\mathcal{D} \bar{q}_i ) \, (\mathcal{D} \hat{\bar{p}}_i )\,  
(\mathcal{D} \phi )\,  
(\mathcal{D} \pi )\,  (\mathcal{D} c_i ) \, (\mathcal{D} p_i^c)\,  
(\mathcal{D} b_i )\, (\mathcal{D} \hat{p}_i^b)\nonumber \\
& \quad \times \exp\; \left[ 
i \int (\mathcal{L}^{(0)} + J_i p_i  + j_iq_i + \sigma\phi )\,
d^2x \, \right]\, ,
\label{eq:4.29}
\end{align}
with
\begin{equation}
\mathcal{L}^{(0)} = p_i \dot{q}_i + \epsilon \hat{\bar{p}}_i 
\dot{\bar{q}\,}_i + \pi\dot{\phi} + \epsilon \hat{p}_i^b \dot{b}_i + 
p_i^c \dot{c}_i - \mathcal{H}_{ext}\,.
\label{eq:4.30}
\end{equation}
It turns out to be very useful to introduce sources $j_i$ and $\sigma$ not 
only for the geometric variables $q_i$ and for the scalar field $\phi$, but 
also for the momenta $p_i$, denoted by $J_i$. 
Integrating out $\hat{\bar{p}}_i$ and $\bar{q}_i$ yields an 
effective Lagrangian with $\bar{q}_i = a_i$ as required by 
(\ref{eq:4.25}). After further trivial integrations with respect to $b_i$ 
and $\hat{p}_i^b$, and finally $c_i$ and $p_i^c$, (\ref{eq:4.29}) simplifies 
to 
\begin{equation}
W = \int\, (\mathcal{D} q_i )\, (\mathcal{D} p_i )\,  
(\mathcal{D} \phi )\,  
(\mathcal{D} \pi)\, \det M \, \exp i L^{(1)}
\label{eq:4.31}
\end{equation}
with
\begin{equation}
L^{(1)} = \int\, d^2 x \; ( p_i\dot{q}_i + \pi\dot{\phi} + 
a_i G_i + J_i p_i + j_i q_i + \sigma\phi )\; .
\label{eq:4.32}
\end{equation}
In the functional matrix
\begin{equation}
M = \left(
\begin{array}{ccc}
\partial_0 & -1 & 0 \\
0 & \partial_0 & 0 \\
K &\quad p_3 U(p_1) &\quad \partial_0 + p_2 U (p_1)
\end{array}
\right) 
\label{eq:4.33}
\end{equation}
the (complicated) contribution $K$ is irrelevant for its determinant: 
\begin{equation}
\det M = \left(\det\partial_0\right)^2\det\, (\partial_0 + p_2 U (p_1))
\label{eq:4.34}
\end{equation}
Indeed, apart from that important contribution to the measure the 
generating functional of Green functions $W$ with the effective 
Lagrangian eq. (\ref{eq:4.32}) is nothing but the ``naive'' result, obtained 
by gauge fixing the Hamiltonian $\mathcal{H}_c$. 

The present approach to quantization may be questioned
because it is not based directly upon the classical physical theory like the 
dilaton action appearing naturally in SRG. However, the classical 
equivalence argument of sect. \ref{se:2.2.2} in the quantum language 
simply means to integrate out\footnote{In the path integral the classical 
elimination procedure is replaced by first integrating the two components of
$\om_\mu$ which appear linearly. The resulting $\de$-functions allow the 
elimination of $X^a$ by means of the relation (\ref{2.75}).
}\label{fn:60} the torsion-independent part of the 
spin connection and of the $X^a$ in (\ref{2.62}). The only delicate 
point is the transformation in the measure of the path integral. 
As shown in ref. \cite{Kummer:1997hy} there exists a gauge ($e_0^- = 
1,\; e_1^+ = 1,\; e_0^+ = 0$) which does not produce a 
Faddeev-Popov-type determinant in the transition to the equivalent dilaton theory (of the form (\ref{2.59})). Thus for all physical 
observables, defined as to be independent of the gauge, that equivalence should also hold at the quantum level. 

The (Gaussian) path integral\footnote{An effective action for a general class
of gauges where this integration is not possible has been proposed too 
\cite{Katanaev:2000kc}.} of momenta $\pi$ in (\ref{eq:4.31}) is 
done in the next step:
\begin{equation}
W = \int\, (\mathcal{D} q_i) (\mathcal{D}p_i) (\mathcal{D}\phi)
(\det q_2)^{1/2} \, \det M \, \exp\, i L^{(2)}
\label{eq:4.35}
\end{equation}
\meq{
L^{(2)} = \int d^2x \Big[ p_i\dot{q}_i+q_1p_2-q_3(V(p_1)+U(p_1)p_2p_3) \\
+F(p_1)\left((\partial_1\phi)(\partial_0\phi)-q_2(\partial_0\phi)^2-q_3f(\phi)
\right)+j_iq_i+J_ip_i+\sigma\phi\Big]
}{eq:4.36}
In the only contributing constraint $G_2$, which is now written explicitly, the 
total derivative $\partial_1 p_2$ has been dropped\footnote{There exists a 
shortcut to obtain (\ref{eq:4.35}) with (\ref{eq:4.36}) and (\ref{eq:4.34})
\cite{Grumiller:2001ea}: instead of (\ref{eq:4.24}) with (\ref{eq:4.26}) one 
can use the gauge fixing fermion $\Psi=p_2^c$ and straightforwardly integrate 
all ghosts and their momenta, without limiting procedure for a quantity $\eps$ as in (\ref{eq:4.26}).}. 

As it stands (\ref{eq:4.35}) lacks a covariant measure for the final 
matter integration. If this is not corrected properly, 
counterterms emerging from a noncovariant measure may obscure the 
quantization procedure. The accepted remedy \cite{Toms:1987sh} 
(cf. also \cite{Fujikawa:1988ie,Bastianelli:1998jb}) is to 
insert the appropriate factor by hand. It arises from the requirement that the 
inner product \[
\int d^2x\sqrt{-g}\phi\psi=\int d^2x \left(\phi\sqrt[4]{-g}\right)\left(\psi\sqrt[4]{-g}\right)
\] be invariant under diffeomorphisms. As a consequence, the Gaussian integral with the invariant measure yields
\begin{equation}
\int (\mathcal{D}\phi  \sqrt[4]{-g}) \exp{\left[i\int\sqrt{-g}\phi\Gamma\phi
\right]} = \left(\det{\Gamma}\right)^{-1/2}\,.
\label{eq:4.37}
\end{equation}
In 
the present gauge (\ref{eq:a2}) with $\sqrt{-g} = (e) = e_1^+ = q_3$ this 
means that $(\det q_2)^{1/2}$ in (\ref{eq:4.35}) should be replaced by 
$(\det q_3)^{1/2}$. 

In the customary approach the next step would be the integral of the momenta $p_i$. However, in a generic $2D$ gravity model, including the physically 
relevant SRG, the $p$-integrals are not Gaussian. On the other hand, the action (\ref{eq:4.36}) is \textit{linear in the geometric coordinates $q_i$}. Even the new determinant in the measure by the identity ($u$ and $\bar{u}$ are anticommuting scalars and $v$ a commuting one) 
\begin{equation}
\left(q_3\right)^{1/2} = \int \left(\mathcal{D}v\right)\left(\mathcal{D}\bar{u}\right)\left(\mathcal{D}u\right)
\exp{i(v^2+\bar{u}u)q_3} \label{eq:4.38}
\end{equation}
may be reexpressed formally as yet another linear contribution (in $q_3$) 
to the action. This suggests to perform the $q_i$-integrals 
\textit{first}, yielding (functional) $\delta$-functions which, 
in turn, may be used to get rid of the $p_i$-integrals 
afterwards\footnote{Historically this exact integrability was realized 
first for the matterless KV-model \cite{Haider:1994cw} where 
the quadratic dependence on $p_i$ allowed the first integration 
to be the (traditional) Gaussian one. The inverted sequence of integrals has 
been initiated in \cite{Kummer:1997hy}.}.

The vanishing arguments of the three $\delta$-functions for the respective 
$(q_1, q_2, q_3)$-integrals yield three differential equations ($h = \bar{u}u +v^2$ from (\ref{eq:4.38}))
\begin{gather}
\partial_0 p_1 - p_2 - j_1 = 0\, , \label{eq:4.39}\\
\partial_0 p_2 - j_2 + F(p_1) (\partial_0 \phi)^2 = 0\, , \label{eq:4.40}\\
(\partial_0 + p_2U(p_1))p_3 + V(p_1) + F(p_1)f(\phi) - h - j_3 = 0 \; . 
\label{eq:4.41}
\end{gather}
The differential operators acting on $p_i$ precisely combine to 
the ones in the determinant $\det M$ of the measure in 
(\ref{eq:4.31}) with (\ref{eq:4.34}). Therefore, $\det M$ will be cancelled 
exactly in the subsequent integration of $p_i$. Eqs.\ 
(\ref{eq:4.39})-(\ref{eq:4.41}) are the classical (Hamiltonian) 
differential equations for the momenta (with sources $j_i$). Matter is 
represented in (\ref{eq:4.40}), (\ref{eq:4.41}) by the terms proportional to 
Newton's constant ($F \propto \kappa$).  

\subsection{Path integral without matter}

It is not possible to obtain an exact solution for $p_i$ for general matter interactions. Therefore, the latter can be treated only perturbatively, and in the first step $F\to 0$ should be considered\footnote{For minimally 
coupled scalars $F= \rm const.$ is independent of $p_i$ so that this term 
can be taken along one more step. In the end, however, one cannot 
avoid that perturbation expansion.}.
Then the solutions of (\ref{eq:4.31})-(\ref{eq:4.39}) can be written as
\begin{align}
p_1 &= B_1 = \bar{p}_1 + \partial_0^{-1}(p_2+j_1)\,, \label{eq:4.42}\\
p_2 &= B_2 = \bar{p}_2 + \partial_0^{-1} j_2 \,, \label{eq:4.43}\\
p_3 &= B_3 = e^{-Q}\left[ \partial_0^{-1} e^Q 
\left( j_3 - V(p_1)\right) + \bar{p}_3\right]
 \label{eq:4.44}
\end{align}
where $\partial_0 \bar{p}_i = 0$ and $\partial_0^{-1}, \partial_0^{-2}$ have 
to be properly defined one-dimensional Green-functions in the genuine realm of 
quantum theory (see below). In the integration of (\ref{eq:4.41}) the 
differential operator $H=(\partial_0+p_2U(p_1))$ has been reexpressed in terms 
of
\begin{equation}
Q = \partial_0^{-1} \, \left( U(B_1)\, B_2 \right)
\label{eq:4.45}
\end{equation}
as $H^{-1} = e^{-Q}\, \partial_0^{-1}\, e^Q$.
Proceeding as announced above, after $\int (\mathcal{D} q)\, 
(\mathcal{D} p)$ we arrive at the exact expression for
the generating functional for Green functions
\beqa
&& W_0 (j,J) = \exp \, i L_{\rm eff}^{(0)}\,,
\label{eq:4.47} \\
&& L^{(0)}_{\rm eff} = \int d^2x \left[ J_i B_i + \tilde{\mathcal{L}}_0 
(j,B)\, \right]\,, \label{eq:4.48}
\eeqa
where $B_i=B_i(j, \bar{p}_i)$ . Here $L^{(0)}_{\rm eff}$ trivially coincides 
with the generating functional of {\em connected} Green's function. 
In (\ref{eq:4.48}) a new contribution 
$\tilde{\mathcal{L}}_0 (j,B)$ has been added. It originates from 
an ambiguity in the first term of the square bracket. In an 
expression $\int dx^0 \, \int dy^0 \, J_{x^0} (\partial_0^{-1}) A$ the 
symbol $\partial_0^{-1}$ means an integral which when 
acting upon $J$ contains an undetermined integration constant 
$\bar{g} (x^1)$. This generates a new term $\bar{g}\, \int A$. 
Applying this to $J_1 B_1 + J_2 B_2$ yields uninteresting 
couplings\footnote{It turns out that the corresponding ambiguous contributions
are fixed uniquely by imposing boundary conditions on the momenta $p_1$ and
$p_2$.}\label{fn:53}. But from $B_3$ with $A = e^Q (j_3 - V)$ together with 
$J_3$ an important contribution to the action follows:
\begin{equation}
\tilde{\mathcal{L}}_{(0)} = \bar{g}\, e^Q \, (j_3 - V)\,
\label{eq:4.49}
\end{equation}
Indeed that term is the only one to survive in the matterless 
case at $J_i = 0$, i.e.\ for vanishing sources of the momenta. On 
the other hand, precisely that action $\tilde{\mathcal{L}}_{(0)}$ had 
been derived in the first exact path integral \cite{Haider:1994cw}
computed for the KV-model \cite{Katanaev:1986wk,Katanaev:1990qm}. There 
$J_i\equiv 0$ had been 
taken from the beginning. It also plays a crucial role for the 
derivation of the solutions for the (classical) e.o.m.-s for the 
geometric variables which simply follow from the ``expectation 
values'' in the matterless case
\begin{equation}
\langle q_i \rangle = \left. \frac{1}{i\, W_0 (0)}\; 
\frac{\delta\, W_0}{\delta j_i} \; \right\vert_{j = J = 0}\; .
\label{eq:4.50}
\end{equation}
These $\langle q_i \rangle$ indeed coincide with the classical solutions 
(\ref{eq:a17})-(\ref{eq:a19}) when constants of integration are adjusted and 
the gauge (\ref{eq:a2}) is assumed.
The usefulness of the sources $J_i$ for the momenta is evident 
when the Casimir function $\mathcal{C}^{(g)}$ of (\ref{eq:a11}) is determined 
from its expectation value:
\begin{equation}
\left\langle\; \mathcal{C}^{(g)} (p)\; \right\rangle = 
\left. \frac{\mathcal{C}^{(g)} \; 
\left( \frac{1}{i}\, \frac{\delta}{\delta J} \right) \; 
W_0 (J)}{W_0 (0)} \; \right\vert_{J=0} = \; \bar{p}_3
\label{eq:4.51}
\end{equation}
The last equality in (\ref{eq:4.51}) follows from introducing the solutions 
(\ref{eq:4.42})-(\ref{eq:4.44}) into $\left\langle\; \mathcal{C}^{(g)} (p)\; 
\right\rangle=\mathcal{C}^{(g)}(B_i^0)$, where  in $B^0_i=B_i(j=0)$ the 
residual gauge is fixed so that $B^0_2 = \bar{p}_2 = 1, B_1^0 
=x^0$. Here, as well as in the solutions $q_i$ of (\ref{eq:4.50}) it is 
evident that $\bar{p}_3$ describes the (classical) 
background. 

It should be emphasized that one encounters the unusual situation 
that in the matterless case the classical theory is expressed by $W_0$ in 
a quantum field theoretical formalism. Therefore, the question arises for the whereabouts of the e.o.m-s which are the counterparts of 
(\ref{eq:4.39})-(\ref{eq:4.41}), but which depend on derivatives 
$\partial_1 p_i$ instead of $\partial_0 p_i$. These relations disappeared 
because of the gauge fixing, much like the Gauss law disappears in the 
temporal gauge $A_0 =0$ for the $U(1)$ gauge field $A_\mu$. Actually, in 
the ``quantum'' formalism  of the classical result they reappear 
as ``Ward-identities'' by gauge variations (diffeomorphisms, 
local Lorentz transformations) of $W_0 (j,J)$. For details refs.
\cite{Kummer:1998zs,Grumiller:2001ea} can be consulted.

There is one most important lesson to be drawn retrospectively 
for the matterless case which may have consequences in a more 
general context than the present one of $2D$ theories of gravity: 
The exact quantum integral of the geometry leading here to the 
classical theory uses, among others, a path integral over 
\textit{all} values of $q_3$ in order to arrive at the classical
equation for the momenta through the $\delta$-function. However, 
$q_3$ in the gauge (\ref{eq:a2}) is 
identical to the determinant $(e)=\sqrt{-g}$. Therefore, a 
summation including negative and vanishing volumes has to be 
made to arrive at the correct (classical) result. 

It is instructive to derive the effective action corresponding to the
generating functional (\ref{eq:4.48}). In terms of the mean fields
$\langle q_i \rangle$, $\langle p_i \rangle$ 
\begin{equation}
\langle q_i \rangle =\frac{\delta L^{(0)}_{\rm eff}}{\delta j_i}\,,
\qquad
\langle p_i \rangle =\frac{\delta L^{(0)}_{\rm eff}}{\delta J_i}\,,
\label{e4mean}
\end{equation}
the effective action $\Gamma (\langle q_i \rangle ,\langle p_i \rangle )$
results from the Legendre transform of $L^{(0)}_{\rm eff}$,
\begin{equation}
\Gamma (\langle q_i \rangle ,\langle p_i \rangle ) 
=L^{(0)}_{\rm eff}(j,J) -\int d^2x \left( j_i \langle q_i \rangle +
J_i \langle p_i \rangle \right) \,, \label{e4Lt}
\end{equation}
where the sources must be expressed through the mean fields.
To economize writing the brackets the notations
\begin{equation}
\langle q_i \rangle =(\omega_1,e_1^-,e_1^+)\,,\qquad
\langle p_i \rangle =(X,X^+,X^-) \,,\label{e4not}
\end{equation}
imply a simple return to the original geometric variables (cf. (\ref{eq:4.2})
and (\ref{eq:4.3})). 

A peculiar feature of the first order formalism is that only the last three
of the equations (\ref{e4mean}) are needed:
\begin{eqnarray}
&&j_1=\partial_0 X-X^+ \,,\label{e4j1}\\
&&j_2=\partial_0 X^+\,,\label{e4j2}\\
&&j_3-V(X)=e^{-Q} \partial_0 \left( e^QX^-\right)
=\left(\partial_0 +X^+ U(X)\right) X^- \,.\label{e4j3}
\end{eqnarray}
The exact effective action (\ref{e4Lt}) for the dilaton 
gravity models without matter immediately follows from (\ref{e4j1})-(\ref{e4j3}):
\begin{eqnarray}
&&\Gamma =\int_{\mathcal{M}} d^2x \left[ \omega_1 X^+ 
-\omega_1 (\partial_0 X) -e_1^-(\partial_0 X^+)
-e_1^+ (\partial_0 X^-) \right. \nonumber \\
&&\qquad \left. -e_1^+ (V(X)+X^+X^-U(X))\right]
\pm \bar{g} \int_{\partial\mathcal{M}} dx^1 e^QX^- \,,\label{e4ea}
\end{eqnarray}
It has been assumed that the manifold $\mathcal{M}$ has the form of a
strip $\mathcal{M}=[r_1,r_2]\times \mathbb{R}$. The upper ($+$) sign
in front of the surface term\footnote{The two omitted surface terms (cf. 
footnote \ref{fn:53}) are just $\int_{\partial\mathcal{M}} dx^1 X^+$ and
$\int_{\partial\mathcal{M}} dx^1 X$. Since both quantities will be fixed by
suitable boundary conditions in the next section we have already dropped 
them.} corresponds to the ``right'' boundary
$x^0=r_2$, and the lower ($-$) sign to $x^0=r_1$. 

The volume term in (\ref{e4ea}) is just the classical action
in the temporal gauge (cf. first line in (\ref{eq:4.36})).
Thus all dilaton gravities without matter are {\it locally 
quantum trivial} \cite{Kummer:1997hy}. Therefore, all eventual quantum effects 
are encoded in the boundary part and are {\it global}.

Except in subsection \ref{admsec} and in eq. (\ref{e4ea}) so far complications
from boundary effects were entirely ignored. Their inclusion in the path 
integral approach is a highly nontrivial problem and, in general, requires the 
introduction of non-local operators at the boundary (cf. e.g. 
\cite{Barvinsky:1987dg,Marachevsky:1996dr,Vassilevich:1995zk,Moss:1997ip,
Esposito:2001rx}). Matterless $2D$ quantum gravity being a special case of 
PSM-s, the discussion can be incorporated into the one of these more general 
models\footnote{We are grateful to L. Bergamin and P. van Nieuwenhuizen for 
discussions on that model.}, if the bulk action (\ref{2.80}) is 
supplemented by
\eq{
\int_{\partial M_2} f(\mathcal{C})X^IA_I\,,
}{eq:4.b1}
where $\mathcal{C}$ is the Casimir function (\ref{2.91}-\ref{2.93}). A 
consistent way\footnote{We mean consistency as defined in 
\cite{Lindstrom:2002}: boundary conditions arise from (1) extremizing the 
action, (2) invariance of the action and (3) 
closure of the set of boundary conditions under symmetry transformations.
In Maxwell theory, e.g., these consistency requirements single out 
electric or magnetic boundary conditions \cite{Vassilevich:1995cz}. 
} 
to implement boundary conditions is fixing $f(\mathcal{C})=0$ and $\de X^I=0$
at time-like $\partial M_2$, which implies $X^I|_{\partial M_2}=X^I(r)$ with 
$r$ being the ``radius''. In fact, this prescription we had imposed tacitly by 
dropping all boundary contributions in (\ref{eq:4.47}) and by choosing fixed 
functions of $x^0$ (which corresponds to a radial coordinate in our gauge) for 
the boundary values of $p_i$.

As pointed out in sect. \ref{se:2.2.3}, gravity theories in $D=2$ without 
matter are special examples of PSM-s. Not surprisingly, the exact path 
integral also is encountered there \cite{Hirshfeld:1999xm}. Recently also an 
``almost closed expression'' for the partition function on an arbitrary 
oriented two-manifold has been presented as well \cite{Hirshfeld:2001cm}.

 It is interesting to 
compare this result with local perturbative calculations 
\cite{Russo:1992yg,Kantowski:1992dv,Elizalde:1994qq,Elizalde:1995zj}.
Results obtained in different gauges must coincide
on-shell only. The effective ``quantum'' actions obtained in these papers
indeed vanish on-shell in full agreement with our non-perturbative
calculations. Hence the non-trivial off-shell counterterms appearing in refs.
\cite{Russo:1992yg,Kantowski:1992dv,Elizalde:1994qq,Elizalde:1995zj}
are pure artifacts of the gauges employed. 
In ref. \cite{Cadoni:2002hg} local quantum triviality of some dilaton models
has been verified with the conformal field theory technique.
Some authors 
\cite{Cavaglia:1999zu,Cavaglia:2001zz} also rely on rather complicated
field redefinitions which, however, as a rule produce Jacobian factors, making 
a comparison with the result \cite{Kummer:1997hy} very difficult. Finally, ref.
\cite{Lombardo:1998zj} should be mentioned where loop calculations in the
presence of the Polyakov term have been performed, although
only part of the degrees of freedom was quantized.

\subsection{Path integral with matter}

\subsubsection{General formalism}

When the geometry interacts with matter the computation must be resumed at the 
generating functional (\ref{eq:4.35}). After performing the integrations 
$(\mathcal{D}p)(\mathcal{D}q)$ as in the matterless case one arrives at
\begin{equation}
W = \int (\mathcal{D}\phi)\exp{iL^{(3)}},
\label{eq:4.52}
\end{equation}
\begin{equation}
L^{(3)}=\int d^2x\left[F(\hat{B}_1)(\partial_0\phi)(\partial_1\phi)+
J_i\hat{B}_i+\si\phi+\tilde{\mathcal{L}}(j_i,\hat{B}_i)\right],
\label{eq:4.53a}
\end{equation}
\eq{
\tilde{\mathcal{L}}(j_i,\hat{B}_i) = \tilde{g} e^{\hat{Q}} 
(h+j_3-V(\hat{B}_1)-F(\hat{B}_1)f(\phi)).
}{eq:4.53}
Here $\hat{B}_i$ are the solutions for $p_i$ from (\ref{eq:4.39})-(\ref{eq:4.41}) and thus are functions of the scalar field $\phi$ as well. The notation $\hat{Q}$ and $\hat{V}$ also indicates the dependence of these
quantities on $\hat{B}_i$ instead of $B_i$ (cf.\ (\ref{eq:4.45})). 
Nevertheless, the action still is seen to be only linear in 
$h = \bar{u}u + v^2$ in $\tilde{\mathcal{L}}$ (\ref{eq:4.53}) and 
$\hat{B}_3$ of (\ref{eq:4.41}). Therefore the identity (\ref{eq:4.38}) may be 
used backwards with $q_3$ replaced by
\begin{equation}
E_1^+ (J,j,\bar{p},f) = J_3\, e^{-\hat{Q}} \partial_0\, 
e^{\hat{Q}}\, \partial_0^{-1}\, e^{\hat{Q}} + 
\bar{g}\, e^{\hat{Q}}\, .
\label{eq:4.54}
\end{equation}
In the ensuing new version of (\ref{eq:4.52})
\begin{align}
W  &= \int (\mathcal{D}\, \phi)\, (\det E_1^+)^{1/2}\, \exp\, i 
L_{\rm eff}\,, \label{eq:4.55} \\
L_{\rm eff} &= \int d^2 x\, \left[ F(\hat{B}_1) (\partial_0\phi)
(\partial_1\phi)+J_i \hat{B}_i \, \vert_{h=0} + 
\tilde{\mathcal{L}}\, \vert_{h=0} + \sigma\phi\, \right]\,,
\label{eq:4.56}
\end{align}
the measure allows an intuitive interpretation. For physically interesting Green functions with $J_3=0$ the penultimate term in (\ref{eq:4.54}) is nothing but $q_3$ again, however expressed in terms of the sources $j$ \textit{and} containing the scalar field. Thus this determinant in the measure duely takes into account back reactions of scalar matter upon the geometry. 

This is how far we can get using non-perturbative methods. The
final matter integration cannot be performed exactly.

\subsubsection{Perturbation theory}
 
In the treatment of matter one now may follow the usual steps  of perturbative 
quantum field theory. In order to avoid cumbersome formulas and, nevertheless,
elucidate the basic principles we restrict the discussion to minimal coupling 
($F(p_1)=1$) and no local self-interaction ($f(\phi)=0$).

First the terms quadratic in $\phi$ of (\ref{eq:4.53a}) are 
isolated by expanding $\hat{B}_i$ and $\tilde{\mathcal{L}}$ 
to this order. Then, they are considered together with the source term 
$\sigma\phi$ in a (Gaussian) path integral. The $\phi$-dependence in the 
measure contributes to 
higher loop order only. Higher order terms $\mathcal{O}(\phi^{2n})$ 
$(n\geq 2)$ in (\ref{eq:4.56}) are interpreted as vertices and taken outside 
the integral, with 
the replacement $\phi\to \frac{1}{i}\,\frac{\delta}{\delta \sigma}$. We 
denote them summarily as $Z(\phi)$:
\begin{align}
W &= \exp{\left[i\, \int Z \left( \frac{1}{i}\,\frac{\delta}{\delta 
\sigma}\right)
\right]}\; \widetilde{W} \label{eq:4.57} \\
\widetilde{W} (j,J,\sigma) &= \int (\mathcal{D} \phi) \; \sqrt{\det E_1^+} \; 
\exp{i \int[(\partial_0 \phi)(\partial_1 \phi) - E_1^- (\partial_0 \phi)^2 + 
\sigma\phi ]}
\label{eq:4.58}
\end{align}
In the coefficient $E_1^-$ the different (nonlocal) contributions 
from the quadratic terms in $(\partial_0\phi)^2$ of (\ref{eq:4.57}) are 
lumped together. Comparing (\ref{eq:4.58}) with the path integral
\begin{equation}
\widetilde{W} = \int (\mathcal{D} \phi \sqrt[4]{-g})\; 
\exp{i \int d^2x \sqrt{-g}\, 
[\frac{1}{2} g^{\mu\nu} (\partial_\mu \phi)(\partial_\nu \phi) + \sigma\phi]}
\label{eq:4.59}
\end{equation}
explains our choice of the symbol $E_1^-$, because it is a generalization of 
the zweibein component $e_1 ^-$ which for the EF gauge would appear in this 
place. 
Here, by construction, $E_1^+$ and $E_1^-$ depend on the external sources and 
not on the scalar field. A more general form (with $F'(p_1)\neq 0$) of the 
``effective zweibein'' will be considered in sect.~\ref{sec:VBH}.

A Gaussian integral 
like (\ref{eq:4.59}) leads to the inverse square root of a functional 
determinant which in $D=2$ may be reexpressed as a Polyakov action\footnote{%
For nonminimal coupling (e.g. $F(p_1)\propto p_1$) its place would be taken by 
a corresponding quantity generalized to depend on the dilaton field 
$p_1=\hat{B}_1$ expressed in terms of the scalar field and external sources.
The effective action proposed in \cite{Kummer:1998dc} indicates the possible 
form of such a generalization.
} 
\cite{Polyakov:1981rd} ($\square = g^{\mu\nu} \nabla_\mu \partial_\nu$)
\begin{gather}
[\, \det \, \square\, ]^{-1/2} = \exp{i L^{\rm (Pol)}} \,, 
\label{eq:4.60} \\
L^{\rm (Pol)} = - \frac{1}{96 \pi}\, \iint \, 
d^2xd^2y \sqrt{-g} R_x \square^{-1}_{xy}\, R_y \, .
\label{eq:4.61}
\end{gather}
Then the full expression $\widetilde{W}$ (\ref{eq:4.59}), written as 
(\ref{eq:4.60}), becomes 
\begin{equation}
\widetilde{W} = \exp \left[i L^{\rm (Pol)}\right] \; \exp{\left[-\frac{i}{2}\, 
\int d^2x \int d^2y\, \sigma_x \Delta_{xy} \sigma_y\right]} \,,
\label{eq:4.62}
\end{equation}
where the propagator $\Delta = \Delta (j,J) = [\, \sqrt{-g}\, 
\square\, ]^{-1}$ by its dependence on $j$ contains the eventual 
interaction with external zweibeine. For minimally coupled 
scalars $(F=1)$ it obeys
\begin{equation}
( \partial_0 \partial_1 - \partial_0 E_1^- \, \partial_0 ) \, 
\Delta_{xx'} = \delta^2 (x-x')\, .
\label{eq:4.63}
\end{equation}
The ansatz
\begin{equation}
\Delta_{xx'} =  \int\limits_{x''} \theta_{xx''} 
( \partial_0^{-1}\, )_{x''x'} d^2x''
\label{eq:4.64}
\end{equation}
allows the formal computation of $\theta_{xx'}$ as 
( $\mathcal{P}$ means path ordering) 
\begin{gather}
\theta_{xx'} = P^{-1}\, \partial^{-1}_1 \, P \,, \label{eq:4.65} \\
P(x) = \mathcal{P}\, \exp\left[- \int_x dy^1  E_1^- 
(x^0,y^1)\, \partial_0\right] \,.
\label{eq:4.66}
\end{gather}
For the classical expressions in the exact path integral the 
meaning of $\partial_0^{-1}$ as an integration (with undetermined 
integration constant) is evident. In the quantum case a more 
careful definition of the Green functions is required which 
implies a UV and IR regularization. A suitable definition is 
$( \partial_0 \to \nabla_0 = \partial_0 - i\mu)$ 
\begin{equation}
\nabla_0^{-1} = - \Theta (y^0-x^0 ) \, e^{i\mu (x^0-y^0)}\; .
\label{eq:4.67}
\end{equation}
The regularization parameter $\mu=\mu_0-i\epsilon\quad 
(\mu_0 \to +0, \epsilon\to +0)$ guarantees  proper behavior at
$x^0 - y^0 \to \pm\infty$. One easily verifies the same property in 
$\nabla_0^{-2}{}_{xy}=\int_z\nabla_0^{-1}{}_{xz}\nabla_0^{-1}{}_{zy}$ 
as well as in higher powers. Only in expressions involving the classical 
background 
like $\nabla_0^{-1}\, \bar{p}_2$ in (\ref{eq:4.42}) when (\ref{eq:4.43}) 
is inserted this rule must be adapted. With $\nabla_0\, \bar{p}_2 = 0$ and 
thus $\bar{p}_2 = \bar{p}_2 (x^1)\, e^{i\mu x^0}$ the expression 
$\nabla_0^{-1}\, \bar{p}_2$ diverges. The solution consists in simply 
going back in these (classical) terms to the classical interpretation 
where $\nabla_0^{-1} = \partial_0^{-1}$ corresponds to integration. 

The formulas for GDTs with nonminimally coupled self interacting
scalars can be derived retaining $F\neq 1$ and $f\neq 0$ in (\ref{eq:4.40}), 
(\ref{eq:4.41}). Then the perturbation expansion in  terms of Newton's
constant requires an expansion in terms of $F$ 
already in the solution of these equations. Together with  the perturbation 
theory outlined in connection with a path integral (\ref{eq:4.58}) this yields 
rather complicated formulas which, therefore, will not be exhibited here. 
Only in connection with the ``virtual BH'' of sect. \ref{sec:VBH} this case will 
be dealt with. 

At the moment there seems to exist only one computation of higher loop 
effects for the simpler case of minimally coupled scalars ($F=1$) 
and a corresponding dilaton theory without kinetic term ($U=0$), but the 
tools are available for arbitrary loop calculations. As noted in 
ref. \cite{Kummer:1998jj} for that specific class 
of theories the whole two-loop effect is just a renormalization of the 
potential $V$.

\subsubsection {Exact path integral with matter}

The JT model (\ref{eq:JT}) is an example of a situation where the path integral
can be calculated exactly even in the presence of minimally coupled
matter fields \cite{Kummer:1997hy}. There the integration $(\mathcal{D}\phi)$ 
produces the Polyakov action. Thus the generating functional for the Green 
functions reads:
\begin{eqnarray}
&&W_{\rm JT}=\int (\mathcal{D}q_i) (\mathcal{D}p_i) \exp iL_{\rm eff}^{\rm JT} 
\,, \label{e4JTW} \\
&&L_{\rm eff}^{\rm JT}
= \int d^2x \Big[ p_i\dot{q}_i+q_1p_2+\Lambda q_3 p_1 \Big]
+L^{\rm (Pol)}(q_2,q_3) \,,\label{e4JTL}
\end{eqnarray}
where source terms $j,J,\si$ and the propagator for the scalar field have 
been dropped. The crucial observation is that (\ref{e4JTL}) is now linear 
in $p_i$. Therefore, when integrating first over the momenta one obtains 
three functional delta functions which may be used to integrate over $q_i$. 
Up to this change, the whole procedure works as before. For details we refer 
to ref. \cite{Kummer:1997hy}. Again, something like local quantum triviality 
occurs. The action (\ref{e4JTL}) already incorporates all quantum effects, 
because in this case no higher loop corrections exist, a feature used in refs.
\cite{Fabbri:2000xh,Fabbri:2000es,Brigante:2002rv} to 
extend the one-loop calculations to 
all orders of perturbation theory.

The method of exact functional integration described here seems to be
a rather general one, although it does not seem possible to formulate general criteria of applicability. We just note that in a similar way an exact path integral has been calculated in a different context, namely the Bianchi IX reduction of Ashtekar gravity \cite{Alexandrov:1998yk}.
As a final remark of this section it should be stressed that in gravity theories -- in contrast to quantum field theory in Minkowski space -- there is no immediate relation between classical and quantum integrability.


\clearpage

\section{Virtual black hole and S-Matrix}
\label{sec:VBH}

Only the interaction with matter in $D=2$ provides continuous physical degrees 
of freedom. Since the asymptotic 
states depend on the model under consideration we do not discuss the most 
general case, but for illustrating the main technical details we focus on an 
explicit example instead: SRG with a non-minimally coupled massless scalar 
field. We also select a situation in which the existence of an $S$-matrix in 
the usual quantum field theoretic sense should be unchallenged, namely 
gravitational scattering of scalars in asymptotically flat space, however, 
without fixing the background before quantization.

In sect.~\ref{se:3.2} we have demonstrated that all geometrical degrees
of freedom can be integrated out exactly. This procedure
yields effective non-local interactions of the remaining fields
(scalars in our case). In this section we calculate explicitly some
lowest order effective vertices and ensuing tree level $S$-matrix elements
corresponding to gravitational scattering of $s$-wave scalars. The
results can be interpreted as an exchange of virtual black holes.

The vertices are extracted from the action (\ref{eq:4.56}) after separation of 
the interaction part $Z$ (cf. (\ref{eq:4.57})). They appear 
as complicated nonlocal expressions with multiple integrals in $x^0$ from 
repeated multiplications of the formal object $\partial_0^{-1}$ present in 
$\hat{B}_i$, the solutions for $p_i$ of eqs. (\ref{eq:4.39})-(\ref{eq:4.41}). 
There exist two classes of vertices (we have 
attached all outer legs in the formulae below), symmetric ones 
\eq{
V^{(2n)}_{a} = \int dx^2_1 \dots dx^2_n v^{(2n)}_a(x_1,\dots,x_n)
\left(\partial_0\phi\right)^2_{x_1}\dots\left(\partial_0\phi\right)^2_{x_n},
}{eq:4.100}
and non-symmetric ones\footnote{For the evaluation of the $S$-matrix one has 
to permute all external legs and thus leg-exchange symmetry is restored.}
\eq{
V^{(2n)}_{b} = \int dx^2_1 \dots dx^2_n v^{(2n)}_b(x_1,\dots,x_n)
\left(\partial_0\phi\partial_1\phi\right)_{x_1}
\left(\partial_0\phi\right)^2_{x_2}\dots\left(\partial_0\phi\right)^2_{x_n}.
}{eq:4.101}
They have the following properties:
\blist
\item They contain an even number of outer legs. Thus, in addition to a
propagator term (cf. e.g. (\ref{eq:4.62}) for the simpler case of minimally 
coupled scalars) there are $\phi^4$-vertices, $\phi^6$-vertices and so on.
\item Each pair of outer legs is attached at one point $x_i$ to the non-local 
vertex.
\item Each outer leg contains one derivative.
\item Non-locality is inherited from the $\partial_0^{-1}$ operators in
the $\hat{B}_i$.
\item The symmetric vertices originate from the term $\tilde{\mathcal{L}}$ in 
(\ref{eq:4.56}).
\item The non-symmetric vertices are produced on the one hand by 
$F(\hat{B}_1)\partial_0\phi\partial_1\phi$ in (\ref{eq:4.56}), on the other 
hand also $\tilde{\mathcal{L}}$ yields such terms. For minimal coupling all
non-symmetric vertices vanish.
\item The information contained in the tree-graphs is classical. Thus, it must 
be possible to extract it by other means. Nevertheless, the path integral 
seems to be the most adequate language to derive scattering 
amplitudes\footnote{As in other well-known examples -- e.g. the Klein-Nishina
formula for relativistic Compton scattering \cite{Klein:1929} -- this 
formalism  seems to be much superior to a classical computation.}.
\elist

The lowest order tree-graph and the ensuing $S$-matrix element has been 
evaluated in ref. \cite{Grumiller:2000ah} for $F(X)= \rm const.$\,. The result 
was trivial,
unless mass-terms $f(\phi)=m^2\phi^2$ had been added (cf. the discussion after
eq. (\ref{vbh:hamil})). Therefore, we will focus in the rest of this
section on the (also phenomenologically more relevant) case of non-minimal
coupling \cite{Fischer:2001vz}. 

\subsection{Non-minimal coupling, spherically reduced gravity}

In principle all effective interactions of the scalars can be
extracted by expanding the non-local action (\ref{eq:4.56}) in a power series 
of $\phi$. At each order the number of integrations increases, and one has to 
fix the ranges appropriately. This becomes cumbersome already at the $\phi^4$ 
level. Fortunately, two observations \cite{Kummer:1998zs}
simplify the calculations considerably. First, instead of dealing with
complicated non-local kernels one may solve corresponding differential
equations. All ambiguities are then removed by imposing asymptotic
conditions on the solutions. Second, instead of taking the $n$-th
functional derivative of the action with respect to bilinear combinations
of the scalar, the matter fields may be localized at $n$
different space-time points. This mimics the effect of
functional differentiation. 

To be more specific, 
the symmetric vertex (\ref{eq:4.100}) may serve as an example
\begin{equation}
v_a^{(2n)}(x_1,\dots ,x_n)\propto
\frac{\delta^n L_{\rm eff}}{\delta ((\partial_0\phi (x_1))^2)
\dots \delta ((\partial_0\phi (x_n))^2)} \,.\label{eq:4.69a}
\end{equation}
By its definition, the functional derivative
is a response to a small localized change of the functional argument
($(\partial_0\phi)^2$ in the present case). Therefore, 
let us choose a specific matter distribution such that 
$(\partial_0\phi)^2$ is localized at $n-1$ points:
\begin{equation}
(\partial_0\phi)^2(x)\propto \sum_{k=2}^n c^{[k]} \delta^2(x-x_k)
\,.\label{eq:4.69b} \end{equation}
Now let $X(x_1)$ and $E_1^-$ be solutions of the classical field equations
in the presence of localized matter (\ref{eq:4.69b}). In
\begin{equation}
v_a^{(2n)}(x_1,\dots ,x_n)\propto 
F(X(x_1))E_1^-(x_1)[x_2,\dots x_n]\vert_{\pi_c}
\label{eq:4.69c}
\end{equation}
we have indicated the dependence of $X$ and $E_1$ on the matter
distribution. The notation $\pi_c$ after the vertical line 
means that one has to expand in $c^{[k]}$ and to select the coefficient in 
front of the product $\prod_{k=2}^n c^{[k]}$. The proof of this statement with
explicit coefficients instead of the proportionality symbol as well
as a corresponding argument for the nonsymmetric vertex (\ref{eq:4.101})
can be found in refs. \cite{Kummer:1998zs,Grumiller:2000ah,Fischer:2001vz}.

Qualitatively, the result (\ref{eq:4.69c}) is rather easy to understand.
In the classical action $(\partial_0\phi)^2$ appears multiplied
by $F(X)e_1^-$. Due to local quantum triviality in the absence of matter 
it is natural to expect ``effective'' quantities of the same nature in the 
vertices. 

The most interesting case is SRG \cite{Fischer:2001vz}, where 
$V=-2,\; U=- (2p_1)^{-1},\;$ $f(\phi)=0,\; F=\frac{p_1}{2},\; j_i=J_i=Q=0$. 
To extract the terms quartic in $\phi$ ($n=2$ in (\ref{eq:4.100}),
(\ref{eq:4.101})) one has to take a matter distribution localized at one
space-time point (cf. (\ref{eq:4.69b}))
\begin{align}
\phi_0 &:= \frac{1}{2}\, (\partial_0 \phi)^2 \to c_0\, \delta^2 (x-y)
\label{eq:4.69}\, , \\
\phi_1 &:= \frac{1}{2}\, (\partial_0 \phi) (\partial_1 \phi) \to c_1\, 
\delta^2\, (x-y)\, ,
\label{eq:4.70}
\end{align}
and to solve the classical e.o.m.-s up to linear order in the 
constants $c_0$ and $c_1$ which just keep track of the number of 
sources. 
The differential equations (\ref{eq:4.39})-(\ref{eq:4.41}), together with 
classical equations for $q_i$ become 
\seq{2.5cm}{
&& \partial_0 p_1 = p_2, \\
&& \partial_0 p_2 = p_1 \phi_0, \\
&& \partial_0 p_3 = 2 + \frac{p_2p_3}{2p_1},
}{4cm}{
&& \partial_0 q_1 = \frac{q_3p_2p_3}{2p_1^2} + \phi_1 - q_2 \phi_0, \\
&& \partial_0 q_2 = - q_1 - \frac{q_3p_3}{2p_1}, \\
&& \partial_0 q_3 = - \frac{q_3p_2}{2p_1}.
}{vbh:eom}\\
Their solutions, to be substituted back into the action with the inverse 
replacement (\ref{eq:4.69}) and (\ref{eq:4.70}), are found easily:
\begin{align}
p_1 (x) &= x^0 - (x^0 - y^0) c_0 y^0 h(x,y), \label{vbh:p1} \\
p_2 (x) &= 1 - c_0 y^0 h(x,y), \label{vbh:p2} \\
q_2 (x) &= 4 \sqrt{p_1} + \left(8c_0y^0\sqrt{p_1}
-2c_0{y^0}^{3/2}-c_1y^0 + (c_1-6c_0{y^0}^{1/2})p_1 \right) h(x,y) \label{vbh:q2} \\
q_3 (x) &= \frac{1}{\sqrt{p_1}}. \label{vbh:q3}
\end{align}
Here $h(x,y) := \theta (y^0 - x^0) \de(x^1-y^1)$ corresponds to one of the 
possible prescriptions introduced in ref. \cite{Grumiller:2000ah} for the 
boundary values at 
$x^0 \to\infty$. It turns out that the vertices below are {\em independent} of 
{\em any} such choice.
The matching conditions at $x^0=y^0$ follow from continuity properties:
$p_1, q_2$ and $q_3$ are $C^0$ and $\partial_0 q_2(y^0+0)-\partial_0 q_2(y^0-0)
= -\left(c_1 - q_2(y^0) c_0\right) \de(x^1-y^1)$. 
The integration constant which would produce an asymptotic (i.e. for $x^0 \to 
\infty$) Schwarzschild term has been fixed to zero. Consistency of 
integration constants with the set of e.o.m.-s 
containing $\partial_1$ automatically yields a vanishing Rindler term.
Furthermore, it relates the asymptotic Schwarzschild term to the asymptotic 
value for the geometric part of the conserved quantity \cite{Kummer:1998zs}. 
Thus, only four integration constants can be chosen independently. We fix 
those in $p_i$ and $q_3$. Because of our particular choice 
$p_3(x^0\to\infty)=0$ a BH may appear only at an intermediate stage 
(the ``virtual black hole'', see below), but should not act asymptotically. 
Due to the infinite range of gravity this is necessary for a proper 
$S$-matrix element if spherical waves are used as asymptotic states for 
incoming and outgoing scalar particles.

\subsection{Effective line element}

The arguments of the previous section suggest that matter interacts
with some effective geometry which solves the classical e.o.m.-s
in the presence of external sources. Moreover, this geometry
can be extracted directly from the vertices. A more formal
(but essentially equivalent) way to see this is to calculate 
the vacuum expectation values of $q_2$ and $q_3$ by varying the exact
path integral (\ref{eq:4.55}) with respect to $j_2$ and $j_3$ in lowest
order of the matter loop expansion and in the presence of external
matter field. The method described in the previous subsection appears to be
more straightforward and considerably simpler.

The matter dependent solutions in the gauge (\ref{eq:a2}) with 
(\ref{vbh:q2}) and (\ref{vbh:q3}) define an effective line element
\eq{
(ds)^2 = 2q_3dx^0 \left( dx^1 + q_2dx^0 \right) = 2 dr du + K(r,u) (du)^2,
}{vbh:dsef} 
with the identifications\footnote{The somewhat unusual role of the coordinates 
should be noted: $x^0$ is asymptotically proportional to $r^2$; thus our 
Hamiltonian evolves with respect to a ``radius'' as ``time''-parameter. 
This also implies that e.g. the 
asymptotic energy density is related to the component $T_{11}$ and not 
$T_{00}$ of the energy momentum tensor.}
\eq{
u = 2\sqrt{2} x^1\,\hspace{0.5cm} r = \sqrt{p_1(x^0)/2}\,. 
}{eq:4.200}
In the asymptotic region by our previous residual gauge fixing the Killing 
norm $\left. K (r,u)\right|_{x^0 > y^0} = 1$ is constant. The line element 
(\ref{vbh:dsef}) then appears in outgoing Sachs-Bondi form. In the VBH 
region the Killing norm
\eq{
\left. K (r,u) \right|_{x^0 < y^0} = \left(1 - \frac{2m}{r} - a r + d\right)
\left(1+ {\mathcal O}(c_0) \right),
}{vbh:kn}
with $m = \de(x^1-y^1)(c_1 y^0 + 2 c_0 {y^0}^{3/2})/2^{7/2}$,
$a = \de(x^1-y^1)(6c_0{y^0}^{1/2}-c_1)/2^{3/2}$ and $d=\de(x^1-y^1)2c_0y^0$ 
has two zeros located approximately at $r = 2m$ and $r = 1/a$ corresponding 
for positive $m$ and $a$ to a Schwarzschild horizon and a Rindler type one. 

\subsection{Virtual black hole}

The geometric part of the conserved quantity (\ref{eq:a11}) in 
our present notation (\ref{eq:a33}) reads
\eq{
{\mathcal C}^{(g)} = \frac{p_2p_3}{\sqrt{p_1}} - 4 \sqrt{p_1}. 
}{vbh:conserved}
As a consequence of the choice of integration constants $\mathcal{C}^{(g)}$ 
vanishes in the asymptotic region $x^0 > y^0$. The functions $p_1$ and $p_3$ 
are continuous, but $p_2$ jumps at $x^0 = y^0$. Thus, $\mathcal{C}^{(g)}$ is 
discontinuous. This phenomenon has been 
called ``virtual black hole'' (VBH) in \cite{Grumiller:2000ah}. It is generic 
rather than an artifact of our special choice of asymptotic conditions. The
reason why we have chosen this name is simple: The geometric part of the
conserved quantity (\ref{vbh:conserved}) is essentially equivalent to the 
so-called mass aspect function, which is closely related to the BH 
mass (cf. sect. \ref{se:5.2}). Moreover, inspection of the Killing norm 
(\ref{vbh:kn}) reveals that for negligible Rindler acceleration $a$ the 
Schwarzschild horizon corresponds to a BH with precisely that mass. It 
disappears in the asymptotic states (by construction), but mediates an 
interaction between them.

The idea that BH-s must be considered in the $S$-matrix together
with elementary matter fields has been put forward some time ago 
\cite{'tHooft:1996tq}. The approach \cite{Fischer:2001vz} reviewed here, for 
the first time allowed to derive (rather than to conjecture) the appearance of 
the BH states in the quantum scattering matrix of gravity.

The solutions (\ref{vbh:p1}) and (\ref{vbh:p2}) establish
\eq{
\left. {\mathcal C}^{(g)} \right|_{x^0 < y^0} = 4 c_0 {y^0}^{3/2} \propto - 
m_{VBH}. 
}{vbh:vbh}
Thus, $c_1$ only enters the Rindler term in the Killing norm, but not 
the VBH mass (\ref{vbh:vbh}). 


\begin{wrapfigure}{r}{3cm}
\epsfig{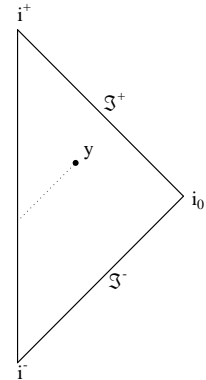}
\caption{CP diagram of the VBH}
\label{fig:cp}
\end{wrapfigure}
The CP diagram corresponding to the line element (\ref{vbh:kn}) as presented 
in figure \ref{fig:cp} needs some explanations: first of all, the effective 
line element is non-local in the sense that it depends not only on one set 
of coordinates (e.g. ${u, r}$) but on two ($x=(u,r), y=(u_0,r_0)$), where 
$r_0$ and $u_0$ are related to $y^0$ and $y^1$ like $r$ and $u$ to $x^0$ 
and $x^1$ in (\ref{eq:4.200}). As discussed previously, this non-locality 
was a consequence of integrating out geometry non-perturbatively. For each 
choice of $y$ it is possible to draw an ordinary CP-diagram treating 
$u_0, r_0$ as external parameters. The light-like ``cut'' in figure 
\ref{fig:cp} corresponds to $u=u_0$ and the endpoint labeled by $y$ to the 
point $x=y$. The 
non-trivial part of our effective geometry is concentrated on the cut.

We do not want to suggest to take the effective geometry (\ref{vbh:dsef})
at face value -- this 
would be like over-interpreting the role of virtual particles in a loop 
diagram. It is a nonlocal entity and we still have to ``sum'' (read: 
integrate) over all possible geometries of this type in order to obtain the 
nonlocal vertices and the scattering amplitude. Nonetheless, the 
simplicity of this geometry and the fact that all possible configurations 
are summed over, are nice qualitative features of this picture. 

The localization of ``mass'' and ``Rindler acceleration'' on a 
light-like cut  (see fig. \ref{fig:cp}) in (\ref{vbh:dsef}) is not an artifact 
of an accidental gauge choice, but has a physical interpretation in terms of
the Ricci-scalar \cite{Grumiller:2001pt}, the explicit form of which is given 
by \cite{Grumiller:2001rg}
\meq{
R^{(VBH)}(u,r;u_0,r_0) = \de(u-u_0)\Bigg[
-\de(r-r_0) \left( \frac{4m_0}{r^2}-\frac{4d}{r}+6a_0 \right) \\
+\Theta(r_0-r) \left( \frac{6a_0}{r}-\frac{2d}{r^2} \right)\Bigg]\,.
}{vbh:Ricci}
As discussed in  ref. \cite{Grumiller:2001rg} certain parallels to Hawking's
Euclidean VBHs \cite{Hawking:1996ag} can be observed, but also essential 
differences. The main one is our Minkowski signature which we deem to be a 
positive feature. 

\subsection{Non-local $\phi^4$ vertices}

All integration constants have been fixed by the arguments in the preceding 
paragraphs. The fourth order vertex of quantum field theory is extracted from
the second line of (\ref{eq:4.36}) by collecting the terms linear in $c_0$ and 
$c_1$ replacing each by $\phi_0$ and $\phi_1$, respectively. 
\begin{figure}
\epsfxsize=7cm
\centerline{\epsfbox[70 210 540 360]{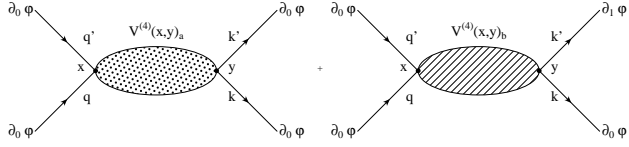}}
\caption{Total $V^{(4)}$-vertex with outer legs}
\label{fig:bothvertices}
\end{figure}
\noindent The tree graphs we obtain in that way (cf. fig. 
\ref{fig:bothvertices}) contain the nonlocal vertices
\eqa{
V^{(4)}_a && = \int_x\int_y \phi_0(x) \phi_0(y) \left. \left( 
\frac{d q_2}{d c_0} 
p_1 + q_2 \frac{d p_1}{d c_0} \right)\right|_{c_i = 0} \nonumber \\
&& = \int_x \int_y \phi_0(x) \phi_0(y) \left| \sqrt{y^0}-\sqrt{x^0} \right| 
\sqrt{x^0y^0} \nonumber \\
&& \quad \left( 3x^0+3y^0+2\sqrt{x^0y^0} \right) \de(x^1-y^1),
}{vbh:s4a}
and
\eqa{
V^{(4)}_b &&= -\int_x\int_y\left. \left(\phi_0(y) \phi_1(x) 
\frac{d p_1}{d c_0} - 
\phi_0(x) \phi_1(y) \frac{d q_2}{d c_1} p_1 \right)\right|_{c_i = 0} 
\nonumber \\ 
&& = -\int_x\int_y \phi_0(x) \phi_1(y) \left|x^0-y^0 \right| x^0 
\de(x^1 - y^1), 
}{vbh:s4b}
with $\int_x := \int\limits_0^\infty dx^0\int\limits_{-\infty}^\infty dx^1$.

\subsection{Scattering amplitude}

In terms of the time variable $t := r + u$ the scalar field asymptotically 
satisfies the spherical 
wave equation. For proper $s$-waves only the spherical Bessel function
\eq{
R_{k0} (r) = \frac{\sin (kr)}{kr}
}{vbh:impeig}
survives in the mode decomposition ($Dk:=4\pi k^2dk$):
\eq{
\phi(r,t) = \frac{1}{(2\pi)^{3/2}}\int\limits_0^{\infty} \frac{Dk}
{\sqrt{2k}} R_{k0} \left[a^+_k e^{ikt} + a^-_k e^{-ikt}\right]. 
}{vbh:asyscalmod}
With $a^\pm$ obeying the commutation relation $[a_k^-,a_{k'}^+] = \de(k-k')/
(4\pi k^2)$, they will be used to define asymptotic states 
and to construct the Fock space. The normalization factor is chosen such that 
the Hamiltonian of asymptotic scalars reads
\eq{
H^{\rm (as)} = \frac{1}{2}\int\limits\limits_0^{\infty}Dr\left[(\partial_t\phi
)^2 + (\partial_r \phi)^2 \right] = \int\limits_0^{\infty} Dk a^+_k a^-_k k.
}{vbh:hamil}

In ref. \cite{Grumiller:2000ah} we had observed a non-physical feature in the 
massless case for (in $D=2$) {\it minimally} coupled scalars: 
Either the $S$-matrix 
was divergent or -- if the VBH was ``plugged'' by suitable boundary conditions 
on $\phi$ at $r=0$ -- it vanished. This implied an effective decoupling of the
plane waves from the geometry. For massive scalars a finite nonvanishing 
scattering amplitude has been found.

In the present more physical case of $s$-waves from $D=4$ GR at
a first glance it may seem surprising that the simple additional factor $X$
in front of the matter Lagrangian induces fundamental changes in the 
qualitative behavior. In fact, it causes the partial differential equations 
(\ref{vbh:eom}) to become coupled, giving rise to an additional vertex 
($V^{(4)}_{b}$). 

After a long and tedious calculation (for details see refs. 
\cite{Grumiller:2001ea,fischervertices}) for
the $S$-matrix element with ingoing modes $q, q'$ and outgoing
ones $k, k'$,
\eq{
T(q, q'; k, k') = \frac{1}{2} \left< 0 \left| a^-_ka^-_{k'} \left(V^{(4)}_a 
+ V^{(4)}_b \right) a^+_qa^+_{q'}\right| 0 \right>\,, 
}{vbh:T}
having restored\footnote{Up to this point the overall factor in (\ref{2.58})
had been omitted.} the full dependence on the gravitational constant 
$\ka=8\pi G_N$, we arrive at 
\eq{
T(q, q'; k, k') = -\frac{i\ka\de\left(k+k'-q-q'\right)}{2(4\pi)^4 
|kk'qq'|^{3/2}} E^3 \tilde{T}
}{vbh:RESULT}
with the conserved total energy $E=q+q'$,
\eqa{
&& \tilde{T} (q, q'; k, k') := \frac{1}{E^3}{\Bigg [}\Pi \ln{\frac{\Pi^2}{E^6}}
+ \frac{1} {\Pi} \sum_{p \in \left\{k,k',q,q'\right\}}p^2 \ln{\frac{p^2}{E^2}} 
\nonumber \\
&& \quad\quad\quad\quad\quad\quad \cdot {\Bigg (}3 kk'qq'-\frac{1}{2}
\sum_{r\neq p} \sum_{s \neq r,p}\left(r^2s^2\right){\Bigg )} {\Bigg ]},
}{vbh:feynman}
and the momentum transfer function $\Pi = (k+k')(k-q)(k'-q)$. The interesting 
part of the scattering amplitude is encoded in the scale independent factor 
$\tilde{T}$. 

With the definitions $k=E\al$, $k'=E(1-\al)$, $q=E\be$, and $q'=E(1-\be)$ 
($\al,\be\in[0,1]$, $E\in\mathbb{R}^+$) a quantity to be interpreted as a 
cross-section for spherical waves can be defined \cite{Fischer:2001vz}:
\begin{equation}
 \frac{d\sigma}{d\alpha}=\frac{1}{4(4\pi)^3}\frac{\kappa^2 E^2 |\tilde{T}
(\alpha, \beta)|^2}{(1-|2\beta-1|)(1-\alpha)(1-\beta)\alpha\beta}.
\label{vbh:crosssection}
\end{equation}
\begin{figure}
\centering
  \epsfig{file=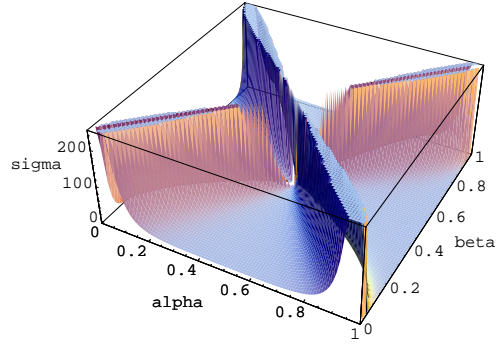,width=.47\linewidth}
  \caption{Kinematic plot of $s$-wave cross-section $d\si/d\al$}
  \label{fig:kin}
\end{figure}
The kinematic plot fig. \ref{fig:kin} contains the relevant physical 
information. The dependence of the cross-section on the total incoming energy
is trivially given by the monomial prefactor $E^2$: it vanishes in the IR 
limit and diverges 
quadratically in the UV limit. At least the last fact is not surprising, 
considering our assumption of energies being small as compared to the Planck 
energy. It simply signals the breakdown of our perturbation theory.

The main results of the detailed discussion 
\cite{Fischer:2001vz,Grumiller:2001rg} are: 
\blist
\item Poles exist in the case of vanishing momentum transfer (forward 
scattering).
\item An ingoing $s$-wave can decay into three outgoing ones. Although this 
 may be expected on general grounds, within the present formalism it is
possible to provide explicit results for the decay rate.
\item Despite the non-locality of the effective theory, the $S$-matrix is 
$CPT$ invariant at tree level.
\item Fig. \ref{fig:kin} appears to exhibit self-similarity. Indeed, by zooming
into the center of that figure one obtains again an identically-looking plot.
However, this self-similarity is only a leading (and next-to-leading) order 
effect and breakes down in the flat regions.
\elist

\subsection{Implications for the information paradox}\label{ipar}

Very roughly, the information paradox \cite{Banks:1995ph}
may be formulated in the following way. Imagine a pure 
quantum state in a non-singular asymptotically
Minkowski space-time. Let this pure state collapse into a BH
which evaporates due to the Hawking effect.
This effect is only understood for a background which does not
change appreciably due to radiation: if it does, it is, nevertheless, assumed
that this evaporation proceeds through an (unknown!) final phase
so that the BH disappears. The
final state of this process will be Minkowski space filled with
thermal radiation which is definitely a mixed quantum state.
Therefore, a pure quantum state seems to evolve into a mixed
one, contradicting basic laws of quantum mechanics. 
One reason already has been given why this picture
is a rather approximate one. In addition the exact thermal Planck spectrum of
the radiation requires infinite time for its formation, i.e.
radiation can be strictly thermal only
if the BH never disappears completely. 
However, the formation of even approximately
thermal final states seems very difficult to master in quantum theory which  
has prompted several very interesting developments in quantum gravity, as
e.g. models of stable BH remnants \cite{Aharonov:1987tp} 
and the $S$-matrix approach of ref. \cite{Stephens:1994an}.

The understanding of the evolution of VBHs is crucial 
for quantum gravity \cite{Hawking:1982dj}. Their evaporation --
or, more exactly, their conversion to mixed states would
inevitably violate either locality or energy-momentum conservation
\cite{Banks:1984by}. Since the non-perturbative approach to 
path integral quantization now also predicts VBHs
in two dimensions, it is important to understand whether there is
an ``information loss'' in these models. Of course, there is none
at the tree level discussed above. However, we also have good grounds to 
believe that the situation will remain the same in higher loops. 

Our first argument is somewhat formal. In the two-dimensional
model we were able to extract the VBH from the degrees of freedom 
already present in the theory rather than to be forced to introduce 
it from the outside. The whole system has been quantized in full
accord with the general principles of the quantization for systems
with constraints. According to general theorems \cite{Henneaux:1992} the 
resulting quantum theory must be unitary, respect causality and energy 
conservation, and must forbid transitions of pure states to mixed ones, as 
long as we are able to refer to a Fock space of the asymptotic states.

Our next argument is more physical. BH evaporation is related
to the condition (\ref{ehmUvac}) which fixes the energy-momentum
tensor at the horizon and thus defines the Unruh vacuum state.
This condition is clearly not applicable to VBHs.
The relevant vacuum state for the scalar field is just the
usual Minkowski space vacuum containing no information about
VBH states which may be formed in quantum 
scattering. Kruskal coordinates for a VBH cannot be associated
with any real observer. Therefore, the argument that the energy-momentum
tensor must be finite at the horizon is not applicable to it.
The only vacuum state which can be defined by a condition at
infinity rather then on a horizon is the Boulware vacuum which does not 
contain Hawking radiation so that VBHs do not radiate anything to infinity. 

It must be admitted that in order to put this argumentation upon a firm basis 
one should calculate the next (one-loop) order in the path integral. Since 
Hawking radiation is a one-loop effect, this order of the perturbation
theory will be actually sufficient.


\clearpage

\section{Canonical quantization}\label{se:reduced}

Canonical quantization methods dominated $2D$ dilaton
gravity during its early years.  They owe their success to the fact that the geometrical sector contains no propagating degrees of freedom, and, therefore, the problem reduces effectively to a quantum mechanical one. 

After the extensive discussion of the path integral in the previous section we intend to be brief for this essentially equivalent approach.
The prehistory of canonical quantization of gravity involves the seminal papers
of Arnowitt, Deser and Misner \cite{Arnowitt:1962}, Wheeler 
\cite{Wheeler:1968} and DeWitt \cite{DeWitt:1967yk} which led to Misner's ``minisuperspace quantization'' program \cite{Misner:1972},
where almost all degrees of freedom were frozen by symmetry requirements.
Kucha{\v{r}} extended these techniques to ``midisuperspace quantization'' for
the explicit example of 
cylindrical gravitational waves \cite{Kuchar:1971xm}, i.e. to
a system with field degrees of freedom, albeit still using symmetry 
requirements in order to simplify the formalism. This seems to 
be the only midisuperspace model that could be treated exactly. The most 
notable example of a non-soluble model is the collapse of spherically 
symmetric matter \cite{Berger:1972pg} (cf. \cite{Unruh:1976db} for an 
essential correction to that paper). A canonical treatment of a complete 
Schwarzschild spacetime under somewhat too strong assumptions provided 
Lund's proof \cite{Lund:1973} of the non-existence of an extrinsic 
time representation for vacuum Schwarzschild BHs \cite{Cangemi:1993up}.

Possibly due to the impact of Hawking's work on semi-classical radiation of 
BHs \cite{Hawking:1975sw} the discussion of genuine quantum gravity
effects was postponed until the CGHS model \cite{Callan:1992rs} rekindled the interest in (exact) quantization of BH-s \cite{Gegenberg:1993rg,Kuchar:1994zk,Kuchar:1997zm,Cangemi:1996yz,Gegenberg:1995pv,Strobl:1994eu,Barvinsky:1996hr,Gegenberg:1997de,Louis-Martinez:1994eh,Cangemi:1992bj,Cangemi:1993sd,Cangemi:1993up,Cangemi:1994ei,Mikovic:1992id,Mikovic:1993vy,Mikovic:1995ub,Varadarajan:1995jj,Vergeles:2001zp}. 
In particular, in ref. \cite{Kuchar:1994zk} an extrinsic time 
representation for the quantized Schwarzschild BH allowed to 
circumvent Lund's no-go theorem by relaxing its premises.

As an explicit example for demonstrating the main points we focus on the 
CGHS model, the Dirac quantization of which has been studied by Jackiw and 
collaborators \cite{Cangemi:1992bj,Cangemi:1993sd,Cangemi:1994ei,
Cangemi:1996yz,Benedict:1996qy}, by Mikovic 
\cite{Mikovic:1992id,Mikovic:1993vy,Mikovic:1995ub}, and later also by other 
authors. 

Our brief summary follows the work of Kucha{\v{r}}, Romano and 
Varadarajan \cite{Kuchar:1997zm}.
The starting point is not really the CGHS action (\ref{esLd}), but its 
conformally related one (\ref{eq:trafopotential}). In this way, one 
eliminates the kinetic term of the dilaton field in (\ref{2.59}) at the 
cost of a singular conformal transformation (\ref{eq:conformalfactor}). This 
action is then cast into canonical form by the
standard ADM decomposition. It has to be supplemented by surface terms invoking
the requirement of functional differentiability\footnote{The physically
most transparent way to impose it is a careful treatment of asymptotic 
conditions on the geometric variables \cite{Beig:1987} (including
lapse and shift; cf. also sect. \ref{admsec}).}. However, the boundary 
action leads to an important caveat: At 
the left and right infinity corresponding to the asymptotic regions of patch 
$\mathcal A$ and patch $\mathcal{B}$ in the figures \ref{fig:2.6} and 
\ref{fig:2.7} arbitrary variations of the lapse are required. 
Otherwise unwanted ``natural'' boundary conditions for the 
BH mass emerge which imply vanishing of the BH mass.
This problem has been resolved for the Schwarzschild BH \cite{Kuchar:1994zk}, 
parameterizing the lapse function at the boundaries
by a proper time function. It turns out that the total action
\eq{
L=\int d^2x \left(\pi \dot{\phi} + \pi_y \dot{y} + p_\si \dot{\si} - NH - 
N^1 H_1 \right) + \int dt \left(\dot{\tau}_L m_L - \dot{\tau}_R m_R \right)\,,
}{eq:cq1}
depends on these two additional parameters $\tau_L$ and $\tau_R$
and on the standard canonical variables: $N$ is the lapse, $N^1$ the shift, 
$H$ the Hamiltonian constraint, $H_1$ the momentum constraint, $\pi,\phi$ 
denotes the matter degree of freedom (the presence of a single minimally 
coupled scalar field is assumed), and in the notation of ref. 
\cite{Kuchar:1997zm} $y,\pi_y,\si,p_\si$ are geometric canonical 
field variables. In the boundary action the indices $L,R$ refer to ``left 
infinity'' and ``right infinity'', $m_{L,R}$ are the -- conveniently 
normalized -- BH masses. The relative sign between the last two terms 
originates from the 
different time orientations one chooses for the two patches in order to match 
the behavior of the Killing time $T$ in the corresponding global diagram. 

There exists a canonical transformation mapping the action (\ref{eq:cq1})
onto a simpler one (in terms of which the constraints become second order
polynomials). One has to be particularly careful with the boundary part. In 
the new variables the constraints can be solved exactly, because they have
the same form as those of a parametrized massless scalar field propagating on
a flat $2D$ background. The main obstacle in replacing the canonical variables
by corresponding operators is a Schwinger term encountered in the commutators
of the energy-momentum tensor operators \cite{Boulware:1967}. This
anomaly converts the classical first class constraints into quantum second
class ones and thus the imposition of the operator constraints on the states
leads to inconsistencies\footnote{This is true at least in the Schr\"odinger picture. In the Heisenberg picture the quantum theory is well-defined and has the same number of degrees of freedom as the classical one. Indeed, also the Heisenberg e.o.m.-s have the same form as the classical e.o.m.-s (of course, one has two additional quantum mechanical degrees of freedom from the two
parameters in the boundary action, but they are just constants of motion).}.
Kucha{\v{r}} proposed a trick to get rid of that anomaly in the Schr\"odinger picture \cite{Kuchar:1989wz,Kuchar:1989bk}: The momentum operators are supplemented by an additional term which does not change the canonical commutation relations but which cancels the anomaly.

To summarize: by performing first a conformal and then a canonical 
transformation the CGHS model was mapped onto a parametrized field theory on a
flat background which could be quantized successfully. Clearly this quantum 
theory of the parametrized field is a standard unitary quantum field theory, 
i.e. no information loss is encountered. However, the interesting
questions are precisely those related to the physical spacetime. Thus, one 
still has to show how to pose such questions in the framework based on the
auxiliary flat background. As emphasized by the authors themselves 
\cite{Kuchar:1994zk,Kuchar:1997zm} it seems that difficult 
problems reemerge which were avoided so far: 
\blist
\item It is not clear how to make sense of the operator version of
the physical line element $(ds)^2_{\rm physical} = (ds)^2_{\rm flat} 
\exp{(-2\rho)}$.
\item The ``correct'' operator ordering of the 
conformal factor is an open question when $\rho$ is expressed in terms of the auxiliary canonical 
variables.
\item The classical dilaton field should remain positive to ensure the correct
signature of the physical metric. In a quantum theory it is highly nontrivial 
to maintain this positivity requirement. There have been attempts to clarify
this issue with a $1+0$ dimensional model \cite{Watson:2000zp}.
\elist
Besides, the presence of an anomaly may add difficulties in implementation 
of the Dirac quantization scheme \cite{Cangemi:1994ei}.

If there is no matter field in the model the canonical quantization
is especially simple. The reduced phase space quantization program\footnote{
A clear explanation of the reduced phase space quantization
can be found in \cite{Faddeev:1988qp}.} can be carried through exactly to the very end, i.e. one can solve the constraints and fix the gauge freedom.
However, the result is essentially trivial for spherically reduced gravity 
\cite{Thiemann:1993jj,Kuchar:1994zk,Cavaglia:1995yc,Cavaglia:1996bb}, 
as well as for the other dilaton models \cite{Cavaglia:1998uc}: the quantum functional only depends on the ADM mass.

Since the Maxwell field in two dimensions does not add new propagating
degrees of freedom an extension of the canonical approach to charged
BHs may be done in a rather straightforward manner 
(cf. \cite{Medved:1998ks,Barvinsky:2001tw}).


Other instances where the programme of canonical quantization has been 
carried through in essentially quantum mechanical models like the collapse 
of spherically symmetric (null-)dust are refs. 
\cite{Bicak:1997bx,Hajicek:2001ky,Hajicek:2001kz,Kouletsis:2001ma,
Vaz:2001bd,Hajicek:2002ny}. 
An example 
for a semiclassical model of BH evolution with time variable is ref. 
\cite{Casadio:2001at}.

\clearpage

\section{Conclusions and discussion}\label{se:conclu}

The last decade has seen remarkable progress in the treatment of $2D$ dilaton gravity models. Having retraced the main historical lines of this development in the Introduction we now confront the main results of this field with a list of well-known problems which quantum gravity, in fact, shares with other quantum theories in which geometry plays a fundamental dynamical role. 

First we summarize the main contributions which dilaton gravity has been able to provide with respect to these questions and which we have described in some detail in this report. 

Dilaton models in $D=2$ possess the basic advantage that their geometric part, in a certain sense, is a ``topological'' theory, albeit one where the solutions are not related to a discrete winding number. An important special case is the theory which arises from spherical reduction of Einstein gravity in $D=4$, i.e. also the treatment of Schwarzschild black holes is covered by it. Other relevant members are the string-inspired dilaton theory and the Jackiw-Teitelboim model.

In the absence of matter the classical solution for all such theories can be given in closed analytic form, a result which appears more naturally in the Eddington-Finkelstein gauge for the metric. In that gauge also a very straightforward procedure allows the construction of the global solution without the necessity to introduce explicitly or implicitly global Kruskal-like coordinates. It is a peculiar feature of effective two dimensions of space-time that the ADM-mass, even in the presence of matter interactions, generalizes to an  "absolute"  (in space and time!) conserved quantity. Technically many new results are related to the complete dynamical equivalence between the standard formulation of dilaton theories by an action expressed in terms of metric and dilaton field on the one hand, and a ``first order''  (``covariant Hamiltonian'') action on the other hand. The latter involves auxiliary fields and the geometry is expressed in Cartan variables (zweibeine and spin connection). This equivalent formulation also contains nontrivial torsion and turns out to represent a special case of the very general concept of  Poisson-Sigma models, a new and rapidly developing field of research with important connections to strings and non-commutative geometry. Certain generalizations, as e.g. Yang-Mills fields or supergravity extensions are covered directly by this formalism.

Strictly speaking, the (semi-classical) treatment of Hawking radiation does not represent an application of quantum gravity but it is formulated with respect to a given classical (Black Hole) background. Nevertheless, in order to justify other $2D$ quantum gravity results derived from an effective $2D$ theory, it should  emerge as well from a treatment of the spherically reduced case. As far as (in $D=2$) minimally coupled scalar fields are concerned all aspects are well understood. In spherically reduced matter (nonminimal coupling in $D=2$) a correct relation between Hawking temperature and Hawking flux has been proposed, however based upon mathematical steps whose justification as yet has not been proved conclusively. For a wide class of twodimensional gravity models relations between Hawking temperature and ADM mass can be obtained which differ from the one in Einstein gravity.

The full impact of the advantages from the first order formulation of dilaton models $D=2$ is revealed in the path integral quantization of such theories. In the temporal gauge for Cartan variables  --  corresponding to the Eddington-Finkelstein gauge of the metric  --  it proved possible to exactly integrate out all geometric degrees of  freedom. This intrinsically nonperturbative result is closely related to the quantum field theoretical ``triviality'' of generic gravity theories without matter interactions in $D=2$. In this derivation the path integral over all positive and negative ``volumes'' is an essential ingredient, thus establishing an important confirmation of the conjecture that this should also be the correct procedure in $D=4$. 

If matter fields are present, still an effective theory is obtained in which geometry is treated in a nonperturbative manner. A perturbation expansion in terms of the  interactions with matter follows standard quantum field theoretical methods.  It is valid as long as the energies are small as compared to Planck's mass. The effective non-local vertices of scalar fields in this formulation can be interpreted as the appearance of an intermediate ``virtual'' Black Hole in certain scattering amplitudes of spherical waves. It should be stressed that this seems to be the first instance where such a virtual Black Hole reflects an intrinsic feature of the theory and is not introduced by any additional assumption. 

As far as the problem of observables in quantum gravity is concerned, the computation of a special (gauge-independent) $S$-matrix element for spherically reduced Einstein gravity seems to be an interesting feature as well. 
Also some progress has been reported regarding the final stages of Black Hole evaporation and the intimately connected  ``quantum information paradoxon''. The very fact that now a formulation of that system exists in the form of a standard quantum field theory implies that  --  also at the very end of its existence --- a Black Hole does not violate quantum mechanical concepts like unitarity.
In that case as in others it has turned out that  --  at least in $D=2$ -- the application of standard quantum field theoretic techniques can go very far, leading to interesting results without the necessity to infer additional concepts. 

This suggests further studies in many directions which have not yet sufficiently been covered so far.

At the classical level a more systematic search for $2D$ gravity theories involving matter interactions, but still allowing exact solutions, seems desirable, e.g. for a  -- perhaps at first only qualitative, but nevertheless exact  --  description of critical behavior in spherical collapse. The same applies for models with additional abelian or nonabelian gauge fields from which the spherically symmetric Black Hole with (nonabelian) charges could be studied. Although the general principle to obtain supergravity extensions from $2D$ dilaton theories is now available, the new comprehensive  approach based upon the Poisson-Sigma structure of such models has posed many new questions.
 
Recently a whole new field of scalar-tensor theories in $D=4$ (quin\-tes\-sence) has been developed. There a dilaton (``Jordan''-) field already appears in the higher dimension. Certain important aspects of these models should be accessible by $2D$-methods when the effective spherically reduced theory is considered.

Within the realm of semi-classical problems despite new insight for the treatment of Hawking radiation starting in the spherically reduced case, still several important questions are open. 

Among the possible directions of research in full $2D$ quantum gravity higher loop corrections could be investigated. The issue of  ``quantum'' observables is closely related  to the treatment of systems with finite boundary and related boundary variables. 

Possibly also new elements for the long discussion of quantum gravity at the Big Bang (quantum cosmology) could emerge. More immediate consequences of the present approach are a generalization of gravitational scattering of scalars described in this report, for scattering off a Black Hole. Another generalization in the quantum case would be the treatment of fermions, either directly introduced in $D=2$ or obtained from $D=4$ by reduction.  Finally, the virtual black hole phenomenon exists for generic dilaton models. It could be interesting to study the $S$-matrix of gravitational scattering of matter in the extended context of generalized dilaton theories.
The range of technically feasible investigations now certainly has been enlarged substantially.


\clearpage

\lhead{ACKNOWLEDGEMENT}
\section*{Acknowledgement}\label{se:ack}

\addcontentsline{toc}{section}{Acknowledgement}

We have profited from numerous enlightning discussions with our previous 
collaborators S. Alexandrov, M. Bordag,
M. Ertl, P. Fischer, F. Haider, D. Hofmann, 
M.O. Katanaev, T. Kl\"osch, S. Lau, H. Liebl, D.J. Schwarz, T. Strobl, 
G. Tieber, P. Widerin, A. Zelnikov
and the members of the Institute for Theoretical 
Physics at the TU Vienna (especially H. Balasin and L. Bergamin).
The exchange of views and e-mails on the quantization part with 
P. van Nieuwenhuizen and R. Jackiw is gratefully acknowledged.
We also thank those authors who suggested supplementary references.

One of the authors (W.K.) is especially grateful to R. Jackiw and G. Segr{\'e}
who in different ways during the 80s kindled his interest in quantum gravity,
especially for models in $D=2$.

The \LaTeX-nical support of F. Hochfellner and E. M\"ossmer has been a
great help for us. Finally, we render special thanks to T. Kl\"osch for 
letting us ``steal'' two of his beautiful {\tt xfig}-pictures.

This work has been supported by project P-14650-TPH of the Austrian Science 
Foundation (FWF) and by project BO 1112/11-1 of the Deutsche 
Forschungsgemeinschaft (DFG).

\vspace{1cm}

{\em Note added after proofs:} Due to the broadness of the topics covered
explicitly or implicitly in this review the cited literature is necessarily
somewhat incomplete. Unfortunately we have omitted some relevant references
which should have been included: the third part of the series of papers of
Kl\"osch and Strobl on classical gravity in $2D$ \cite{Klosch:1997md} which
deals with solutions of arbitrary topology and a series of papers of
Chamseddine et al.\ who was among the pioneers of $2D$ dilaton gravity, 
discussing it mostly from a stringy perspective \cite{Chamseddine:1989yz,
Chamseddine:1989tu,Chamseddine:1990wn,Chamseddine:1991hr,
Chamseddine:1991fg,Chamseddine:1992qu}.
Additionally, we should mention that quantization of PSMs (cf.\ end of
sect.\ 2.3) was done before Hirshfeld and Schwarzweller, e.g. in
\cite{Schaller:1994pm}.

{\em Addendum in January 2008:} More recent results are contained in another review \cite{Grumiller:2006rc}.

\clearpage

\lhead{\leftmark}
\begin{appendix}

\section{Spherical reduction of the curvature 2-form} \label{app:A}

In a $D$-dimensional Pseudo-Riemannian manifold ${\mathcal M}$ with 
Lo\-rentz\-ian signature $(+, -, -, ..., -)$ and spherical 
symmetry\footnote{I.e. the isometry group of the metric has a group isomorphic 
to $SO(D-1)$ as subroup with $S^{D-2}$-spheres as orbits.} the 
coordinates describing the manifold can be separated in a two-dimensional 
Lorentzian part spanning the manifold $L$ and a $(D-2)$-dimensional Riemannian 
angular part constituting an $S^{D-2}$. In adapted coordinates the line
element reads
\begin{equation}
ds^{2}_{\mathcal M} = g_{\mu\nu}dx^\mu\otimes dx^\nu = g_{\al\be}dx^{\al}\otimes dx^{\be}-
\Phi^{2}\left(x^\al\right)g_{\rho\si}dx^{\rho}\otimes dx^{\si},
\end{equation}
using letters from the beginning of the alphabet ($\al,\be,\dots$; $a,b,\dots$)
for quantities connected with $L$, letters from the middle 
of the alphabet ($\mu,\nu,\dots$; $m,n,\dots$) for quantities connected with 
${\mathcal M}$ and letters from the end of the alphabet for quantities connected with 
$S^{D-2}$ ($\rho,\si,\dots$; $r,s,\dots$). Indices will be lowered and raised
with their corresponding metrics.
 
In the vielbein-formalism\footnote{The notation 
and the meaning of all quantities appearing here is explained in sect. 
\ref{se:2.1.1}.} 
$ds^2_L = \eta_{ab}\bar{e}^a\otimes\bar{e}^b$, 
$ds^2_S = \de_{rs}\bar{e}^r\otimes\bar{e}^s$ and comparing with 
$ds^2_M=\eta_{mn}e^m\otimes e^n=ds^2_L-\Phi^2ds^2_S$ yields
\newline
\parbox{4cm}{\begin{eqnarray*}
&& e^{a} = \bar{e}^{a}, \\
&& e^{r} = \Phi \bar{e}^{r}, 
\end{eqnarray*}} \hfill
\parbox{4cm}{\begin{eqnarray*}
&& e_{a} = \bar{e}_{a}, \\
&& e_{r} = \Phi^{-1} \bar{e}_{r}.
\end{eqnarray*}} \hfill
\parbox{1cm}{\begin{equation} \label{app:vieltrafo} \end{equation}} \hfill
\newline
Metricity and torsionlessness for the connection 1-forms on $M, L$ and $S$ 
leads to
\begin{equation}
\omega^a{}_b = \bar{\omega}^a{}_b,\hspace{0.5cm} 
\omega^r{}_s = \bar{\omega}^r{}_s,\hspace{0.5cm} 
\omega^r{}_a = \left( \bar{e}_a \Phi \right) \bar{e}^r,\hspace{0.5cm}
\omega^a{}_r = \eta^{ab}\de_{rs}\left( \bar{e}_b \Phi \right) \bar{e}^s,
\end{equation}
using the relations (\ref{app:vieltrafo}).

From Cartan's structure equation (\ref{2.23})
the curvature 2-form on $M$ follows:
\begin{eqnarray}
&& R^a{}_b = \bar{R}^a{}_b, \\
&& R^r{}_s = \bar{R}^r{}_s+\eta^{ab}\left( \bar{e}_{a}\Phi \right) 
\left( \bar{e}_{b}\Phi \right) \bar{e}^{r}\bar{e}_{s}, \\
&& R^a{}_r = \eta^{ac}\left( \bar{e}_{b}\bar{e}_{c}\Phi \right) \bar{e}^{b}
\bar{e}_{r}+\left(\bar{e}_{b}\Phi \right) \bar{\omega}^{ab}
\bar{e}_{r}, \label{app:cur3} \\
&& R^r{}_a = \left( \bar{e}_{b}\bar{e}_{a}\Phi \right) \bar{e}^{b}
\bar{e}^{r}-\left(\bar{e}_{b}\Phi \right) \bar{\omega}^b{}_a \bar{e}^{r},
\label{app:cur4}
\end{eqnarray}
where $\bar{R^a{}_b}$ and $\bar{R}^r{}_s$ are the curvature
two forms on $L$ and $S$, respectively.

Contracting the vector indices with the form indices and using 
$\bar{R}^r{}_s=\bar{e}^{r} \bar{e}_{s}$ yields the curvature scalar
\begin{equation}
R = R^L-\frac{\left( D-2\right) \left( D-3\right) }
{\Phi ^{2}} \left[ 1+\left( \nabla_{\al}\Phi \right) \left( \nabla^{\al}\Phi 
\right) \right] -2\left( \frac{D-2}{\Phi }\right) \left( \square \Phi 
\right),
\label{app:redcurv}
\end{equation}
where $\nabla$ is the covariant derivative with respect to the metric on $L$ 
and $\square=\nabla_\al\nabla^\al$. This -- together 
with $\sqrt{|g^M|}=\Phi^{(D-2)}\sqrt{-g^L}$ -- is the starting point of 
spherically reduced gravity formulated by a $2D$ effective 
action. Note that the generalization to continuous and negative dimensions $D$ 
is possible in (\ref{app:redcurv}) which leads to the subclass $b=a-1$ of the
models of (\ref{e23eact}).

Characteristic classes are independent of the metrical structure since they
depend solely on the topology, but typically they can be expressed as 
integrals over local quantities using index theorems. As an example we treat 
Euler and Pontryagin class in $D=4$. The latter can be expressed as
\eq{
P_4 = \frac{1}{8\pi^2}\int_M R^{mn}R_{mn} = 0,
}{app:pontryagin}
and it vanishes because $R^{ab}R_{ab}=0=R^{st}R_{st}$ and with 
(\ref{app:cur3}), (\ref{app:cur4}) also $R^{as}R_{as}$ yields no contribution.

The Euler class
\eq{
E_4 = \frac{1}{2(4\pi)^2}\int_M R^{kl}R^{mn} \eps_{klmn}
}{app:euler}
is non-trivial in general and can be expressed as a $2D$ integral over $L$: 
\meq{
E_4 = \frac{1}{2\pi}\int_L \eps_a{}^b \Big[R^a{}_b\left(1+\eta^{cd}
(\bar{e}_c\Phi)(\bar{e}_d\Phi)\right) \\
+2\left(\eta^{ad}(\bar{e}_c\bar{e}_d\Phi)\bar{e}^c+(\bar{e}_c\Phi)
\bar{\om}^{ac}\right)\left(\eta^{be}(\bar{e}_c\bar{e}_e\Phi)\bar{e}^c+
(\bar{e}_c\Phi)\bar{\om}^c{}_b\right)\Big].
}{app:redeuler}

\section{Heat kernel expansion}\label{appB}

Some basic properties of the heat kernel expansion are collected here
which are needed in the main text. More details
can be found in the monographs 
\cite{Gilkey:1994,Esposito:1997,Avramidi:2000,Kirsten:2001}.

In most of the quantum field theory problems one deals with
an operator of Laplace type. In a suitable basis such an
operator can be represented as:
\begin{equation}
A=-(g^{\mu\nu}\nabla_\mu\nabla_\nu +E) \,,\label{ebA}
\end{equation}
where $\nabla_\mu$ is a covariant derivative, and
$E$ is an endomorphism of a vector bundle (or, in simpler terms, a matrix
valued function). The connection in the
covariant derivative and the matrix $E$ may have gauge and
spin indices. We consider the oparator $A$ in arbitrary dimension
$D$.

The smeared heat kernel is defined by the equation
\begin{equation}
K(f,A,t)={\rm Tr} (f\exp (-tA)) \,,\label{ebK}
\end{equation}
where $f$ is a function, but more complicated cases with $f$ being
a differential operator may be considered as well \cite{Branson:1998ze}.
If the underlying manifold ${\mathcal{M}}$ has no boundary, there exists an
asymptotic series as $t\to +0$
\begin{equation}
K(f,A,t)\simeq \sum_{n=0}^\infty  a_n(f,A) t^{-\frac D2 +n} \,,
\label{ebas}
\end{equation}
where the coefficients $a_n$ are locally computable. This means that
they can be expressed as integrals of local polynomials constructed
from the Riemannian curvature, $E$, gauge field strength, and
covariant derivatives. On manifolds with boundaries half-integer
$n$ are also admitted.
A very important property is that numerical
coefficients in front of a monomial depend on the dimension $D$ 
via an overall factor $(4\pi)^{-D/2}$ only \cite{Gilkey:1994}. 
This last statement
follows immediately from writing the general form of such a coefficient
on a product manifold ${\mathcal{M}}={\mathcal{M}}_1\otimes S^1$ and assuming
complete triviality in the $S^1$ direction.

The heat kernel coefficients $a_n$ are known for $n \leq 5$
\cite{vandeVen:1997pf}. We find it instructive to present here the
calculation of $a_0$ and $a_1$ in order to make our
review self-contained, and to advertise a very powerful method
of such calculations. The first step is to write down all
possible invariants of an appropriate dimension. The mass dimension
of the operator $A$ is given by ${\rm dim}\, A=+2$. Therefore, 
${\rm dim}\, t = -2$.
The volume element has the dimension $-D$. All geometric invariants
(like e.g. curvature) have positive dimensions. The lowest dimension
($-D$) involves just the integral of the smearing function over the volume.
This explains why the expansion (\ref{ebas}) starts with $t^{-D/2}$.
Thus, the first two terms in (\ref{ebas}) must read  
\begin{eqnarray}
&&a_{0}(f,A)=(4\pi)^{-D/2}\int_{\mathcal{M}} d^Dx \sqrt{g} {\rm tr}
(\alpha_0 f) \,,\label{eb0gen} \\
&&a_1(f,A)
         =(4\pi)^{-D/2}\int_{\mathcal{M}} d^Dx \sqrt{g} {\rm tr} \left(
f(\alpha_1E+\alpha_2R)\right).\label{eb2gen}
\end{eqnarray}
where  ${\rm tr}$ denotes the finite-dimensional matrix trace. At this point 
$\alpha_i$ still are unknown constants which will be defined
by particular case calculations or through functional relations between
the heat kernels for different operators. The constant 
$\alpha_0$ follows from the well-know solution
of the heat equation in flat space:
\begin{equation}
\alpha_0=1 \,.\label{ebalp0}
\end{equation}

Let us consider now how the heat kernel changes under the conformal
transformations of the operator $A$ and by the shift by a function.
\begin{eqnarray}
&&\left. \frac{d}{d\epsilon} \right|_{\epsilon =0}
a_n(1,e^{-2\epsilon f}A)=(D-2n) a_n(f,A)\,.\label{ebv1}\\
&&\left. \frac{d}{d\epsilon} \right|_{\epsilon =0}
a_n(1,A-\epsilon F)=a_{n-1}(F,A)\,.\label{ebv2}
\end{eqnarray}
Here $f$ and $F$ are arbitrary functions. The
proof of these two properties is purely combinatorial. It uses differentiation 
of an exponential (\ref{ebK}) and commutativity under the trace.
From the equations (\ref{ebalp0}) and (\ref{ebv2})
\begin{equation}
\alpha_1=1\label{ebalp1}
\end{equation}
follows. A combination of the two transformations, 
$A(\epsilon,\delta )=e^{-2\epsilon f}(A-\delta F)$, allows to prove that
for $D=2(n+1)$
\begin{eqnarray}
&&0=\left. \frac{d}{d\epsilon} \right|_{\epsilon =0}
a_{n+1}(1,A(\epsilon,\delta ))\,,\nonumber \\
&&0=\left. \frac{d}{d\delta } \right|_{\delta =0}
\left. \frac{d}{d\epsilon} \right|_{\epsilon =0}
a_{n+1}(1,A(\epsilon,\delta ))
=\left. \frac{d}{d\epsilon} \right|_{\epsilon =0}
\left. \frac{d}{d\delta } \right|_{\delta =0}
a_{n+1}(1,A(\epsilon,\delta )) \nonumber \\
&&\qquad \qquad =
\left. \frac{d}{d\epsilon} \right|_{\epsilon =0}
a_n (e^{-2\epsilon f}F, e^{-2\epsilon f} A) \,.\label{ebv3}
\end{eqnarray}
The conformal transformations of the individual
invariants which may enter (\ref{ebv3}) must be defined.
They are perfectly standard in the ``geometry'' part:
\begin{eqnarray}
&&\left. \frac{d}{d\epsilon} \right|_{\epsilon =0} \sqrt{g} =
D f\sqrt{g} \,,\nonumber \\
&&\left. \frac{d}{d\epsilon} \right|_{\epsilon =0} R =
-2fR-2(D-1)\nabla^2 f \,.\label{ebcv1}
\end{eqnarray}
$E$ is transformed such that the operator $A$ is conformally covariant:
\begin{equation}
\left. \frac{d}{d\epsilon} \right|_{\epsilon =0} E = -2fE 
+\frac 12 (D-2)\nabla^2 f \,.\label{ebcv2}
\end{equation}
Note, that for the standard conformal (Weyl) transformations
the ``potential'' term $E$ transforms homgeneously, i.e. the second
term on r.h.s. of (\ref{ebcv2}) is absent.

Finally, the general expression (\ref{eb2gen}) is substituted
in the variational equation (\ref{ebv3}) for $D=4$. The result
\begin{equation}
\alpha_2 =\frac{1}{6} \label{ebalp2}
\end{equation}
completes the calculation of $a_1$.

Heat kernel methods became standard in quantum field theory
after the famous works by DeWitt \cite{Dewitt:1965}\footnote{
For the first time the heat kernel (proper time) methods were
used in quantum theory by Fock \cite{Fock:1937}.} where a different
calculation scheme was used. The approach we have presented here
goes back to the paper by Gilkey \cite{Gilkey:1975}. This approach appears
somewhat simpler, although is less ``algorithmic''
since one has to invent new functional relations appropriate for
a particular problem. The full power of this method has been demostrated
on manifolds with boundaries \cite{Branson:1990} (cf. also 
\cite{Vassilevich:1995we}   for minor corrections). 
With no other method a complicated calculation as the one for $a_{5/2}$ for 
mixed boundary conditions \cite{Branson:1999jz} is possible.

The last topic to be addressed is the relation between
the heat kernel and the zeta function of the same operator. It is clear
from the definitions (\ref{ehmdW}) and (\ref{ebK}) that
\begin{equation}
\zeta (s|f,A)=\frac 1{\Gamma (s)} \int\limits_{0}^{\infty} dt t^{s-1}
K(f,A,t) \,.\label{ebzK}
\end{equation}
This relation can be inverted,
\begin{equation}
K(f,A,t) =\frac 1{2\pi i} \oint ds \Gamma (s) \zeta (s|f,A) t^{-s} \,,
\label{ebKz}
\end{equation}
where the integration contour encircles all poles of the integrand.
The coeffcient in front of $t^p$ in the asymptotic expansion (\ref{ebas})
corresponds to the residue of $\Gamma (s) \zeta (s|f,A)$ at the point
$s=-p$. In particular,
\begin{equation}
a_{D/2}(f,A)={\rm Res}_{s=0} (\Gamma (s) \zeta (s|f,A))=
 \zeta (0|f,A) \,.\label{ebD2}
\end{equation}
For $D=2$, $A=-\Delta$, $E=0$ the equations (\ref{eb2gen}), (\ref{ebalp2})
and (\ref{ebD2}) provide the relation (\ref{ehmz0}) of the main text.








\end{appendix}

\clearpage



\lhead{REFERENCES}

\providecommand{\href}[2]{#2}\begingroup\raggedright\endgroup

\end{document}